\RequirePackage{fix-cm}

\documentclass[a4paper,reqno,12pt]{amsart}
\usepackage[text={5.9in,8.5in},centering,top=30mm]{geometry}
\usepackage{mathrsfs}

\DeclareSymbolFont{cmsymbols}{OMS}{cmsy}{m}{n}
\SetSymbolFont{cmsymbols}{bold}{OMS}{cmsy}{b}{n}
\DeclareSymbolFontAlphabet{\mtc}{cmsymbols}

\usepackage{graphicx}
\usepackage{mathptmx}      

\usepackage{bm}

\usepackage{amsmath,amsthm,amssymb}

\usepackage{times}

\usepackage{enumerate}

\usepackage{tabularx}

\usepackage[mathscr]{euscript}
\usepackage{eufrak}

\usepackage{stmaryrd}

\usepackage{pgf}
\usepackage{tikz}
\usetikzlibrary{patterns}

\usepackage[colorlinks=true,
            linkcolor=blue,
            urlcolor=black,
            citecolor=blue]{hyperref}

\providecommand{\D}{\mathbb}

\newcommand{\ii}{\mathrm{i}}

\newcommand{\dist}{\mathrm{dist}}
\newcommand{\Ord}[1]{\mathrm{O}\left(#1\right)}
\newcommand{\ord}[1]{\mathrm{o}\left(#1\right)}

\newcommand{\diam}{\mathrm{diam}}

\newcommand{\supp}{\mathrm{supp\,}}

\newcommand{\one}{\mathbf{1}}

\def\mcA{\mtc{A}}

\def\mcC{\mtc{C}}

\def\mcJ{\mtc{J}}

\def\mcX{\mtc{X}}

\def\mcZ{\mtc{Z}}

\newcommand{\cO}{\mathcal{O}}
\newcommand{\cXbar}{\overline{\mathcal{X}}}

\newcommand{\hal}{\widehat{\alpha}}

\newcommand{\hmu}{\widehat{\mu}}

\newcommand{\hp}{\widehat{\mathrm{p}}}

\definecolor{redd}{rgb}{0.95,0.2,0.2}
\definecolor{gris}{rgb}{0.9,0.9,0.9}
\definecolor{greenn}{rgb}{0.1,0.6,0.2}

\definecolor{cmgray}{rgb}{0.7,0.7,0.7}

\definecolor{cmblue}{rgb}{0.2,0.5,0.8}

\newcommand{\be}{\begin{equation}}
\newcommand{\ee}{\end{equation}}
\newcommand{\ba}{\begin{array}{l}}
\newcommand{\ea}{\end{array}}

\newcommand{\bal}{\begin{aligned}}
\newcommand{\eal}{\end{aligned}}

\newcommand{\baln}{\begin{align}}
\newcommand{\ealn}{\end{align}}

\newcommand{\ble}{\begin{lemma}}
\newcommand{\ele}{\end{lemma}}

\newcommand{\bco}{\begin{cor}}
\newcommand{\eco}{\end{cor}}

\newcommand{\bpr}{\begin{proposition}}
\newcommand{\epr}{\end{proposition}}

\newcommand{\bre}{\begin{remark}}
\newcommand{\ere}{\end{remark}}

\newcommand{\btm}{\begin{thm}}
\newcommand{\etm}{\end{thm}}

\newcommand{\bde}{\begin{definition}}
\newcommand{\ede}{\end{definition}}

\newcommand{\lcite}[2]{{\cite[#2]{#1}}}

\newcommand{\myset}[1]{\left\{ \, #1 \, \right\}}

\newcommand{\eu}{{\mathrm{e}}}

\newcommand{\ffi}{\varphi}

\newcommand{\all}{\forall\,}

\newcommand{\half}{\frac{1}{2}}
\newcommand{\shalf}{{\textstyle{\frac{1}{2}}}}

\newcommand{\eps}{\epsilon}

\DeclareSymbolFont{newfont}{OML}{cmm}{m}{it}
\DeclareMathSymbol{\Epsilon}{3}{newfont}{15}
\DeclareMathSymbol{\Varrho}{3}{newfont}{37}
\DeclareMathSymbol{\rro}{3}{newfont}{37}

\newcommand{\Const}{\mathrm{Const\,}}
\newcommand{\const}{\mathrm{const\,}}

\newcommand{\pr}[1]{\mathbb{P}\left\{ #1 \right\}}

\newcommand{\prsub}[2]{\mathbb{P}_{#1}\left\{ #2 \right\}}

\newcommand{\BIGpr}[1]{\mathbb{P}\Bigg\{ #1 \Bigg\}}

\newcommand{\esm}[1]{\D{E}\left[\, #1\, \right]}
\newcommand{\Bigesm}[1]{\D{E}\Big[\, #1\, \Big]}

\newcommand{\vertii}[1]{{\big\vert\kern-0.25ex\big\vert #1
    \big\vert\kern-0.25ex\big\vert\kern-0.25ex}}

\newcommand{\vertiii}[1]{{\big\vert\kern-0.25ex\big\vert\kern-0.25ex\big\vert #1
    \big\vert\kern-0.25ex\big\vert\kern-0.25ex\big\vert}}

\newcommand{\kap}{\varkappa}
\newcommand{\lam}{\lambda}
\newcommand{\om}{\omega}

\newcommand{\Bom}{\boldsymbol{\omega}}

\newcommand{\Bfq}{\pmb{\mathfrak{q}}}

\newcommand{\tth}{\theta}

\newcommand{\tthu}{\theta_{\Bu}}

\newcommand{\bp}{\beta_{+}}

\def\zp{\zeta_{+}}
\def\zm{\zeta_{-}}

\newcommand{\Lam}{\Lambda}
\newcommand{\Om}{\Omega}

\newcommand{\tom}{\widetilde{\omega}}

\newcommand{\tA}{\widetilde{A}}

\newcommand{\tV}{\widetilde{V}}

\newcommand{\Unif}{{\mathrm{Unif}}}

\newcommand{\ball}{\mathrm{B}}

\newcommand{\fa}{\mathfrak{a}}
\newcommand{\fb}{\mathfrak{b}}

\newcommand{\fm}{\mathfrak{m}}

\newcommand{\fp}{\mathfrak{p}}
\newcommand{\fq}{\mathfrak{q}}
\newcommand{\fs}{\mathfrak{s}}

\newcommand{\fB}{\mathfrak{B}}

\newcommand{\fF}{\mathfrak{F}}

\newcommand{\fu}{\mathfrak{u}}

\newcommand{\Bzeta}{\bm{\zeta}}

\newcommand{\Bt}{\mathbf{t}}

\newcommand{\Bc}{\mathbf{c}}
\newcommand{\Bd}{\mathbf{d}}

\newcommand{\BU}{\mathbf{U}}

\newcommand{\cF}{\mathcal{F}}

\newcommand{\cX}{\mathcal{X}}

\newcommand{\Bx}{\mathbf{x}}

\newcommand{\Bz}{\mathbf{z}}
\newcommand{\Bu}{\mathbf{u}}

\newcommand{\rc}{\mathrm{c}}
\newcommand{\rd}{\mathrm{d}}

\newcommand{\ry}{\mathrm{y}}

\newcommand{\rA}{\mathrm{A}}

\newcommand{\rY}{\mathrm{Y}}

\newcommand{\rC}{\mathrm{C}}

\newcommand{\rL}{\mathrm{L}}

\newcommand{\rP}{\mathrm{P}}

\newcommand{\rQ}{\mathrm{Q}}

\newcommand{\DC}{\mathbb{C}}
\newcommand{\DP}{\mathbb{P}}
\newcommand{\DR}{\mathbb{R}}

\newcommand{\DZ}{\mathbb{Z}}
\newcommand{\DT}{\mathbb{T}}
\newcommand{\DN}{\mathbb{N}}

\newcommand{\etal}{\emph{et al.}\xspace}

\newcommand{\cond}{\,\big|\,}

\newcommand{\tto}[1]{\smash{\mathop{\,\,\,\, \longrightarrow \,\,\,\, }\limits_{#1}}}

\newenvironment{hyp}[1]{
\vskip3mm\par\noindent
$\blacklozenge$\;\textbf{Hypothesis #1.}
\noindent}

\newcommand{\bhy}[1]{\begin{hyp}{#1}}
\newcommand{\ehy}{\end{hyp}}

\newtheorem{thm}{Theorem}[section]

\newtheorem{cor}{Corollary}

\newtheorem{proposition}[thm]{Proposition}

\newcommand{\beal}{\begin{equation}\begin{aligned}}
\newcommand{\eeal}{\end{aligned}\end{equation}}

\newcommand{\moins}{\setminus}

\newcommand{\gea}{\gtrsim}
\newcommand{\lea}{\lesssim}

\newcommand{\tlam}{\widetilde{\lambda}}

\def\llb{\llbracket}
\def\rrb{\rrbracket}

\newcommand{\sigal}{$\sigma$-algebra\xspace}

\newcommand{\btau}{{\Bar{\tau}}}
\newcommand{\oball}{{\overline{\ball}}}
\newcommand{\ooball}{{\overline{\overline{\ball}}}}

\newcommand{\ttzeta}{{\zeta}}

\newcommand{\esmk}[2]{\D{E}_{#1}\left[\, #2\, \right]}

\newcommand{\fr}{\mathfrak{r}}

\newcommand{\fukap}{\overline{\mathfrak{u}}^{(\varkappa)}}

\newcommand{\gamkap}{\gamma_\varkappa}

\newcommand{\tH}{\widetilde{H}}

\DeclareMathOperator*{\essup}{ess\,sup}

\def\mcH{\mtc{H}}

\def\mcS{\mtc{S}}

\def\hmcX{\widehat{\mtc{X}}}

\newcommand{\hy}{{\widehat{y}}}

\newcommand{\tW}{\widetilde{W}}
\newcommand{\tWin}{W}

\newcommand{\tWort}{W_L^{\perp}}

\newcommand{\trpar}[1]{\mathrm{tr}\left[\,#1\,\right]}

\newcommand{\prcXL}[1]{\prsub{\!\!\!\cX_{L}^{\!\!\perp}}{#1}}

\newcommand{\ra}{\mathrm{a}}
\newcommand{\rb}{\mathrm{b}}

\newcommand{\hlam}{\widehat{\lam}}

\newcommand{\bsigma}{{\Bar{\sigma}}}

\newcommand{\hLam}{{\widehat{\Lam}}}

\newcommand{\ttx}{\tilde{x}}

\newcommand{\fmbar}{\overline{\mathfrak{m}}}

\usepackage{amssymb}
\usepackage{amsmath}
\usepackage{mathrsfs}

\usepackage{xspace}

\usepackage{epsfig}
\usepackage{enumerate}
\usepackage{graphicx}

\usepackage[title]{appendix}

\numberwithin{equation}{section}

\newtheorem{lemma}{Lemma}[section]

\newtheorem{remark}{Remark}[section]

\newtheorem{definition}{Definition}[section]

\begin{document}

\title[Smoothness of DoS under long-range interactions]
{Universality of smoothness of Density of States
\\in arbitrary higher-dimensional disorder
\\under non-local interactions I.
\\From Vi\'{e}te--Euler identity to Anderson localization}



\author{Victor Chulaevsky}

\address{Universit\'{e} de Reims\\
D\'{e}partement de math\'{e}matiques\\
51687 Reims Cedex, France}

\email{victor.tchoulaevski@univ-reims.fr}



\date{\today}

\maketitle

\begin{abstract}
It is shown that in a large class of disordered systems
with non-degenerate disorder, in presence of non-local interactions,
the Integrated Density of States (IDS) is
at least H\"{o}lder continuous in one dimension and universally infinitely differentiable
in higher dimensions. This result applies also to the IDS
in any finite volume subject to the random potential induced by an ambient, infinitely extended disordered
media. Dimension one is critical: in the Bernoulli case,
within the class of exponential interactions, the IDS measure undergoes continuity phase transitions,
from absolutely continuous to singular continuous behaviour (the singularity in the latter case was known before).
The continuity transitions do not occur for
sub-exponential or slower decaying interactions, nor for $d\ge 2$.
Technically, the case of polynomial decay is the simplest one.

The proposed approach provides a complement to the classical Wegner estimate which says, essentially,
that the IDS in the short-range models is at least as regular as the marginal distribution of
the disorder. In the models with non-local interaction the IDS is actually much more
regular than the underlying disorder, which can even be discrete, due to the smoothing effect of multiple convolutions.
In turn, smoothness of the IDS is responsable for a mechanism complementing the usual Lifshitz tails phenomenon.

It is also shown that the  disorder can take various forms
(e.g., substitution or random displacements) and
need not be stochastically stationary (as in Delone--Anderson or trim\-med/crooked Hamiltonians, for example);
this does not affect the main phenomena observed already in the simplest setting.

Contrary to the situation with the usual lattice Bernoulli--Anderson Hamiltonians,
the proof of Anderson localization in the models with infinite-range interaction follows in a fairly
standard way from the main bounds on the finite-volume IDS; taking into account a considerable
size of the current text, the proof of localization is now presented in a separate work.
Another distinction from
the approach developed by Bourgain and Kenig for the continuous Bernoulli--Anderson Hamiltonians,
and later extended by Germinet and Klein to arbitrary (locally IID) disorder,
is that all nontrivial marginal distributions
are treated in a unified way, via  harmonic analysis and without reduction to an
embedded Bernoulli model, thus keeping potential
benefits from less singular forms of underlying disorder.

Long-range models have an amazingly large number of connections to several classical
problems of harmonic analysis, probability theory, dynamical systems and number theory.

\vskip7mm

\end{abstract}

\setcounter{tocdepth}{3}
\tableofcontents

\emph{Compared to the first version of this text (28.04.2016), a number of changes have been made recently.
The most significant one concerns the proof of smoothness of the finite-volume IDS
(in the original version we proved smoothness of the marginal distribution of
the cumulative potential $V(x,\om)$ and Wegner estimates). The section on ILS estimates
was rendered more complete, while the localization analysis
was moved to a separate manuscript \cite{C16f} due to a considerable size of the present paper.}

\section{Introduction}
\label{sec:intro}

Since the discovery of the phenomenon of quantum localization by Philip W. Anderson \cite{A58},
a certain number
of "simplifying" assumptions were made both in physical and mathematical models of disordered media. Probably,
the most important of all are those concerning the nature of interactions between the quantum objects involved.
Specifically, one has to distinguish two kinds of interactions:

\begin{itemize}
  \item between the mobile objects (e.g., charged particles), and the "external" sources of disorder (e.g., heavy ions);

  \item between the above mentioned mobile objects themselves.
\end{itemize}

The latter interactions have been the main subject of recent physical and mathematical works over the last
decade; cf. e.g. \cite{GMP05,BAA06,CS08,CS09a,CS09b,AW09,CBS11,CS13,FW15,CS16,C16b}.
Importance of this new problematic initiated, on the mathematical level,
in 2003 at the Isaac Newton Institute (Cambridge University, UK)
has been recognized at the XVIth Congress on Mathematical Physics (2010)
and reported by Aizenman and Warzel \cite{AW10}.
Speaking of rigorous aspects of the problem, a number of questions in this new area of
spectral theory of random operators still remain wide-open and challenging; they will not be discussed
in the present paper.

\vskip2mm

The main topic of this work is the impact of the non-local physical nature of interactions
between the mobile quantum objects and an ambient disordered classical environment,
on the qualitative characteristics of the Integrated Density of States (IDS)
and, where applicable, the Density of States (DoS). The existence of the DoS
(and more generally, regularity of the IDS) is usually derived in higher-dimensional models from
local regularity of the marginal distribution of disorder, via Wegner's estimate and its generalizations.
Nothing can ever prevent mathematicians from assuming anything, yet it is quite natural to ask:
"\textbf{Where does a \emph{regular} disorder come from?}" A short answer to this question, which was at the
origin of the present work, is that it is simply hard to avoid, for it emerges in a fairly universal
way from virtually any kind of disorder more or less evenly distributed in the configuration space.
The main mechanism is the smoothing effect of multiple convolutions,
and the principal, very convenient mathematical tool for analyzing this effect is harmonic analysis
of probability measures.

\vskip2mm

The one-dimensional systems are set apart in this respect, since the regularity of the IDS in the case
of strongly singular local disorder (e.g. 1D Bernoulli) is known to follow by Hilbert transform
from that of the Lyapunov exponents, but deep inside, one finds the same regularizing
effect of multiple convolutions. The Riccati dynamics for the so-called Pr\"{u}fer phase is of course nonlinear,
but the Pastur--Figotin argument \cite{PF92} shows that in the particular case of weak disorder
the linearized dynamics alone leads to an asymptotically exact formula for the positive Lyapunov exponents,
and even in the correlated case \cite{CSp95} the asymptotic behavior can be derived, in fact,
from the linear harmonic analysis. All this can be done for the local models of disorder: the Lyapunov solutions,
from which the Green functions (GFs) are built, provide in 1D convenient "test functions" accumulating the effect
of multiple local fluctuations (be those linearized or not). Revealing a similar mechanism in higher dimension,
with local (e.g., lattice IID) disorder seems much harder a problem, but the situation changes radically
as soon as we turn to physics and recall ourselves that fundamental interactions are NOT local. In particular,
the Coulomb interaction between charged particles has infinite range.

Sometimes I refer
to the mobile particles as electrons, but the detailed discussion of the real physical processes,
especially on the level of second quantization, is most certainly beyond the scope of this
paper, and occasional use of physical terminology is intended for terminological brevity only.
The reality is of course much richer. Depending upon a specific physical model, the mobile objects
may carry charge and spin either together or separately.

Speaking of electrically charged particles
subject to electrostatic interactions (which can of course be complemented by magnetic fields),
the fundamental Coulomb interaction is extremely slowly decaying, but in a large sample
of a heterogeneous media, composed of a huge number of more or less mobile charges,
it actually manifests itself only in a dampened, "screened" form.
The screening effects in solid state media have been since several decades
an inalienable part of any physical work realistically describing quantum many-body systems.
On the other hand, a vast majority of mathematical papers on Anderson localization operate with
local models of disorder, starting with the pioneering papers on localization in one dimension
(Goldsheid, Molchanov and Pastur \cite{GMP77} in $\DR^1$; Kunz and Souillard \cite{KS80} in $\DZ^1$)
and in higher dimension (Fr\"{o}hlich and Spencer \cite{FS83}: exponential decay of Green functions;
Fr\"{o}hlich, Martinelli, Scoppola and Spencer \cite{FMSS85}: pure point spectrum with exponentially decaying
eigenfunctions).

In a few exceptions, the extended, non-local nature of the potential
has been in prior works more of a nuisance, or perhaps an additional technical challenge, overcoming which would
warrant certain sacrifices in the strength of the localization results one aimed to obtain.
In fact, the first rather general result on correlated Gaussian potentials was obtained by von Dreifus and Klein
\cite{DK91} shortly after their reformulation of the energy-interval, or variable-energy, MSA (VEMSA)
in the frequently cited paper \cite{DK89}. Later on, Kirsch et al. \cite{KirStoStolz98a}
considered more general (non-Gaussian) marginal distributions. See also  recent works
\cite{Krug12,KarPeyTautVes15,Ves10,Ves14} and references therein. It seems only fair to
explore the true role of non-local interactions (apparently, the only ones known in physics
of solid state) in a broader context. This is precisely the main goal of the project
the first part of which is presented in this paper. We shall see that the infinite range
of interaction is indeed much deeper a subject than a  technical mathematical nuisance.

\vskip2mm

We argue that a traditional reduction of the environment of a
finite volume to the PDE-type boundary conditions hides a significant part
of the story, and that the picture becomes substantially more complete in the traditional setting
of statistical mechanics, where the environment acts as a thermal bath. It proves
fairly instructive to decompose the integrated density of states into two components which, for
the lack of better words and following a mechanical metaphor, we call respectively "tidal" and "ripple"
components.  In the daily movements of the ocean's level on a sloping shore, local
perturbations (wind, irregularities of the beach) determine a perceptible profile of the water surface,
but the principal movements themselves are the result of incommensurably weaker fluctuations of gravitational forces
from extremely remote sources; forces which would be unable to move water, say, by a meter or two in vertical direction,
in a strictly isolated
container of the size comparable to the beach, e.g., in a  lake, so they have to act through a much
larger external volume. Imperceptible per se,
those gravitational forces produce an easily perceptible by eye movement of water, back and forth, on an almost
horizontal, yet sloping beach.
Weak or not, it is the tidal mechanism which determines a considerable periodic evolution of the coastal area,
that a local wind could not produce.

This is more than just a qualitative metaphor: the gravitational potential, like Coulomb, is a slowly
decaying function in the sense that its gradient decays faster than the function itself,  so
distant sources produce ``almost flat'' (yet nowhere flat) fluctuations.
We discuss this aspect briefly in Section \ref{sec:two-point} dedicated to the analysis of regularity of the
two-point correlation measures,
but it may have important implications for the many-body localization phenomena,
in the light of some technical issues raised in \cite{C16b}.

\vskip3mm

$\blacklozenge$ This text is quite long and probably not easy to read; perhaps it is worthwhile to single out two
techniques most useful from a pragmatic point of view. First, for a relatively simple proof of H\"{o}lder-type
Wegner estimates, the integral bounds based on the techniques due to Wiener and Wintner
\cite{WW38} (closely related to \cite{Wien1924})) are quite useful.
In a broader context, these techniques have been used and further developed by Strichartz
\cite{Stri90b,Stri90a}).
Secondly, in the most relevant models with weak (polynomial)
screening, Wintner's technique \cite{Wint1935} is both very simple and efficient for the proof of smoothness
estimates. $\blacktriangleleft$

\vskip3mm

$\blacklozenge$ Bourgain and Kenig \cite{BK05} proved a remarkable eigenvalue concentration (EVC)
estimate for Bernoulli--Anderson
Hamiltonians in $\DR^d$, which was later extended to arbitrary nontrivial disorder
by joint efforts of Aizenman, Germinet, Klein and Warzel \cite{AGKW09}.
Their approach was based on a combinatorial argument (the Sperner lemma), but a reflex to saying "$\sqrt{N}$"
among probabilists would certainly be even more Pavlovian than to
saying "\emph{Jingle ...}" in a preschool, on some 23rd December. In a way, the analysis given below
("thermal bath estimates") justifies that reflex.
Observations made here evidence that the interactions of infinite range
(the only physically relevant ones, anyway) actually provide a music easy to sing to, especially
with the help of harmonic analysis.
Singing the same lyrics but \emph{a capella} remains an intriguing mathematical challenge,
regardless of any physical applications. $\blacktriangleleft$

\vskip1mm

We thus come to a more quantitative discussion of a model of the forces originating far away
from a locus where their effects are to be studied.

\vskip1mm
The strongest form of screening occurs in 3D systems when charged particles
are highly mobile, e.g., in plasma; the Debye--screened Coulomb potential originating from a given
local source decays exponentially fast at large distances $r$, which have to be larger than
some characteristic length, so as to enable several layers of induced waves of concentration
of positive and negative charges to be created around the aforementioned remote source.
The simplest approach relies on the classical statistical physics.
Even in this case, screened Coulomb potential in
dimension $d\le 2$ is slowly decaying.

It was realized by physicists that the classical approximation results in an oversimplified and
even qualitatively inaccurate picture of screening, particularly in solid state media.
In more accurate models, the correction terms are no longer obtained
by "commutative" probabilistic analysis but require a quantum description, the choice between Fermi--Dirac
and Bose--Einstein quantum statistics, and complex diagrammatic techniques.
Also, a \emph{quantum} charged particle is not a point charge, and linear approximation
to the Gibbs distribution is only an approximation.
More importantly, one has to consider a
full-fledged quantum many-body problem to achieve a good agreement with experiment.
As a result, one has not a universal behaviour, but various forms of screening.
The response of a large sample to a single source causes the so-called
Friedel oscillations (cf. \cite{Fried1952,Fried1958,LanVos1959,KohnVos1960}), observable
experimentally,
and the quantitative parameters, first of all the decay rate of the screened electrostatic
potential from a given source, strongly depend upon the shape
of the Fermi surface
of the mobile particles responsible for the screening.

With these observations in mind, we shall explore various decay rates of the effective (screened)
potential produced by heavy "ions" forming a spatial grid, periodic or not; these will range
from the strongest (exponential) to the slowest power-law ones, just barely
summable.

Below I am going to focus mainly on media of spatial dimension strictly higher than one, for two principal reasons.

\vskip2mm

$\bullet$ Firstly, from the perspective of applications to Anderson localization, the one-dimen\-sional
models are understood to a much greater extent than their higher-dimen\-sional counterparts.
The specifically 1D mechanisms, treated in terms of "back-scattering" in physical
approaches, or with the help of products of random matrices, in the rigorous mathematical works,
result in a much more complete and clear picture.

\vskip2mm

$\bullet$ Secondly, from the perspective of the continuity phase transition of the IDS which we are going
to describe, it will become clear that for many intents and purposes, sufficiently "thick"
quasi-one-dimensional media, i.e., those extended in one direction and having a finite cross-section,
are much closer to higher-dimensional samples than to single-channel linear chains. More precisely,
the dimensional threshold for the continuity phase transition -- for a given exponential decay rate
of the screened potential -- is encountered already within the class of quasi-1D strips, for
the cross-sections large enough. From this point of view, the macroscopic wires already have
a cross-section very large in microscopic (atomic) units.

\vskip2mm

However, 1D systems are certainly worth a thorough investigation. Even a brief familiarization with
physical literature, theoretical and experimental, is useful and can be recommended, to see
that some mathematical issues, requiring in this paper a fair amount of space and efforts,
are perhaps just that -- mathematical ghosts from the land of might-have-been.
Specifically, it seems logical to investigate the IDS continuity
phase transition in low-dimensional media (viz. in 1D or in $1 < d \approx 1$), but our analysis evidences
that the transition to singularity requires nothing less than \emph{exponential} screening, and sufficiently
\emph{strong} one. The reader can see, e.g., in the
works by Gabovich \etal \cite{GIPR78} or
by Petrashov \etal \cite{PetAntNil91}
how ``strong'' a 1D screening can be in physical reality ...

At the same time, I would like to stress that the non-local tidal effects of disorder on
the regularity of the DoS (or IDS) suggest that
the dimensionality parameter has to be properly defined in models with heterogeneous
structure, e.g., in the localization problem in a surface (or a specially designed
internal) layer on a 3D substrate, or in a quasi-1D channel on the surface or inside
a 2D/3D sample. While the quantum tunneling effects for the mobile agents my be limited to a linear sub-manifold
of lower dimension (or a thin neighborhood thereof), the tidal DoS may or might be strongly influenced
by the disorder in the ambient sample of higher dimensionality.

\vskip2mm

In models with a low concentration of "ions" creating a specific, gap-isolated energy band,
this concentration  itself may also become an important parameter near the critical
point of the continuity phase transition: the decay rate of the screened potential is to be
compared to the typical distance between nearest relevant loci.

\vskip2mm

More generally, the predictions concerning the continuity phase transition of the IDS are scale-dependent,
as they result from a renormalization group (RG) type analysis, hence the effects become perceptible
and sharp only beyond some minimal scale. In some mesoscopic systems, their size may or might be insufficient
for the RG limit to give the right answer.

\vskip2mm

$\blacklozenge$ In physics, there is a number of characteristic lengths related to exponentially decaying
functions: instead of $f(x) = \eu^{- a \|x\|}$ with $a>0$, it is customary to write
$f(x) = \eu^{-\|x\|/\xi}$, where $\xi$
describes the distance such that "\emph{for $\|x\|\gg \xi$, $f(x)$ is essentially nothing}".
This may be true in many realistic situations, but the key equations of this work, \eqref{eq:Euler}
and \eqref{eq:Bernoulli}, provide an instructive example of what  can be
the difference between an infinite series with
uniformly bounded, exponentially  decaying terms and \emph{any} of its partial sums.
Somehow, a measure supported by a finite number of atoms is ``slightly'' less regular than an a.c. measure;
in higher dimensions, as we shall see, the latter even has a bad habit to become infinitely smooth.

\vskip3mm
\noindent
\textbf{\emph{Disclaimer.}}
\emph{This paper focuses primarily on the fluctuations of the finite-volume IDS induced by
a random media and on the regularity of their probability distribution, rather than on
an exact form of the IDS. The main subject is therefore not the same as in many physical works.}

\emph{The physical mechanisms of screening are not analyzed.
The effects of a given interaction $\fu$ are studied regardless of whether or not it can
occur in realistic models of a given dimensionality $d$, although the
$d$-dependence of regularity properties of the cumulative potential and of the IDS is studied.
The main goal is to find out how  the most regular forms of disorder can emerge from the most singular ones,
under the most difficult conditions.
}

\emph{The list of bibliographical references, although it is rather long, is quite possibly incomplete, despite
my best efforts. In three words, an explanation but not an excuse, is: paid online access.  }

\section{Vi\'{e}te--Euler identity and smoothness of IDS: there and back again}
\label{sec:Euler.Viete}

\subsection{Introductory remarks. Notation. Alloy transform}

In presence of non-local interactions, one has to distinguish two kinds of potentials:
\begin{itemize}
  \item the "source" potential, described by the amplitudes at the origin points;

  \item the "target", cumulative potential registered at each point of the space.
\end{itemize}
We do not discuss the self-consistent, many-body models, so the basic disorder comes from
the immobile sources, the configuration of which is to be determined in the framework of statistical physics;
we assume the sources to be stochastically independent.
One possible model is the so-called alloy potential (displacements models will be briefly
discussed; they can be treated in essentially the same way),
$$
x \mapsto \sum_{y \in \mcZ} \fq_y \fu(x-y),
$$
where $\mcZ$ is a countable subset of the configuration space $\mcX$, e.g., of $\mcX=\DZ^d, \DR^d$.
For definiteness, we will consider $\mcX=\DZ^d$. The registered
cumulative potential, as a function on $\mcX$, is defined through a linear mapping,
which can be called alloy transform,
\be
\label{eq:def.BU}
\BU:\, \Bfq \mapsto \BU[\Bfq] = V\,, \;\; V:\,\DZ^d \to \DR \,,
\ee
where
\be\label{eq:def.BU.2}
V(x) = (\BU[\Bfq])(x) = \sum_{y\in\DZ^d} \fu(y-x) \fq_x.
\ee
The interaction potential $\fu(\cdot)$ will always be assumed absolutely summable,
$$
\sum_{x\in\DZ^d} |\fu(x)| \le C < \infty,
$$
and
nonnegative\footnote{In physical models, correlations can be sign-indefinite. We usually deal with
absolute amplitudes, but even these can be somewhere closer to $0$ than in average. It will be clear from
our analysis that exclusion of some radii is harmless for the main phenomena.
A more detailed analysis will be carried out in a forthcoming work.},
chosen from the class of power-law or (sub-)exponential functions, for we shall need
lower bounds on the decay of $\fu(\cdot)$, too. These notations will be used in the context of deterministic estimates
and statements, to avoid confusion with probabilistic arguments where $\fq$ are replaced by random variables,
assumed IID in this paper, forming a random field on $\DZ^d$ relative to a probability space $(\Om,\fF,\DP)$.
To keep parallels with the deterministic setup, we denote $\om = \{\om_x, \, x\in\DZ^d\}$ samples
of the random field of amplitudes $\om_x$;  the potential registered at a site $x$
has the same linear-algebraic form $V(x;\om) = (\BU[\om])(x)$, with $\om$ replacing $\Bfq$ in
\eqref{eq:def.BU.2}.

\vskip2mm

To reduce the number of auxiliary constants in intermediate statements and calculations,
we often use a standard notation $f(s) \asymp g(s)$ for functions of an integer
or real parameter $s$, usually in the context where $s\uparrow +\infty$ or $s\downarrow 0$, meaning that
$C_1 g(s) \le f(s) \le C_2 g(s)$ for some $C_1, C_2\in(0,+\infty)$.
Respectively, $f(s) \gtrsim g(s)$ will stand for $f(s) \ge C_2 g(s)$.

\vskip2mm

Fixing a point $x\in\mcX$, we come to the analysis of regularity of the probability distribution $\nu_x$ of
$V(x,\om)$. Assuming that $\om_y$ are IID with common probability measure $\mu$,
$\nu_x$ is the image of a transform of $\mu$ parameterized by
\begin{enumerate}
  \item a countable subset $\mcZ\subset \mcX$,

  \item a function $\fu:\,\DR_+\to \DR$,
\end{enumerate}
which provides an interesting generalization, and not just an abstract one, of the theory of random series, closely
related to the theory of self-similar measures. Considering all $x\in\mcX$ at once, or in a bounded domain,
we encounter an even more intriguing problem for the $(\mcZ,\fu)$-parameterized transform of a
measure $\mu$ into a random field $V(x,\om)$ on $\mcX$. Discussion in Section \ref{sec:two-point} barely scratches the surface
of the latter problematic.

\vskip1mm

As functions on the lattice, both $\Bfq$ and $\om$ will be assumed uniformly bounded; in the case of $\Bfq$
this is a non-ambiguous statement (in other words, $\Bfq$ will be assumed elements of $\ell^\infty(\DZ^d)$,
and even of a finite ball at the origin thereof), while $\om$ requires a bit of formalities:
the uniform boundedness is to be assumed with probability one. Alternatively, we can simply define
$\Om = [0,1]^{\DZ^d}$.


With these remarks, $\BU$ is well-defined on all admissible $\Bfq$ or $\om$, considered as elements of
$\ell^\infty(\DZ^d)$. We will have to control the dependence of the image
$\BU(\Bfq)$ (or, respectively, $\BU(\om)$) upon the values $\fq_x$ (resp., $\om_x$) inside and outside
some finite balls $\ball$ in $\DZ^d$. To this end, we canonically inject
$\ell^\infty(\ball), \ell^\infty(\ball^\rc) \hookrightarrow \ell^\infty(\DZ^d)$ by zero-extensions, and note that
for $\Bfq = \Bfq_{\ball} + \Bfq_{\ball^rc}$, where $\Bfq_{\ball} \cdot \Bfq_{\ball^rc} \equiv 0$
as function on $\DZ^d$,   one has $\BU(\Bfq) = \BU(\Bfq_{\ball}) + \BU(\Bfq_{\ball^\rc})$, but of course
there is no reason in general for $\BU(\Bfq_{\ball}) \cdot \BU(\Bfq_{\ball^\rc}) = 0$. Indeed, assuming
for example that $\fu$ is strictly positive everywhere, one has $\BU[ \delta_x]$ also strictly positive everywhere,
with $\delta_x$ being the lattice delta-function at an arbitrary point $x$.


The infinite range of the single-point potentials (scatterers) is certainly a double-edged sword, as the reader
will see on a number of occasions. However, one thorny problem of rigorous Anderson localization --
incomplete covering and an inevitable recours to some form of the Unique Continuation Principle, alas,
unavailable in general discrete models -- simply has no \emph{raison d'\^{e}tre} in presence of
realistic, long-range
scatterers.


The reverse of the medal starts with the non-local dependencies between the events referring to
localization (insufficient/no localization) in distant finite domains, or proximity
of local spectra in distant domains, possibly leading to a long-distance tunneling.


The former issue had been addressed long ago by Kirsch \etal \cite{KirStoStolz98a} who proposed one possible
way around this problem.
As to the latter, this is one of the instances where the infinite range of interaction
proves salutary, and transforms even the most singular nontrivial disorder distribution into a highly regular one,
thanks to multiple convolutions.

\subsection{Vi\'{e}te--Euler identity, Bernoulli alloys, and dynamical systems}

In the sixteenth century\footnote{According to different sources, in 1579 or in 1593.}
Fran\c{c}ois
Vi\'{e}te\footnote{Fran\c{c}ois Vi\'{e}te, or Fran\c{c}ois Viette, or Franciscus Vieta (1540--1603). His last
mathematical work
"\emph{Opera mathematica ...}" had remained unfinished, and was published only in 1646 by Frans van Schooten.}
discovered a remarkable identity
\be
\frac{2}{\pi} = \prod_{k=1}^{\infty} \cos\left( \frac{\pi}{2^k}\right)
\ee
which was generalized two centuries later by Leonhard Euler:
\be\label{eq:Euler}
\frac{\sin x}{x} = \prod_{k=1}^{\infty} \cos\left( \frac{x}{2^k}\right);
\ee
the latter follows by simple arguments
from $\sin x = 2 \sin\left( \frac{x}{2}\right) \cos\left( \frac{x}{2}\right)$.
It is the opening topic of Mark Kac' book \cite{Kac59}.
The Vi\'{e}te--Euler identity provided long ago a bridge between two areas of mathematical analysis to emerge much later:
harmonic analysis  and fractal measures closely related to dynamical systems. Moreover, it makes unnecessary
a delicate analysis of a critical model we are going to discuss a bit later.

Curiously, another remarkable elementary identity, figuring in Mark Kac' book as Problem 5 (Chapter 1), is
closely related to one of the
cornerstones of a simple and very efficient smoothing technique used in an uncountable number of
works on asymptotic formulae for the probability distribution functions and/or densities of normalized sums of independent
random variables (identical or not); a topic we shall also come across in the discussion of "smoothness"
of finite-volume IDS (see Section \ref{sec:concluding}). The asymptotic expansions obtained in this way
are widely used in statistics in general and in risk management, in particular.

As we shall see, answers to some tough questions appearing in spectral analysis of random Hamiltonians
with long-range interactions can be found in the harmonic
analysis of probability measures, and the specific form
of some questions may bring new motivations to this classical area of mathematical analysis.

A  prototypical form of the main mechanism we are going to exploit can be seen
from the usual dyadic expansion of a real number
\be\label{eq:Bernoulli}
[0,1] \ni \om = \sum_{k=1}^\infty \frac{ \om_k}{2^k}
\ee
establishing an "almost" bijective isomorphism between $[0,1]$ equip\-ped with the Lebesgue
measure and the set of infinite words $(\om_1, \om_2, \ldots) \in \Om = \{0,1\}^\DN$ endowed with
the structure of a measure space $(\Om, \fB^{\Om})$ ($\fB^{\Om}$ is the cylinder sigma-algebra rendering
measurable all projections $\om \mapsto \om_i$), with the product measure $\DP^\Om$ characterized by
$$
\all n\ge 1 \quad \DP^\Om\{ \om:\, \om_n = 0\} = \DP^\Om\{ \om:\, \om_n = 1\} = \half.
$$

The RHS of \eqref{eq:Bernoulli} can be interpreted as an alloy-type potential on $\DN$ with
exponentially decaying scatterer potential $\fu:\, r \mapsto 2^{-(r+1)}$, symmetric $(\half, \half)$ Bernoulli
distribution of the scatterers amplitudes, and evaluated at the origin. The above mentioned isomorphism
transforms therefore the most singular nontrivial local disorder distribution into a perfect
Lipschitz continuous one, with compactly supported density bounded by $1$.
The LHS of the Vi\'{e}te--Euler identity \eqref{eq:Euler} is the characteristic function (= inverse Fourier transform) of the
probability distribution of $\om$ (i.e., of the Lebesgue measure on $[0,1]$), while its RHS
expresses it as the characteristic function of the sum of independent r.v.
related to $2^{-k}\om_k$ from \eqref{eq:Bernoulli} by a simple affine transformation:
$
\tom_k = 2\om_k - 1 \in\{-1, +1\}.
$
A more symmetrical
alloy model, on the entire lattice $\DZ^1$, produces by independence a convolution of two uniform
distributions, resulting in an even better -- globally continuous -- compactly supported density.

The relation between the admissible values of the individual amplitudes, $0$ and $1$
(or rather the distance between them) and the precise decay exponent of the
function $\fu:\, k\mapsto 2^{-k}$, is crucial for the regularity of the induced single-site measure.
Taking $0$ and $a$  with $a>1$ results
in a Cantor set supporting the infinite convolution measure, for there are obviously gaps in the set of values
of the sums $\sum_n \om_n 2^{-n}$. For example,
$$
0 \cdot \half + \sum_{n\ge 2} \frac{a }{2^{n}} < a \cdot \half +  \sum_{n\ge 2} \frac{0}{2^{n}}   \,.
$$
However, even in such a case the resulting measure is
(singular\footnote{Of course, it is not the presence of gaps by itself which implies singularity; a Cantor set may have
positive Lebesgue measure. Here the gaps are "too big", so the support has zero Lebesgue measure.}) continuous,
even H\"{o}lder continuous. Moreover, a well-known example
(cf., e.g., \lcite{Feller66}{v.2, Section V.4(d)}) shows that the convolution of two singular  Cantor
measures can be a.c. (Lebesgue measure on an interval).

The problem of decay (and where appropriate, decay rate) at infinity of the Fourier
transform/coefficients\footnote{The term "coefficients" was actually introduced by Fran\c{c}ois Vi\'{e}te.}
of a probability measure on $\DR$ (or on $[0,1]$) has a long and rich  history.
It all starts in 1854 with Riemann's proof of decay at infinity of the Fourier coefficients of any
periodic Riemann-integrable function on $\DR$; Lebesgue extended this result to Lebesgue-integrable functions.
A systematic study of Borel \emph{measures} on the torus $\DT = \DR/\DZ \cong [0,1)$ with
decaying Fourier coefficients,
$$
\hmu(n) = \int_\DT \eu^{-\ii n t} \, d\mu(t) \,,
$$
was carried out in 1920s by Rajchman \cite{Rajch22,Rajch28};
such measures have been called Rajchman measures; this class
contains all a.c. measures.

Actually, Menshov ("Menchoff" in the French-style transliteration used in many of his works)
constructed in Ref. \cite{Men1916} an example of a singular Rajchman measure in 1916, precisely one century ago,
although the term "Rajchman measure" was not coined yet at that time.

Shortly after that  (in 1918), Riesz introduced in \cite{Riesz1918} what is called today
Riesz products,
$$
x \mapsto -x + \lim_{n\to\infty} \; \int_0^x \prod_{k=1}^n \big( 1 + \alpha_k \cos(m_k t) \big) \, dt,
$$
with $\alpha_k\in[-1,1]$. When $(m_k)_{k\ge 1}$ is a rapidly growing sequence of positive integers,
the Fourier
coefficients
are not $\ord{n^{-1}}$.

In 1920 Neder \cite{Neder1920}, answering a question raised by Riesz \cite{Riesz1918},
proved that every Rajchman measure is continuous.

Iva\v{s}\"ev--Musatov \cite{IvM52,IvM57} proved that the Fourier coefficients $\hmu(n)$ of
continuous measures $\mu$ mutually singular with
the Lebesgue measure are dominated by all functions of the form
$$
r(n) = \big( n \, \ln n \, \cdots \, \underbrace{\ln \ln \cdots \ln}_{p\text{ times}} n  \big)^{-1/2} \,, \quad
p\ge 1.
$$

By Jessen--Wintner theorem \lcite{JW1935}{Theorem 11}, an infinite product
\be
\label{eq:WW1935}
\ffi(t) = \prod_{k=1}^\infty \cos\left( r_k t \right),
\ee
giving the characteristic function of a random series $S(\om) = \sum_{k\ge 1} r_k X_k(\om)$,
with IID symmetric Bernoulli r.v. $X_k(\om)\in\{-1, +1\}$, is well-defined under
the assumption $\sum_k r_k^2 \equiv \sum_k \esm{ (r_k X_k)^2} < \infty$ (cf. Kolmogorov's three-series
theorem \cite{KhiKolm1925}, e.g. in \lcite{Feller66}{Section IX.9}),
and in this case $S(\cdot)$ has either purely s.c. or a.c. distribution.

Jessen and Wintner \lcite{JW1935}{Section 6} give an instructive set of
examples of random series
of scaled symmetric Bernoulli r.v. with the characteristic functions \eqref{eq:WW1935}.
In particular, Example 4 corresponds to the series $\sum_{k\in\DZ} 2^{-k} X_k$, with compactly supported bounded
density $\rho(x) = \left(\half - \frac{1}{4}x\right) \one_{[-2,2]}(x)$, and in Example 5 one has a series over
$(\DN^*)^2$,
$$
S^{(2)}(\om) = \sum_{k=1}^\infty \sum_{l=1}^\infty 2^{-k-l} X_{k,l}(\om),
$$
again with IID symmetric Bernoulli $X_{k,l}$. The authors of \cite{JW1935} point out that
$$
\ffi_{S^{(2)}}(t) = \prod_{k=1}^\infty \ffi_{S^{(1)}}(2^{-k}t)
$$
with $S^{(1)}(\om) = \sum_{l=1}^\infty 2^{-l} X_{l}(\om)$,
thus the fast decay of $\ffi_{S^{(2)}}(t)$ at infinity implies that $S^{(2)}$ has density
$\rho_{S^{(2)}}\in\mcC^\infty(\DR)$.

The case of polynomial decay did not escape their attention, either, although they consider
(in Example 7)
the situation where the series $S(\om) = \sum_{k=1}^\infty r_k X_{k}(\om)$, $r_k = k^{-1}$
converges in mean square but not absolutely, which gives rise to an unbounded r.v. with
density $\rho\in\mcC^\infty(\DR)$.

Wintner \cite{Wint1934,Wint1935} proposed a very natural and elementary upper bound for the
characteristic functions, proving in the case of polynomially decaying $k \mapsto r_k$ infinite
derivability\footnote{Jessen and Wintner \lcite{JW1935}{Section 6, Example 7} pointed out that
for $\fu(r) = 1/r$ the cumulative distribution has an analytic density in $\DR$ (actually, even
in higher-dimensional convolution models), which is impossible whenever the series at hand converges
to a bounded r.v., having necessarily a compactly supported probability distribution.}
of the respective probability density.
In fact, his technique from \cite{Wint1935} alone suffices for a good half of main results of this paper, and
applies to the most realistic physical models  of disordered solid state media
(with power-law screened interactions).
On the other hand, for the reasons coming from the main application
of this paper (to Anderson localization),
I intentionally avoid below the discussion of the case where
$\sum_k r_k = +\infty$ (but $\sum_k r^2_k <+\infty$) and the method from \cite{Wint1935}
is quite efficient.

Return to the characteristic function and write
\be
\label{eq:ln.ffi.b.1}
\bal
\ln \big| \ffi(t) \big| = \sum_{k=1}^\infty \ln \big| \cos\big( b^{-k} t \big) \big| \,.
\eal
\ee
By parity, we can assume $t>0$.
Let
\be
\label{eq:def.K.t}
K_t = m^{-1}\ln t \,,  \quad m = \ln b \,,
\ee
so that $t \, b^{-K_t} =  t \,  \eu^{-m c \ln t} = t^{1-m m^{-1}}=1$.
For $k> K_t$, $\cos\big(b^{-k} t \big) \le 1 -\frac{t^2}{4 b^{2k}}$
thus
$$
 \ln \big| \cos\big(b^{-k} t \big) \big|^{-1} \ge \ln \big( 1 -\frac{t^2}{4 b^{2k}} \big)^{-1}
\ge C t^2 \, b^{-2k} \,.
$$
Decompose \eqref{eq:ln.ffi.b.1} into two sums:
$$
\bal
\ln \big| \ffi(t) \big|^{-1} &= -\sum_{k=1}^{K_t} \ln \big| \cos\big( b^{-k} t \big) \big|^{-1}
+ \sum_{k>K_t} \ln \big| \cos\big( b^{-k} t \big) \big|^{-1}
\\
&
=: S_1(t) + S_2(t) \ge S_1(t) + C t^2 \sum_{k > K_t} b^{-2k}
\ge S_1(t) + C' t^2 b^{-2 K_t} \,.
\eal
$$
Further,
$$
S_2(t) \ge C' t^2 b^{-2 K_t} \ge C' t^2 \eu^{-2 m m^{-1}\ln t} = C' \,,
$$
which provides no decay to $|\ffi(t)|$, so we turn to $S_1(t)$. Recalling $b^{-K_t} t = 1$,
\begin{align}
\notag
S_1(t) &= - \sum_{k=1}^{K_t} \ln \big| \cos\big( b^{-k} t \big) \big|
= -\sum_{n=1}^{K_t} \ln \big| \cos\big( b^{n} \cdot b^{-K_t} t \big) \big|
\\
\label{ripple.sum}
&
= -\sum_{n=1}^{K_t} \ln \big| \cos\big( b^{n} t \big) \big| \,.
\end{align}
The last expression  certainly calls for an ergodic theorem, namely the one
for the fractional parts of $(\pi^{-1}b^n t)_{n\ge 1}$.
Indeed, the threshold $K_t$ is chosen so as to ensure that for $k\ge K_t$ one has $t \, b^{-k} \lesssim 1$,
so in the reversed time scale, $K_t, K_t - 1, \ldots, 1$, we have a growing sequence of arguments
$b^{-K_t}t \cdot b^{k} \sim C b^{k}$ of the periodic function $\cos$.

Unlike the "tidal" sum with a fixed $t$ and large $r$ leading to the Gaussian micro-scaling asymptotics,
nothing precise can be said in general about any individual term in \eqref{ripple.sum}, but
there are many of them; are they more or less evenly distributed or concentrated in the vicinity of
$\pi \DZ$?

The equidistribution is often established with the help
of Weyl's criterion \cite{Weyl1916}, applicable to a large variety
of dynamical systems on tori.
See also the works by Koksma \cite{Kok1935},  Dubickas \cite{Dubick06}, the monograph by
Cornfeld, Fomin and Sinai \cite{CFSin82} and further references provided therein.

Indeed, Kac, Salem and Zygmund \cite{KacSalZyg47} considered
the equidistribution problem and noticed that (cf.  \lcite{KacSalZyg47}{Section 5})
the expectation value
$$
\frac{1}{2\pi} \int_0^{2\pi} \ln \big| \cos \theta \big|\, d\theta = \ln 2
$$
suggests that for $b=2$, by \eqref{eq:def.K.t} with $m=\ln 2$,
$$
|\ffi(t)| \lesssim \eu^{ - K_t \ln 2 } = \eu^{ - \frac{\ln 2}{m} \ln |t| }
= \left(\ln |t|\right)^{ - \frac{\ln 2}{m} } = |t|\,.
$$
Of course, a simple integration does not suffice here (and neither was it used alone
to infer rigorous consequences in \cite{KacSalZyg47}), for we deal with
the \emph{logarithm} of $|\ffi(t)|$, so the fluctuations cannot be taken lightly in
the equidistribution\footnote{In a forthcoming paper, the equidistribution mechanism
will receive a proper treatment, based on a great wealth of results accumulated in this area.
Cf. e.g. \cite{HarLit1914,Dubick06}, a more recent monograph \cite{CFSin82}
and references therein. The fractional parts
of $(q^n t)_{n\ge 1}$ are equidistributed for a.e. $q>1$ or a.e. $t\ne 0$.
}
argument. Quite fortunately,
Vi\'{e}te and Euler had solved for us the critical model a few centuries ago. Had they not,
the absolute continuity of the critical measure would follow immediately from the
dyadic expansion \eqref{eq:Bernoulli}, but in the non-critical cases the Fourier analysis proves more versatile.

In a more general context, the conditions for absolute continuity of infinite convolutions of Bernoulli measures
(ICBM) have also been studied; cf., e.g., Erd\"{o}s \cite{Er1939}.

Kahane and Salem \cite{KahSal1958} proved the following nice result.
For any $b>1$, let
$$
\DN \ni q := \left\lceil \frac{\ln 2}{\ln b} \right\rceil
\quad \text{($\lceil \cdot \rceil$ stands for rounding up)},
$$
then the measure $\cF[\ffi]$ is H\"{o}lder continuous of order
\be
\label{eq:KS79}
\DR_+ \ni \alpha := \frac{\ln 2}{\ln b} \,.
\ee
In particular, for any integer $q\ge 1$, the measure with scaling factor $b = 2^{1/q}$ is
a.c. with density $\rho\in\mcC^{q-1}(\DR)$. This can be considered
as a generalization of \lcite{JW1935}{Section 6, Example 4}.

Special values of the exponent give rise to interesting number-theoretic problems;
see the papers by Hardy and Littlewood \cite{HarLit1916}, Mahler \cite{Mahler1932}, and
more recent papers by Dubickas, e.g.,  \cite{Dubick06}.

Erd\"{o}s \cite{Er1939} proved that if $b>1$ is a so-called Pisot--Vijayaraghavan (PV) number,
then $\ffi$ does not vanish at infinity,
i.e., $\cF[\ffi]$ is not a Rajchman measure.

Salem \cite{Sal1943} proved the converse of the result by Erd\"{o}s. Thus $\cF[\ffi]$ is a Rajchman
measure for Lebesgue-a.e. $b>1$.

Levin \cite{Lev79eng} (the original Russian version published in 1979) proved
that fractional parts of $(b^n)$ are completely equidistributed; this notion includes
estimates for the deviations from equidistribution. (Cf. also Franclin \cite{Fran63}.)

However, it is to be emphasized that most of these results apply to an exactly exponential
decay rate of $a_{|x|} = \fu(|x|)$, and this is not the case in dimension $d>1$ with Euclidean
distance $|x-y|$, even for periodic lattices; models with integer-valued distances $|x-y|_1$
and $|x-y|_\infty$ are sometimes simpler.
On the other hand, exponential screening in dimension $d\le 2$ is physically questionable,
and the case of screening weaker than exponential is technically easier.

Replacing the fractional parts of an exponential sequence with trajectories of a
skew shift on the torus, one comes to the equidistribution problem for
$\{n^2\alpha\}$); see, e.g. a paper by Rudnick \etal \cite{RudSarZah01}. In our problem, this corresponds
to a polynomially decaying potential $\fu$.
See also a review by Lyons \cite{Lyons95}, the works by Strichartz \cite{Stri90a,Stri90b,Stri00}
and references therein.

\bre
Notice that for the measures satisfying the so-called Condition (C) introduced by Cram\'{e}r, viz.
$\limsup_{|t|\to\infty} \, |\ffi_\mu(t)| \le \zeta < 1 $,
the analysis of equidistribution is unnecessary for the lower bounds on the sum $S_1(t)$ in
\eqref{ripple.sum}, as $-\ln|\ffi_\mu(t)|\ge \ln \zeta^{-1}>0$. This might seem like a very weak hypothesis,
yet it rules out almost periodicity
of the Fourier transform, since $\ffi_\mu(0)=1$.
The convolution analysis for these "poor man's Rajchman measures"
(or rather Cram\'{e}r's measures) is elementary and pleasant.
\ere

Naturally, the absolute integrability of $\ffi(t)$ is neither required for non-singularity of the
convolution measure $\mu=\cF[\ffi]$ nor observed, e.g., in the critical case $b=2$: the function
$\frac{\sin t}{t}$ is not absolutely integrable, but square-integrable, and the density of the respective
measure $\mu$, the indicator function $\one_{[0,1]}$, is an exemplary element of $\rL^2(\DR)$,
albeit neither smooth nor even continuous.

In this connection, recall that
Wiener and Wintner \cite{WW38},
answering a question raised by Nina Bary \lcite{Bary1927}{p. 113},
proved that $\kappa^*  = \half$ is the critical decay exponent
for the Fourier transforms of singular measures, in the following sense:
\begin{enumerate}[(i)]
  \item no singular measure $\mu$ can have $|\hmu(t)|\le C (1+|t|)^{-\kappa}$ for $\kappa>\kappa^*$;

  \item for any $\eps>0$ there are examples where
$$
\hmu(t) = \Ord{ (1+|t|)^{-(\kappa^* - \eps)}} \,.
$$
\end{enumerate}
Item (i) is due to the $\rL^2$-isometry of the Fourier transform, so \cite{WW38} is essentially
a construction of examples for (ii), complementing those by Menchoff \cite{Men1916} and Kershner
\cite{Ker1936}.

Convolution products are always at least as "regular" as the best of the factors involved.
In the case of a.c. measures with nice densities, one can forget about measures and deal directly
with their densities viewed as functions. The classical example of B-splines (convolutions of interval
indicators) shows that each additional factor, starting from $n=3$ factors, brings one more
derivative (Sobolev scale). The picture is however much more complex for singular measures.
Again, the explanation is provided by harmonic analysis.

Convolution \emph{powers} of a singular measure need not become absolutely continuous,
as shows the example of integer-scaled Bernoulli measures \cite{Wien1924}.
Specifically, when $\DN \ni M \ge 3$,
\be
\label{eq:Bernoulli.product.M}
\ffi(t) = \prod_{k=1}^\infty \cos\left( \frac{t}{M^k} \right)
\ee
obeys
$
\limsup_{|t|\to\infty} \big| \ffi(t)\big| >0,
$
and for large $R$,
$$
\frac{1}{2R} \int_{-R}^R \big| \ffi(t)\big|^2 = \Ord{ R^{\frac{\ln 2}{\ln 3}} } \,.
$$
Hu and Lau \cite{HuLau02} recently found that
$$
\bal
\limsup_{|t|\to\infty}  \ffi(t) & \quad
\begin{cases} = \ffi(\pi), & M=2n+1 \,,
\\
               \le \ffi(\pi), & M = 2n \,,
\end{cases}
\\
\liminf_{|t|\to\infty}  \ffi(t) & = -\ffi(\pi), \;\; 3\le M \in\DN \,.
\eal
$$
The classical example\footnote{I had learned it long ago as a part of probabilistic folklore,
but do not know who found it first.}
\lcite{Feller66}{v.2, Section V.4(d)}, on the other hand, shows that convolution products of
\emph{non-identical} singular measures can be more inclined to become more regular than either
of the convolution factors. The explanation is simple: taking a product of two \emph{different} infinite products
of the form \eqref{eq:Bernoulli.product.M}, one may sometimes overlap the unit (or nearly unit)
factors from one product with very small factors from another product, which is impossible
for \emph{identical} products. A similar phenomenon is encountered in the theory of asymptotical
expansions of sample distribution functions of sums of independent variables, where
a number of results are proved differently (or available at all) with and without the so-called
non-lattice distribution condition (cf. \cite{GKolm54,Feller66}).

However, there are many examples of singular measures on locally compact abelian groups with a.c.
convolution powers; cf. Hewit and Zuckerman \cite{HZ66,HZ67}, Saeki \cite{Sae77}, Karanikas and Koumandos
\cite{KK91}. Again, an explanation is provided by harmonic analysis: if (and this is of course a big if)
the Fourier transform $\ffi_\mu$ of a measure $\mu$ does actually have a power-law decay,
$|\ffi_\mu(t)|\le C |t|^{-\eps}$, $|t|\ge 1$, $\eps>0$, then $\cF^{\pm 1}[\mu^{*n}](t) \le C |t|^{-n\eps}$,
so it suffices to take $n > 1/\eps$. In view of the above mentioned Wiener--Wintner result \cite{WW38},
there are s.c. measures with $\half < \eps < 1$, so even squares of some s.c. measures are a.c.
The problem in general is that an s.c. can be not from Rajchman class, i.e. with no decay at all, let alone
power-law rate.

An important particularity of the measures appearing in higher-dimensional non-local alloy models is that one
has there \emph{infinite products} of Fourier transforms, expressed themselves via infinite products coming
from 1D chains filling a $d$-dimensional grid, like in \lcite{JW1935}{Section 6, Example 4}.

\bre
\normalfont
The above discussion is closely related to another topic which I only mention in passing here:
improvement of regularity of the cumulative potential (hence, of the IDS/DoS) of multi-layer
quasi-1D or quasi-2D media of finite cross-section $W$, as $W\to\infty$. Such models are sensitive
to the geometric properties of the scatterers support (periodic/aperiodic grid of scatterers)
and of the metric $\rd(\cdot\,,\cdot)$ in the configuration space $\mcX=\DZ^d,\DR^d$ figuring in the potential
$V(x;\om) = \sum_y \fu\big( \rd(x,y)\big) \om_y$. For example, taking the distance $|x-y|_\infty$, we would have
in a strip of width $W$ in $\DZ^2$ a convolution of $W$ identical measures (cf. the paragraph preceding
\eqref{eq:Bernoulli.product.M}). The Euclidean distance gives rise to an irrational and nonlinear scaling
when we pass from one layer to another:
$$
r = \sqrt{x^2 + y^2} \rightsquigarrow \sqrt{x^2 + (y+1)^2} = r \sqrt{1 + \frac{2y+1}{r^2}} \,,
$$
hence to a convolution of non-identical measures. Its quantitative
analysis becomes, therefore, geometry-specific and less universal; it does not belong in this paper.
In particular, the extension to Delone--Anderson Hamiltonians would not be automatic.
What is clear, is that varying the width $W$ of a strip, the grid spacing and the decay exponent $c>0$ of
a potential $\fu(r) = \eu^{-cr}$, one can rig the model so as to recover the classical example
\lcite{Feller66}{v.2, Section V.4(d)} with an a.c. convolution of s.c. Cantor measures.
Therefore, the continuity phase transition is encountered already within the class of quasi-1D systems
with exponential screening (be it possible or not in physical models). The distances $|\cdot|_1$
and $|\cdot|_\infty$ on $\DZ^d$, $d\ge 2$, on the other hand, give rise to interesting artificial models where
some analytic aspects are simpler than for $|\cdot|_2$. $\blacktriangleleft$
\ere

\bre
\normalfont

I have consciously avoided using any ergodicity arguments regarding the spatial grid $\mcZ$ of
the scatterers. For definiteness, it is assumed that $\mcZ=\DZ^d$, but the actual calculations in Eqn.
\eqref{eq:log.ffi.thermal.1} evidence that the exact periodicity is unnecessary, provided the
number of sites $x\in\mcZ$ per sufficiently large ball is bounded from below (even that can be
slightly relaxed). Upper boundedness
is only required for a uniform convergence and boundedness of cumulative potential, but not
for the regularity as such. Pushing by force a spatially nonhomogeneous environment into the framework of
ergodic systems proves quite useful in the analysis of Delone--Anderson Hamiltonians (cf. e.g.
\cite{RoM12,RoMVes13,GerMulRoM15} and references therein; see also closely related  works
\cite{Kle13,ElKl14} on "crooked/trimmed" random operators). In the present context, however, any reference
to spatial ergodicity would raise suspicions about a possible smuggling of an additional regularity
in a disguised form in the first place;
an old and efficient trick of some smart alchemists of the past centuries. Indeed,
who says "\emph{ergodicity}" says "\emph{with probability one}" (except perhaps for the case of unique ergodicity), but
it is obvious from our analysis that taking a "typical" grid $\mcZ = \{\Bc_x, x\in\DZ^d\}\subset \DR^d$,
where $\Bc_x = x + \Bd_x$ and $\Bd_x\in\DR^d$ are IID r.v. with a bounded density, would beat hands down
any singularity of the amplitudes $\om_x$ (literally: you can even take $\om_x \equiv 1$ !)
and produce a $\mcC^\infty$ PDF of the cumulative potential
without breaking a sweat. Therefore, in order not to raise  any doubts, when using the randomness
of the amplitudes $\om_x$, I stick to a completely "quenched" spatial
order/disorder of the grid $\mcZ$.

On the other hand, in the pure displacements model (with $\om_x \equiv 1$), it suffices to have
$\Bd_x$ taking two different values, as long as $\DR^d \ni x \mapsto \fu(|x-y|)$ is sufficiently non-flat.

I believe that a physically realistic modeling ought to take into account some additional continuity
of the distribution of the "scatterers", first of all via the displacement degrees of freedom
probably provided by the statistical-mechanical description. However, one has to be careful
with the choice of the tools, since we deal here with very fine effects, and the classical,
not quantum, statistical mechanics may be adequate or not.
$\blacktriangleleft$
\ere


From a utilitarian point of view, pointwise decay bounds on the Fourier transforms
of (possibly) singular measures, quite handy when available, are far from being necessary
for the proofs of H\"{o}lder
continuity\footnote{Such EVC estimates allow in principle for the strongest localization results
an MSA based method could provide today,
just as the classical Wegner estimate for Lipschitz continuous marginal disorder.}
(of some positive order) of the infinite convolutions at hand.
Fortunately enough, integral estimates are both easier to establish
and available for a large class of measures.
A fairly explicit and constructive characterization of continuous measures whose Fourier coefficients
decay in Cesaro sense follows from Wiener's results \cite{Wien1924} obtained in 1924:
for a measure $\mu$ on $[0,1]$ with the set of nonzero atoms denoted $\Sigma_{p.p.}(\mu)$,
$$
\lim_{|n|\to\infty} \frac{1}{2n+1} \sum_{|k|\le n|} \big|\hmu(n)\big|^2
=  \sum_{\lam \in \Sigma_{p.p.}(\mu)} \big| \mu(\lam) \big|^2  \,,
$$
and for a measure on $\DR$ one has,
as is well-known, by the same arguments
(cf., e.g., \lcite{RS3}{Theorem XI.114})
$$
\lim_{ T\to+\infty} \frac{1}{2T+1} \; \int_{-T}^{T} \big| \hmu(t) \big|^2 \, dt
= \sum_{\lam\in \Sigma_{p.p.}(\mu)} \big| \mu(\lam) \big|^2\, ,
$$
with $\hmu(t) := \int_\DR \eu^{-\ii t x } \, d\mu(x)$. As is equally well-known, this is a basis for the
celebrated RAGE (Ruelle \cite{Ruel69}, Amrein and Georgescu \cite{AmrGeo73}, and Enss \cite{Enss78}) theorem.

Strichartz \cite{Stri90b} established analogs of Wiener's theorem for expansions
in eigenfunctions of various Schr\"{o}dinger operators, including the Hermite polynomials.
Higher $L^p$: cf. \lcite{Stri90b}{Corollary 5.4}.

Perhaps, the reader might find unwarranted the amount of attention given above to the singular measures,
since this paper focuses mainly on the smooth and \emph{multi}-dimensional case. Indeed, just mentioning all these works,
mostly related to 1D models,
is like opening yet another Pandora's
box\footnote{This is one way to put it. A reader familiar with F. J. Dyson's paper published
in "\emph{Physics today}" in 1967 knows another metaphor, which would sound today much
less "politically correct".
Amazingly, the Soviet censorship authorities, otherwise paranoic, happened to overlook it in the well-known monograph by Lifshitz,
Gredescul and Pastur \cite{LGP88},
on the first page of Chapter II.},
with all kinds of mathematical distractions from the physically most
relevant situations.

It is worthwhile emphasizing the following points:

\begin{itemize}

  \item Large values of $b$ are not the only possible mechanism leading to singularity of the
  infinite convolutions of Bernoulli measures: apart from creating "large" gaps in the Cantor-type support,
  "squeezing" the unit "mass" to a family of intervals of smaller and smaller total length (in the inductive
  construction), one can achieve a singular concentration by making Bernoulli measure
 asymmetric\footnote{See e.g. a discussion in the work by Strichartz \lcite{Stri90a}{Section 2} where it is shown
that in a more general context of self-similar measures $\mu$, related to
contractive maps, the dimension of $\supp\, \mu$ is maximized by the so-called natural weights figuring in definition of $\mu$. In the
Bernoulli case, with identical contraction exponents, the probabilities $p$ and $1-p$ have to be equal to maximize
the dimension of the support.}\footnote{See also \lcite{JW1935}{Section 8, Example 2}.}.
Specifically,
  any measure $\mu_p$ on the interval $[0,1]$ pulled-back by the isomorphism to the set of semi-infinite words
  $\{0,1\}^{\DN}$ with $p = \pr{\om_i=0} \not\in \{0, 1/2, 1\}$ has of course the full support $[0,1]$, but for
  any $p \ne p'$, the measures $\mu_p$ and $\mu_{p'}$ are mutually singular, as follows from the Law of Large Numbers
  for the limiting frequency of the digits $\om_i = 0$. This shows that one should not expect "the" critical point
  for the continuity phase transition(s), but a number of critical parameter zones, where some important
  parameters are functional (PDF of a measure).

  \item A sub-exponential or power-law decay of the potential $\fu$ results in a higher regularity than just
  continuity or mere absolute continuity with bounded density. A polynomial decay of \emph{any} summable
  order gives rise to a compactly supported $\mcC^\infty$-density.

  \item Whenever $\fu$ is a ``slowly decaying'' function (in the sense that its derivative $\fu'$ decays
  faster than $\fu$ itself), a local analysis of the induced single-site random potential reveals
  in \emph{any} dimension $d\ge 1$
  a Gaussian-like nature, due to a convolution of many independent contributions of comparable amplitudes,
  resulting in CLT (Central Limit Theorem) type approximations. It is shown below that some simple two-point
  Gaussian approximations are also possible to obtain. A full-fledged multivariate CLT
  in arbitrary finite domains is more difficult to establish. However, I conjecture that this is possible.

  In any case, the last argument suggests that Gaussian models of disorder, understood within a suitably
  defined Gaussian "micro-scaling" limit, also should have a fairly universal value, at least as useful guides
  to more accurate models.

\end{itemize}

$\blacklozenge$ \textbf{The cumulative potential and the IDS.}
The relations between the regularity of the single-point marginal probability $\mu$ of the potential
registered at individual sites (cumulative potential, in the alloy models) and the IDS are not quite straightforward.
Even in the simplest of the two directions (from regularity of the potential to that of the IDS), it took
some time to extend the original Wegner's result \cite{W81} to probability distributions (in the IID case)
with an arbitrary continuity modulus. The turning point was the spectral averaging  developed by
Simon and Wolf \cite{SW86} and the dimensional reduction via the Birman--Schwinger identity
\cite{Bir61,Schw61}
to a commutative, one-dimensional probabilistic analysis closely related to the Boole identity (1857)
\cite{Boole1857}.
The one-dimensional models provide a good laboratory for studying these relations,
particularly for deriving singularity of the IDS from that of the potential. Since dimension one
is critical for the main phenomenon explored in the present paper, it is certainly worthwhile recalling
some known key facts about the IID potentials in 1D.

Due to a result by Simon and Taylor \cite{SimTay85},
if $\mu$ has a compactly supported density $\rho_\mu\in \rL^\alpha(\DR)$, $\alpha>0$, then the DoS
(in 1D !) exists and is
in $\mcC^\infty(\DR)$. As to arbitrary measures not supported by a single point (in the IID case),
the IDS is always continuous, by Pastur's result \cite{Pas80}. Craig and Simon \cite{CrSim83a,CrSim83b}
established log-H\"{o}lder continuity of the IDS in any dimension (on a lattice), using the duality between
the IDS and the Lyapunov exponents in one-dimensional or quasi-one-dimensional systems. On the other hand,
in the 1D Bernoulli--Anderson model Simon and Taylor \cite{SimTay85} conjectured that, in some parameter zones,
IDS cannot be H\"{o}lder continuous of any order higher than some critical exponent $\hal<1$.
For a Bernoulli measure with values $0$ and $b$, Lipschitz continuity is ruled out for $b$ large enough.

Carmona \etal \cite{CarKleMar87} proved Anderson localization in 1D lattice models with
arbitrary nontrivial disorder; they also proved the above mentioned Simon--Taylor conjecture,
making rigorous the heuristic argument outlined in \cite{SimTay85} and based upon
an adaptation of Temple's inequality \cite{Temp28} and on a result by Halperin \cite{Hal67}.
Formally, it relies on the independence of the values of the potential. However, the key mechanism is the existence
of EFs deterministically localized on single impurities embedded into the ambient constant potential, and
this mechanism is robust enough to produce singular IDS at least for suitably chosen parameters
of the long-range cumulative potential generated by the underlying Bernoulli disorder. Therefore,
\emph{the dimension one harbors indeed transitions in the IDS measure, from absolute continuity to
singular H\"{o}lder continuity}.

Klein \etal \cite{KleMarPer86} used the supersymmetry approach to prove H\"{o}lder continuity in 1D, under a relatively
weak assumption on power-law decay of $\hmu(t)$ at infinity. By comparison, the Fourier analysis used in our
method easily proves $\mcC^\infty$-regularity of the DoS under the same hypotheses
\lcite{KleMarPer86}{Eqs. (1.1)--(1.2)}, for even a much weaker decay of $\hmu(t)$
(viz. a mere fact that $\mu$ is a Rajchman measure) is a dreams-come-true
scenario in the regularity problem for infinite convolutions of (suitably scaled for convergence)
singular measures. But recall that \cite{KleMarPer86} deals with a harder, short-range disorder problem.
The irony is that the assumption of finite range of interaction had been initially made in Anderson-type models
in order to "\emph{simplify}" their analysis!

The authors of \cite{KleMarPer86} conjectured that some hypotheses similar to
\lcite{KleMarPer86}{Eqns. (1.1)--(1.2)},
i.e., relatively weak decay of the characteristic function $\hmu(t)$, should be sufficient for
(at least) H\"{o}lder continuity of the IDS in any dimension. In fact, their conditions refer to
a nontrivial component $\mu_1$ in a mixture $\mu = s \mu_1 + (1-s) \mu_2$ with $s>0$, regardless of
$\mu_2$. The present paper only sheds \emph{some} light on their general conjecture, since
\par\vskip1mm\noindent
$\bullet$ infinite range of interaction $\fu$ is vital for our proofs;
\par\noindent
$\bullet$  the case of a mixture is not considered here.
\vskip1mm

However, I conjecture that the infinite convolution mechanism is akin to the one observed in
one-dimensional models, and that the lattice Bernoulli--Anderson Hamiltonian should have
smooth DoS in dimension $d>1$. (A mere log-H\"{o}lder continuity
would suffice for Anderson localization.)
The arguments in favor of this hypothesis are as follows.
Regularity of the IDS in 1D is derived by Hilbert transform from that of the Lyapunov exponent(s);
the latter come from the transfer-matrix analysis and ultimately from the Lyapunov solutions
(those which grow exponentially), hence from the Green functions constructed from the Lyapunov solutions.
Further, the Green functions traveling across a random media accumulate, like a sort of test functions,
the random site-wise fluctuations, which results in multiple nonlinear convolutions. Moreover,
an asymptotically sharp analysis by
Pastur and Figotin \cite{PF92} (IID case),
and by Chulaevsky and Spencer \cite{CSp95} and Schulz-Baldes and Seidel \cite{SchBSei08}
(correlated case)
evidences that in the weak disorder, the linearized convolutions give rise to the ergodic theorem and
CLT approximation providing the leading order of magnitude of the Lyapunov exponent, under weaker
assumptions that IID (fast decay of correlations). The potential at a site $y$ contributes to the
value of a Green function at $x$ with a weight which is at worst exponentially small in $|x-y|$, but
our analysis suggests that, \emph{were that contribution linear in $V$}, it would suffice for
infinite derivability in any dimension $d>1$.

\vskip1mm
To avoid any misunderstanding, let me stress: Anderson localization in a short-range Bernoulli disorder
on a lattice $\DZ^d$ has been a natural and very tempting conjecture floating in the air
ever since the publication
of \cite{BK05} in 2005. The conjecture I have mentioned above concerns

\textbf{(i)} the suggested mechanism of regularity of the IDS, and

\textbf{(ii)} infinite derivability of the DoS in higher dimension,
\par\noindent
for  compactly supported interactions $\fu$.

\vskip1mm

$\blacklozenge$ \textbf{``Thin'' tails.}
Another important aspect of infinite smoothness, combined with a.s. boundedness of the probability distribution,
is that near every edge  $E_*$ (there may be gaps in its support) it features the decay
$\Ord{|E-E^*|^{\infty}}$. This qualitative result follows from
the infinite derivability without calculations or application of the large deviations theory.
In this connection, recall that Exner, Helm and Stollmann \cite{ExHSto07} used earlier a very simple argument
in the proof of the initial
length scale (ILS)
estimate for the MSA, based on the \emph{hypothesis} of edge decay of the IDS of sufficiently high
polynomial order and replacing the Lifshitz tails estimate. Such "ultra-thin" tails  are therefore universal,
above the critical point of the continuity phase transition for the exponential Bernoulli disorder.
From the perspective of Anderson localization, they are most valuable in the case of sign-definite interactions,
but their "thin" nature is universal and does not require interaction to be sign-definite.

It is to be emphasized that the two kinds of "tails" have different nature and refer to different,
albeit related, phenomena. The Lifshitz tails refer to upper, lower, or asymptotic bounds on the IDS,
and as such do not presume any local regularity property of the IDS measure.
The "thin" tails mentioned above are a direct consequence of $\mcC^\infty$-smoothness of the
compactly supported density of
the cumulative potential. Further, Lifshitz tails asymptotics results from
a collective behaviour of a sufficiently large sub-sample, hence manifesting itself in a sufficiently large
finite volume, while the infinite derivability is a pointwise property, and as was demonstrated
by Exner \etal \cite{ExHSto07}, it can be used in a ball $\ball$ without requiring it to be large. It is an individual,
site-wise response to the collective behaviour of a large number of \emph{remote} sources from the "thermal bath"
surrounding the finite volume at hand. "Freezing" the bath outside some finite ball ("jacuzzi") may destroy
continuity, let alone smoothness, but useful probabilistic upper bounds may be preserved (as are
the Lifshitz tails unrelated to the thermal bath).

\vskip3mm
Summarizing, Lifshitz tails are a "ripple" phenomenon while the "thin" tails are tidal. Yet, both give rise
to robust mechanisms of the onset of Anderson localization, without a physically questionable
condition of strong disorder.

\section{Main results}
\label{sec:main.results}

The word \emph{potential} used alone refers below to the scatterer potential $\fu:\,\DZ^d\to\DR$,
generated by the elements of the disordered
media (sources), and the sum of potentials from all the sources registered at $x$
is called the \emph{cumulative potential}.

\subsection{Smoothness of disorder}
\subsubsection{Polynomial potentials.
Theorem \ref{thm:thermal.bath.F.V.polynom}}

\btm
\label{thm:thermal.bath.F.V.polynom}
Consider a random field $V(x,\om)$ on $\DZ^d$ of the form
\be
\label{eq:def.V.x}
V(x,\om) = \sum_{y\in\DZ^d} \fu(y-x)\, \om_y \,,
\ee
where  $\fu(r) = r^{-A}$, $A>d$,
and $\{\om_x, \, x\in\DZ^d\}$ are bounded IID random variables with nonzero variance.
Then the following holds true:

\begin{enumerate}[\rm(A)]
  \item The common characteristic function $\ffi_V(\cdot)$ of the identically distributed random
variables $V(x,\om)$, $x\in\DZ^d$, satisfies an upper bound
\be
\label{lem:thermal.bath.bound}
\all t\in\DR \quad \big| \ffi_V(t)\big| \le C \, \eu^{-|t|^{d/A}} \,.
\ee

  \item Consequently, the common probability distribution function $F_V(\cdot)$ of the
  cumulative potential at sites $x\in\DZ^d$ has the derivative $\rho_V \in\mcC^\infty(\DR)$.

  \item Denote $v_* := \inf\, \supp \rho_V$, then $F_V(v_*+\lam) = \ord{|\lam|^\infty}$.
\end{enumerate}
\etm

See the proof in Section \ref{sec:char.f.polynom}.

As the reader may expect, the assertions of Theorem \ref{thm:thermal.bath.F.V.polynom}
are not specific to periodic lattices and can be reformulated for the models
in Euclidean spaces $\DR^d$ or in metric graphs. The fact that the above random variables
$V(x,\om)$ are identically distributed requires of course $\DZ^d$-periodicity of the
grid of the scatterers, but the infinite smoothness of their individual measures
is not contingent upon periodicity, as will become clear from the proof given
in Section \ref{ssec:thermal.bath.ch.f}.

Similar remarks can be made regarding several results stated below, as evidence their proofs.

Except for the explicit upper bound \eqref{lem:thermal.bath.bound}, the specific power-law form of the
potential is not crucial
to the result on infinite derivability of the probability density $\rho_V$. We always assume absolute convergence
of the random series $\sum_x \fu(x) \om_x$, but it is not crucial, either: see the discussion in
Section \ref{sec:Euler.Viete}
of a result by Jessen and Wintner \lcite{JW1935}{Section 6, Example 7} establishing analyticity of $\rho_V$
in the one-dimensional case with $\fu(r) = 1/r$ and $\esm{\om_x} = 0$;
in this case, series $\sum_x \fu(x) \om_x$ converges in mean square. The assumption on absolute convergence
allows us to freely shift all $\om_x$ by any constant, so we can switch from $\esm{\om_x} = 0$ to $\om_x\ge 0$
when necessary.

\subsubsection{Exponential and sub-exponential potentials.
Theorem \ref{thm:thermal.bath.F.V.exp}}

\btm
\label{thm:thermal.bath.F.V.exp}
Consider the potential $\fu(r) = \eu^{- a r^\delta}$, $m>0$, $\delta\in(0,1]$ and let $d \ge 1$.
Then the characteristic functions of the random variables
\be
\label{eq:Vx.exp}
V_x(\om) = \sum_{y\in\DZ^d} \fu(|y-x|) \om_x
\ee
obey the upper bound
$$
\bal
\big| \ffi_{V_x}(t)\big|  = \left| \esm{ \eu^{\ii t V_x(\om)} } \right|
&
\le  \big(1 + |t|\big)^{ - \frac{C}{m} \ln^{\frac{d}{\delta} -1} |t|}  \,.
\eal
$$
Consequently:
\begin{enumerate}[\rm(A)\leftmargin=0cm]
  \item for any $d>1$ and $\delta\in(0,1]$, as well as for
  $d=1$ and $\delta\in(0,1)$, the r.v. $V_x$ have probability  densities $\rho_x \in \mcC^\infty(\DR)$;

  \item for $d=\delta=1$ and $a>0$ small enough, $\rho_x$ have $Q\ge \in \mcC^\infty(\DR)$
  with $Q \ge C' \lfloor  a^{-1}\rfloor$;

  \item for $d=\delta=1$ and any $a>0$, the PDF of $V_x$ are H\"{o}lder continuous of some order $\alpha>0$.

\end{enumerate}

\etm

\subsection{Smoothness of DoS and Wegner estimates}

$\,$

First I would like to make a general terminological remark. Usually one means by the
IDS the \emph{limiting} eigenvalue distribution for a family of finite-volume restrictions
$H_{\ball_L(0)}(\om)$ as $L\to +\infty$. Respectively, the DoS is the density of the IDS
w.r.t. the Lebesgue measure (whenever it exists). As was mentioned in the introductory sections,
we emphasize the influence of the exterior configuration $\om_{\DZ^d\moins\ball_L(0)}$ on the spectrum of
$H_{\ball_L(0)}(\om)$. For this reason, the terms IDS and DoS refer in the present paper to
the eigenvalue distribution (starting with individual eigenvalues) of finite-volume Hamiltonians
subject to a infinite or finite "thermal bath". Taking into account a possible strong singularity
of the probability measure of the amplitudes $\om_x$, of particular interest is the smoothness
of the finite-volume DoS in an infinite "bath".

\subsubsection{Staircase potentials with power-law decay.
Theorems \ref{thm:smooth.DoS.pol.staircase}--\ref{thm:Wegner.fukap}}

We start with a somewhat artificial class of polynomially decaying potentials
featuring an infinite number of plateaus. While these potentials, admittedly, are not realistic
and constitute only a toy model, they may provide a good laboratory for further improvements
of our results on regularity of long-range disorder. The main motivation for introducing this
class is that the harmonic analysis of the regularity of the corresponding finite-volume
eigenvalue distributions is much more transparent than for the potentials
$\fu(r) = r^{-A}$. As to the regularity of the cumulative potential for the staircase model,
omitted in the present paper for brevity,
it can be established in virtually the same way as for the genuine polynomial potentials
$\fu(r) = r^{-A}$, owing to linearity of the alloy transform.


\vskip1mm

Let $\varkappa>1$, introduce the integer sequence
$\fr_n = \fr_n(\varkappa) := \lfloor n^\varkappa \rfloor$, $n\ge 0$,
and consider the piecewise constant potential
\be
\label{eq:def.fukap}
\fukap(r) := \sum_{n=0}^\infty \fr_n^{-A} \, \one_{[\fr_n, \fr_{n+1})}(r)\,,  \quad r\ge 0.
\ee

Since $\fr_{n+1} - \fr_n \asymp (n+1)^\varkappa - n^\varkappa = \ord{n^\varkappa}$, it follows that
\be
\label{eq:def.fuk.fu}
\fukap(r) = \big(1+\ord{1} \big)r^{-A} .
\ee

The regularity bound is essentially the same as for the one-point marginal measures
of the cumulative potential, which is not surprising, since suitably chosen fluctuations induced by a staircase
potential on a fixed ball $\ball\subset\DZ^d$ act through a random multiple of the identity operator in $\ell^2(\ball)$.

\btm
\label{thm:smooth.DoS.pol.staircase}
Consider the potential $\fukap$ of the form \eqref{eq:def.fukap}, and let
$\ball = \ball_L(u)$. Then $H_\ball(\om)$ admits the representation
$H_{\ball}(\om) = \tH_\ball(\om) + \xi_\ball(\om) \one_\ball$,
where the random variable $\xi_\ball$ is independent of $\tH_\ball(\om)$
and has bounded, compactly supported  probability density $\rho_\ball\in\mcC^\infty(\DR)$.
Consequently, all eigenvalues of $H_\ball(\om)$ have the form
$\lam_j(\om) = \tlam_j(\om) + \xi_\ball(\om)$ with
$\tlam_j$ independent of $\xi_\ball$, hence the probability distributions of $\lam_j$
have densities $\rho_j\in\mcC^\infty(\DR)$.
\etm


\btm[Wegner estimate, staircase potentials]
\label{thm:Wegner.fukap}
Let $\varkappa>1$ and consider the same potential $\fu$ of the form \eqref{eq:def.fukap}
as in Theorem \ref{thm:smooth.DoS.pol.staircase}.
Let $L\in\DN$ and
\be
\label{eq:def.gamkap}
\tau > \gamma = \gamkap := \frac{\varkappa}{\varkappa-1} \,.
\ee
%
Denote by $\fF_{\ball_{L^\tau}(u)}^\perp$ the \sigal generated by
  $\myset{\om_x:\, x \not\in \ball_{L^\tau}(u)}$.
  Then for all $\eps \ge L^{- A\tau }$
and any eigenvalue $\lam_j(\om)$ of $H_{\ball_L(u)}(\om)$
\be
\label{eq:claim.staircase.DoS.lam.j.eps.R}
\pr{ \lam_j(\om) \in I_\eps  \cond \fF_{\ball_{L^\tau}(u)}^\perp}
\lea L^{\left(A - \frac{d}{2} \right) \gamkap} \eps.
\ee
%


\etm

See the proof in Section \ref{sec:DoS.staircase}.

\subsubsection{Wegner estimate for polynomial potentials. Theorem \ref{thm:Wegner.polynom.with.Bernstein}}

\btm[Wegner estimate]
\label{thm:Wegner.polynom.with.Bernstein}
Consider the model with the potential $\fu(r) = r^{-A}$, $A>d$, and the random Hamiltonian
$H_\ball$ relative to a cube $\ball=\ball_L(x)$. Assume that its eigenvalues $\lam_j(\om)$ are measurably
enumerated in ascending order. Fix real numbers $\theta>0$, $\tau>1$, and let
$\btau/\tau > \max\left[\frac{A}{A-d}, \, 1+\theta\right]$. Denote $\oball = \ball_{L^\btau}(x)$
and decompose $\Bom = \Bom_\oball + \Bom^\perp_{\oball}$, with
$\Bom_\oball = \om|_{\ball_{L^{\btau}}}$ and $\Bom^\perp_\oball = \om|_{\ball^\rc_{L^{\btau}}}$.
There exists $\beta>0$ such that for each eigenvalue $\lam_j(\cdot)$ and for any
$\eps \ge \eps_L := L^{- A(1+\theta)\tau}$
\beal
\label{eq:thm.Wegner.polynom.cond.E.1}
 \qquad
\sup_{E\in\DR} \;
\pr{ \Bom_\oball \cond \exists\,\Bom_\oball^\perp: \;\;
   \lam_j(\Bom_\oball + \Bom_\oball^\perp) \in [E, E+\eps] \le \eps }
\lea L^{ (A + \beta )\tau } \, \eps \,.
\eeal
Consequently,
\be
\label{eq:thm.Wegner.polynom.cond.E.2}
\pr{ \Bom_\oball \cond  \exists\,\Bom_\oball^\perp : \;\;
   \dist\left( \Sigma(H_\ball(\Bom_\oball + \Bom_\oball^\perp), \, E \right) \le \eps_L }
\lea L^{ -\left(A\theta + \beta \right)\tau +d} \,.
\ee
%

\etm

\subsubsection{Eigenvalue comparison for polynomial potentials.
Theorem \ref{thm:EVComp.with.Bernstein.polynom}}

The next result is an eigenvalue comparison for Hamiltonians in two distant finite volumes $\Lam'$ and $\Lam''$,
rather than an eigenvalue concentration estimate for one isolated finite-volume  Hamiltonian, and it is
proved without usual stochastic decoupling arguments. In fact, a bona fide decoupling in the context of
Theorem \ref{thm:EVComp.with.Bernstein.polynom}
is not strong enough to ascertain its claim, so we allow the eigenvalues in $\Lam'$ and $\Lam''$ to be
coupled in a non-negligible way and quantitatively compare their sensitivities to properly selected,
common random fluctuations. Such an approach has been used earlier in our works on multi-particle Anderson
Hamiltonians \cite{C12,CS13,CS16,C16b}, where it was crucial to the proof of localization in the
physically natural symmetrized norm-distance.

\btm[Eigenvalue comparison estimate]
\label{thm:EVComp.with.Bernstein.polynom}
Under the assumptions and with notations of Theorem \ref{thm:Wegner.polynom.with.Bernstein},
consider two cubes $\ball' = \ball_{R'}(x)\subset \ball_{L}(x)$,
$\ball'' = \ball_{R''}(x) \subset \ball_{L}(y)$
with $|x-y|=L^\sigma$ and some $R', R''\le L$\,,
and the Hamiltonians $H_{\ball'}(\om)$ and $H_{\ball''}(\om)$
with the interaction potential $\fu(r) = r^{-A}$, $A>d$. Then for some $\beta>0$
\be
\label{eq:thm.EVComp.with.Bernstein.claim}
\pr{ \om_\ooball: \;  \inf_{\om_\ooball^\perp} \;
\dist\left[ \Sigma\left(H_{\ball'} \right), \, \Sigma\left(H_{\ball''} \right) \right] \le 2\eps }
\le L^{ - \left( A \theta + \beta \right)\sigma + 2d } \,.
\ee

\etm

See the proof in Section \ref{ssec:EV.compare.polynom}.


\subsubsection{Sub-exponential and exponential potentials.
Theorems
\ref{thm:smooth.exp.sum-norm.and.max-norm}--\ref{thm:Wegner.exp.max-norm}--\ref{thm:EVC.two.balls.exp}}

Introduce in $\DR^d$ the norms
\be
|x|_p := \left\{
           \begin{array}{ll}
            \big(|x_1|^p + \cdots + |x_d|^p\big)^{1/p} , & \hbox{ $p\in[1,+\infty)$,} \\
             \max\limits_{1\le i \le d} |x_i|, & \hbox{$p=\infty$,}
           \end{array}
         \right.\ee
and consider the interactions
\be
\label{eq:def.fu.max-norm}
\fu_p(x)=\eu^{-a|x|_p} \,, \quad a>0.
\ee

We will focus on two particular cases: $p=1$ and $p=\infty$; for these values of $p$, the distance induced
on $\DZ^d \hookrightarrow \DR^d$ by the norm $|\cdot|_p$ is integer-valued.

\vskip2mm
\noindent
$\bullet$ \emph{The max-norm distance model.}

Unlike the Euclidean norm $|\cdot|_2$ in $\DR^d$,
the function $x\mapsto |x-y|_\infty$,
with any fixed $y \in \ball_L(0)$, takes constant values
on the boundaries of cubes $\ball_{L'}(0)$ with $L'> L$.

A particularity of this potential, important to our analysis,
is expressed by the following geometrical property
(see Fig.~5 in Section \ref{ssec:smooth.max.norm}).
Given a cube $\ball = \ball_L(0)$, the restrictions on $\ball$
of the interaction potentials $y \mapsto \eu^{-a|x-y|_\infty}$ with $x$ ranging in any
$(d-1)$-dimensional subset of the form
$$
\{x = (r, x_2, \ldots, x_d):\, r\ge 2L, \, \max_{i=2, \ldots d} |x_i| \le r - L \}
$$
are all proportional to the $x$-independent function $U_L(y) := \eu^{-a|y - \hy|_\infty}$,
where $\hy = (L, 0, \ldots, 0)$. This substantially simplifies the analysis of the
random cumulative potential, due to the one-dimensional nature  of convolutions.
If $|x|_\infty$ were replaced by the Euclidean distance $|x|_2$, we would have to deal
with multi-dimensional convolutions of measures on $\DR^\ball$.

\vskip2mm
\noindent
$\bullet$
\emph{The sum-norm distance model} A similar property holds for the Hamiltonians with the
interaction potential $\fu_1(r)$. See the discussion in Section \ref{ssec:proof.thm:EVC.two.balls.exp.sum-norm}.

\btm[Smoothness of DoS, exp.]
\label{thm:smooth.exp.sum-norm.and.max-norm}
Consider the model with a potential of the form $\fu_p$, $p\in\{\infty, 1\}$, and introduce the Hamiltonian
$H_{\ball}(\om)$, $\ball = \ball_L(u)$.
Then $H_\ball$ admits a representation
$$
H_\ball(\om) = \tH_\ball(\om) + S_\ball(\om) U^{(p)}_{\ball}(x), \;\;\min_{x\in\ball} U^{(p)}_{\ball}(x) >0,
$$
where $S_\ball(\om)$ is independent from $\tH_\ball(\om)$.
Furthermore:
\vskip1mm
\noindent
{\rm(A)} If $d>2$, then $S_\ball(\cdot)$ has probability density $\rho^{(p)}_{\ball}\in\mcC^\infty(\DR)$. Consequently,
the probability distribution of each eigenvalue $E^\ball_j(\om)$ of $H_\ball(\om)$ has density
$\rho^\ball_j\in\mcC^\infty(\DR)$.
\vskip1mm
\noindent
{\rm(B)} If $d=2$, then there exist $a_*>0$ and $C\in(0,+\infty)$ such that for $a\in(0,a_*]$,
$S_\ball(\cdot)$ has probability density $\rho^{(p)}_{\ball}\in\mcC^Q(\DR)$
with $Q\ge C a^{-1} \ge 1$.
\etm

\btm[Wegner estimate, exp.]
\label{thm:Wegner.exp.max-norm}{thm:smooth.exp.sum-norm.and.max-norm}
Consider the model with the potential $\fu_p$ with $p\in\{\infty, 1\}$.
Consider a ball $\ball=\ball_L(u)$,
the random Hamiltonian $H_\ball$, and the lattice subset
$\mcX_{L,R} = \cup_{r=2L}^{10L} \mcX_r$, where
$$
\bal
\mcX_{L,R} &:=
\bigcup_{r=2L}^{10L} \Big\{ x = u + (r, x_2, \ldots, x_d):\, \max_{i\ge 2} |x_i| \le r - L \Big\}\,.
\eal
$$
and
$$
\mcX_r :=
\left\{
  \begin{array}{ll}
    \big\{ x = u + (r, x_2, \ldots, x_d):\, \max_{i\ge 2} |x_i| \le r - L \big\}, & \hbox{ for $p=\infty$;}
\\
   u + (L, \ldots, L) + \myset{ x \in \DN^d:\, |x|_1 = r } , & \hbox{ for $p=1$.}
  \end{array}
\right.
$$

Let $\fF_{L}$ be the $\sigma$-algebra generated by $\{\om_x,\, x \not\in \mcX_{L,10L} \}$.
Then for any $\eps\in[\eps_L, 1]$ with $\eps_L := \eu^{ - 10 aL }$, and any interval
$I_\eps\subset\DR$ of length $\eps$ one has
\be
\esm{ \trpar{ \rP_{I_\eps}\big( H_{\ball}(\om) \big)}  \cond \fF_{L}} \le C \,|\ball_L|\, \eu^{6aL} \eps \,.
\ee

Next, decompose $\Bom = \Bom^-_L + \Bom^+_L$, with
$\Bom^-_L = \om|_{\ball_{20L}}$ and $\Bom^+_L = \om|_{\DZ^d\setminus \ball_{20 L}}$. Then
\be
\esm{ \sup_{\Bom^+_L}  \;\; \trpar{ \rP_{I_\eps}\big( H_{\ball}(\Bom^-_L + \Bom^+_L) \big)} } \le C \,|\ball_L|\, \eu^{6aL} \eps \,.
\ee

\etm

\btm[Eigenvalue comparison estimate, exp.]
\label{thm:EVC.two.balls.exp}
Let be given two balls $\ball_{L}(u')$, $\ball_{L}(u'')$ and subsets $\Lam'\subset\ball_{L}(u')$,
$\Lam''\subset\ball_{L}(u'')$. Consider the Hamiltonians $H_{\Lam'}(\om)$ and $H_{\Lam''}(\om)$
with the interaction potential $\fu(r) = \eu^{-ar}$, $a>0$. Then
there exist constants $C, C' \in(0,+\infty)$ such that if $|u' - u''|\ge C L$, then
\be
\label{eq:thm.EVC.two.balls.exp}
\pr{ \dist\left( \Sigma\big(H_{\Lam'}(\om)\big), \, \Sigma\big(H_{\Lam''}(\om)\big) \right) \le \eps }
\lea \big| \Lam'\big| \, \big|\Lam''\big| \, \eu^{8aL} \, \eps \,.
\ee

\etm

See the proof in Section \ref{ssec:proof.thm:EVC.two.balls.exp.sum-norm}

\vskip3mm

Now we turn to the proofs of the main results.

\section{Smoothness of disorder under polynomial screening}
\label{sec:char.f.polynom}

Now we turn to the potentials $\fu(r) = r^{-A}$, $A>d$, in dimension $d$, aiming essentially at $d\ge 2$,
as $d=1$ calls for more efficient, specifically one-dimensional techniques.
Allowing $A>d$ to be arbitrarily close to $d$ may look artificial, but note that for example
in dimension $d=2$ the decay exponent of a screened Coulomb potential
can be in some models\footnote{Cf. e.g. Gabovich \etal. \cite{GIPR78}.}  $A=\frac{5}{2} = d + \half$.
In any case, let us check that any summable power-law decay can be tolerated in the framework
of our general approach.

\subsection{A convolution lemma}
\label{ssec:conv.lemma.A.B}

Our main results are formulated for individually scaled identically distributed random variables,
but it is clear from the proofs that the assumption of identical distribution can be substantially relaxed.
In many cases, it suffices to have a global positive lower bound for their variances and some decent
bounds on their moments of order $3$. Some finer results require a uniform boundedness which
probably can be relaxed to finiteness of some exponential moment.

To reduce the number of auxiliary constants in intermediate statements and calculations,
we often use a standard notation $f(s) \asymp g(s)$ for functions of an integer
or real parameter $s$, usually in the context where $s\uparrow +\infty$ or $s\downarrow 0$, meaning that
$C_1 g(s) \le f(s) \le C_2 g(s)$ for some $C_1, C_2\in(0,+\infty)$.
Respectively, $f(s) \gtrsim g(s)$ will stand for $f(s) \ge C_2 g(s)$.

The first result, Lemma \ref{lem:Main}, applies to a large class of marginal measures not concentrated
on a single point. This class contains all compactly supported measures, i.e., those of bounded
(and non-constant) random variables, but boundedness can be relaxed to finiteness of the third moment.
In this connection, it is worth recalling  the results by Jessen and Wintner
\lcite{JW1935}{Section 6, Example 7} which show that some random series with unbounded terms and
merely converging in mean square (i.e., in $\rL^{\!\!2}(\Om)$)
may even have an analytic probability density of their sum (defined
almost surely, as an unbounded  random variable, but not pointwise).


Below we often use a formal convention $\ln 0^{-1} = +\infty$ convenient for lower bounds.

\ble
\label{lem:Main}
Let be given a family of IID random variables
$$
X_{n,k}(\om), \; n\in \DN, \;\; 1 \le k \le K_n \,, \;\; K_n \asymp n^{d-1}, \;\; d\ge 1.
$$
Assume that $\esm{X_{n,\bullet}}$ exists and is zero,
and the common characteristic function $\ffi_X(t) = \esm{ \eu^{\ii t X_{n,\bullet}}}$ obeys
\be
\label{eq:Main.Lemma.ffi.t0}
\ln \, \big| \ffi_X(s) \big|^{-1}  \gea s^2, \;\; |s|\le s_0\in(0,+\infty).
\ee
Let
$$
\bal
S(\om) &= \sum_{n\ge 1} \; \sum_{k=1}^{K_n} \fa_{n,k} X_{n,k}(\om), \;\; \fa_{n,k} \asymp n^{-A} \,, \;\; A>d\,,
\eal
$$
and
$$
\bal
S_{M,N}(\om) &= \sum_{n=M}^N \; \sum_{k=1}^{K_n} \fa_{n,k} X_{n,k}(\om), \;\; M \le N \,.
\eal
$$
Then the following holds true.
\begin{enumerate}[\rm(A)]
  \item\label{item:Main.Lemma.A}
There exists $c\in(0,+\infty)$ such that
$$
\all t\in\DR  \quad \left| \esm{\eu^{\ii t S(\om)}} \right| \lea \eu^{- c|t|^{d/A}} \,.
$$

  \item\label{item:Main.Lemma.B}
   For $N \ge (1+c')M \ge 1$ with $c'>0$,  and $|t| \le N^{A}$,
\be
\label{eq:Main.Lemma.B}
\ln \left| \esm{ \eu^{\ii t S_{M,N}(\om)}} \right|^{-1}
\gea  M^{-2A+d} \,t^2 \,.
\ee


  \item\label{item:Main.Lemma.C}
  For any $\eps\ge N^{-A}$
\be
\label{eq:Main.Lemma.C}
\sup_{a\in\DR} \; \pr{ S_{M,N}(\om) \in [a,a+\eps] } \lea M^{ A-\frac{d}{2} } \, \eps \,.
\ee
\end{enumerate}
\ele



Validity of the assumption \eqref{eq:Main.Lemma.ffi.t0} can be established for a large class of
probability measures; it is well-known (cf., e.g., \cite{Feller66}) that it suffices to
require $\esm{|X_{n,\bullet}|^3}<+\infty$, but in applications we always assume
$\| X_{n,\bullet}\|_{\rL^\infty(\Om)} = \essup |X_{n,\bullet}|<+\infty$.

\proof

\par\noindent
\textbf{(A)} By the IID property of the family $\{ X_{n,k} \}$ we have
$$
\bal
\ffi_{S}(t) &:= \esm{ \exp\left( \ii t \sum_{n\ge 1} \sum_{k=1}^{K_n} \fa_{n,k} X_{n,k}(\om) \right)}
= \prod_{n\ge 1} \;\; \prod_{k=1}^{K_n} \esm{ \eu^{ \ii t \fa_{n,k} X_{n,k}}}
\\
&
= \prod_{n\ge 1} \;\; \prod_{k=1}^{K_n} \ffi_ X\big( t \fa_{n,k}\big)
= \prod_{n\ge 1}  \left(\ffi_ X\big( t \fa_{n,k}\big)\right)^{K_n} \,.
\eal
$$
When $t\ne 0$ is fixed, we have to distinguish between "small" and "large" values of $n$, making use of
the local bound \eqref{eq:Main.Lemma.ffi.t0} for $s = t \fa_{n,k}$ with $n$ large enough
(i.e., with $\fa_{n,k}$ small enough).
To this end, introduce an integer threshold $N_t = C|t|^{1/A}$ with $C$ chosen so that
\be
\label{eq:def.Nt.polynom}
\forall n \ge N_t \quad n^{-A} |t| \le N_t^{-A} |t| \in[0, s_0] \,.
\ee
Then for the logarithm of $\ffi_S(t)$ we have
\be
\label{eq:Main.Lemma.ln.ffi.b.1}
\bal
\ln \big| \ffi_S(t) \big|^{-1} &= \sum_{n\ge 1} K_n  \ln \big| \ffi_X\big( \fa_{n,k} t \big) \big|^{-1}
\gtrsim \sum_{n\ge 1} n^{d-1}  \ln \big| \ffi_X\big( \fa_{n,k} t \big) \big|^{-1}
\\
&
= \left(\sum_{n=1}^{N_t} + \sum_{n>N_t} \right) n^{d-1}  \ln \big| \ffi_X\big( \fa_{n,k} t \big) \big|^{-1}
\\
&
=: \mcS_1(t) + \mcS_2(t) \,,
\eal
\ee
where all terms in $\mcS_1$ and in $\mcS_2$ are non-negative, since $|\ffi_X(t)|\le 1$ for any $t$.
We focus on $\mcS_2(t)$. By definition of the threshold $N_t$ and the hypothesis \eqref{eq:Main.Lemma.ffi.t0},
$$
\bal
\mcS_2(t)
\gea \sum_{n > N_t} \ln \big| \ffi_X\big( \fa_{n,k} t \big) \big|^{-1}
&
\gtrsim\, t^2 \,  \sum_{n > N_t}  n^{d-1} \fa^2_{n}
\gtrsim \, t^2 \,  \sum_{n > N_t}  n^{-2A+d-1}
\\
&
\gtrsim t^2 \, N_t^{-2A+d}
\gtrsim
t^2 \, |t|^{-\frac{2A - d}{A}}
\\
&
=
|t|^{d/A} \,,
\eal
$$
which proves assertion \eqref{item:Main.Lemma.A}.

\noindent
$\blacktriangleleft$ As the matter of fact, the sum $\mcS_1(t)$ can be assessed with the help of Wintner's approach
\cite{Wint1934} based on a simple equidistribution type result by P\'olya and Szeg\"{o}
\lcite{PolSzego25}{Section II.4.1, Problem 155}.
The final result would be similar : applying Wintner's method to $\mcS_1$,
we would obtain $\mcS_1(t)\gea |t|^{d/A}$, so
discarding the "ripple" contribution $\mcS_1(t)$ to the characteristic function does not result in a noticeable
loss. $\blacktriangleright$

\par\noindent
\textbf{(B)} Proceed as above:
$$
\bal
\sum_{n=M}^{N} \ln \big| \ffi_X(\fa_{n,k} t)\big|^{-1}
&
\gtrsim t^2 \sum_{n=M}^{N} n^{-2A+d-1}
\gtrsim t^2  \int_{n=M}^{N} s^{-2A+d-1} \, ds
\\
&
\gtrsim \, t^2 \, M^{-2A+d}
\\
&
\gtrsim |t|^{d/A} \,.
\eal
$$
Observe that for $N=M$, or close to $M$, we would have a weaker lower bound by
$C |t|^{\frac{d-1}{A}}$, but with $d>1$ it is still good enough for the proof of infinite
derivability. This also works when $d>1$ is non-integer and arbitrarily close to $1$. One possible setting
where this observation can be useful is
a subset of $\DZ^d$ with the rate of growth of balls $r\mapsto r^{1+\delta}$, $\delta>0$.

\vskip2mm

The proof of Assertion (C) requires a technical detour, so we postpone it to Section
\ref{ssec:proof.assertion.C.Berry.theorems}. In fact, it can be proved directly with the help of
arguments similar to those used in the proof of Assertion (B). It can also be easily inferred from a Berry's
theorem. Such a derivation brings up an interesting question that we address in the present paper only
in passing: how far can one really go in asymptotic expansions for the "finite bath" induced probability measures
of the cumulative potential, given its size $R<\infty$? While a mere CLT-type Gaussian asymptotics turns out to
be sufficient for most pragmatic purposes in this paper, one can wonder if one can prove some
higher-order asymptotics, of the type going back to the classical works by Chebyshev \cite{Cheb1887},
Berry \cite{Berry1941}, Cornish and Fisher \cite{CF38}, Edgeworth \cite{Edgew1905},
Feller \cite{Feller66}, Gnedenko and Kolmogorov \cite{GKolm54}
and many other researchers in probability and statistics.


\subsection{Proof of Assertion (C) of Lemma \ref{lem:Main}}
\label{ssec:proof.assertion.C.Berry.theorems}

We need to assess the integrals of the probability measure of $S(\om)$ over intervals $I_\eps\subset \DR$
of length $\Ord{\eps}$. It will be clear from the calculations given below that it suffices to consider
the case where
$I_\eps$ is centered at the origin; a shift results in factors of unit modulus, so we stick to
$I_\eps = [-\eps,\eps]$ to have less cumbersome formulae. Further, since the main estimate will be achieved
in the Fourier representation, it is customary to work with a smoothed indicator function
instead of $\one_{I_\eps}$. A very convenient choice is made in a number of works in the theory
of asymptotic expansions for limiting distributions of the sums of IID random variables
(cf. \cite{Cram37,Ess45,Feller66}).
As is well-known from standard courses of probability theory (cf., e.g., \cite{Feller66}),
for any $T>0$ the compactly supported function
\beal
\label{eq:def.smoothing.g.T}
g_{T}(t) := \left(1 - T^{-1}|t| \right)\one_{[-T,T]}(t) \,, \;\; t\in\DR \,,
\eeal
is the characteristic function of the probability measure with density
%
\beal
\label{eq:def.smoothing.p.T}
p_{T}(x) &= \frac{1 - \cos(T x)}{\pi T x^2}
= T \, \frac{1 - \cos(T x)}{\pi (T x)^2}
\,.
\eeal
For any $B>0$ and $T = \eps^{-1}$,
$$
\bal
p_{\eps^{-1}}(B\eps) & = \eps^{-1} \, \frac{1 - \cos(\eps^{-1} \cdot B\eps)}{\pi (\eps^{-1}B\eps)^2}
= \eps^{-1} \, \frac{1 - \cos B }{\pi B ^2} \,.
\eal
$$
In particular, one finds by a numerical calculation that for $B=\pi/3$
$$
\all x\in[-B\eps, B\eps] \quad
\eps p_{\eps^{-1}}(x)
\ge   \eps\cdot \eps^{-1} \, \frac{1 - \cos(\pi/3)}{\pi^3/9} > \frac{1}{8}.
$$
Since $\pi/3>1$, we have
\beal
\one_{[-\eps, \eps]}(x) \le \one_{[-B\eps, B \eps]}(x) \le 8 \eps p_{\eps^{-1}}(x).
\eeal
Therefore by the Parseval identity,
\beal
\label{eq:proof.C.J1.J2}
\mu_S(I_\eps) &= \int_\DR \one_{I_\eps}(x) \, dF_S(x)
\le 8\eps \int_\DR p_{\eps^{-1}}(x)  \, d \mu(x)
\\
&
\le 8\eps \int_\DR \big| \hp_{\eps^{-1}}(t) \big| \, \big|\ffi_\mu(t)\big| \, dt
= 8\eps \int_{|t|\le \eps^{-1}} \left( 1 - \eps^{-1} |t| \right)  \, \big|\ffi_\mu(t)\big| \, dt
\\
&
\le 8\eps \int_{|t|\le \eps^{-1}} \big|\ffi_\mu(t)\big| \, dt \,.
\eeal

Recall that we have assumed the lower bound \eqref{eq:Main.Lemma.ffi.t0}.
Define a mapping $n \mapsto T_n$ by
\be
\label{eq:def.T.N.pol}
T_n = \sup\myset{ t>0:\, \fa_n t \le s_0 }
= \sup\myset{ t>0:\,  n^{-A} t \le s_0 } = C(s_0) n^A \,,
\ee
and let
$$
\mcX_n := \myset{x\in\DZ^d:\, |x|\in[n,n+1)}\,, \quad n\in\DN,
$$
then
\be
\forall\, t\in[-T_n, T_n] \qquad
\sum_{x\in\mcX_n} \ln \big| \ffi_X(\fa_n t)\big|^{-1} \gea t^2 n^{-2A+d-1} \,.
\ee

$\bullet$ For $|t| \le T_M$, assuming $N \ge (1+c)M$ with $c>0$,
we have
\beal
\label{eq:polynom.t<T.M}
\ln \big| \ffi_{S_{M,N}}(t)\big|^{-1} &\ge \sum_{n=M}^{N} K_n \ln \big| \ffi_{X}(\fa_{n,k} t)\big|^{-1}
\gea t^2 \, \sum_{n=M}^{N} n^{d-1} \fa^2_n
\\
&
\gea t^2 \, M^{-2A+d} \,.
\eeal

$\bullet$ For $T_M \le |t| \le T_N$, hence for $|t| \in \big[ T_M, \, \eps^{-1} \big]$,
we have, setting $N_t := C' |t|^{1/A}$, as long as $(1+c)N_t \le N$ with some $c>0$,
\beal
\label{eq:polynom.T.M<t<T.N}
\ln \big| \ffi_{S_{M,N}}(t)\big|^{-1} &\ge \sum_{n=N_t}^{N} K_n \ln \big| \ffi_{X}(\fa_{n,k} t)\big|^{-1}
\gea t^2 \, \sum_{n=N_t}^{N} n^{d-1} \fa^2_n t
\\
&
\gea t^2 \, N_t^{-2A+d} \gea |t|^{2 + \frac{d-2A}{A} }
= |t|^{ \frac{d}{A} } \,.
\eeal
By assumption, $N \lea M^C$, so for any $\sigma>0$, $C_1,C_2\in(0,+\infty)$, and $M$ large enough
\be
\label{eq:TM.TN}
\eu^{ - C_1 M^\sigma} \le N^{-C_2}.
\ee
Thus writing
\be
\label{eq:J1.J1+.J1-}
\bal
8\eps \int_{-\eps^{-1}}^{\eps^{-1}}  \big|  \ffi_S(t) \big| \, dt
 =  8\eps \int_{-T_M}^{T_M}  \big|  \ffi_S(t) \big| \, dt
+ 8\eps  \int_{ T_M\le |t|\le \eps^{-1} }  \big|  \ffi_S(t) \big| \, dt
 =: J_- + J_+ \,,
 \eal
\ee
we have for $J_+$, on account of \eqref{eq:TM.TN},
\beal
\label{eq:bound.J1+}
J_+ &\lea 8\eps  \int_{ T_M}^{\eps^{-1}} \eu^{-c|t|^{d/A} } \, dt
= \ord{\eps} ,
\eeal
while for $J_-$ we can use a Gaussian-type bound:
\be
\label{eq:bound.J1.pol}
\bal
J_- &\le 8\eps  \int_{-T_M}^{T_M} \big|\ffi_S(t) \big| \, dt
\lea \eps \int_{\DR} \eu^{- t^2 M^{-2A+d} }  \, dt
\lea M^{A- \frac{d}{2}} \, \eps
\,.
\eal
\ee
This completes the proof of assertion (C) of Lemma \ref{lem:Main}: for $\eps \ge N^{-  A }$
$$
\mu_{S_{M,N}}(I_\eps) \lea M^{A- \frac{d}{2}} \, \eps + \ord{\eps}
\lea M^{A- \frac{d}{2}}\, \eps \,.
$$
\qed

\subsection{Auxiliary estimates for the characteristic functions}
\label{ssec:aux.char.funct}


The condition \eqref{eq:Main.Lemma.ffi.t0} has been \emph{assumed} in Lemma \ref{lem:Main}, but
for the intended applications we have to derive it from some moment inequalities.


The next result is a standard tool used in the proof of the Gaussian limit for the sums of
independent random variables with identical or comparable variances. The details of the proof
can be found, e.g., in \lcite{Feller66}{Section XV.4}.
\ble
\label{lem:bound.ln.ffi}
Assume that $\fm_3 := \esm{|X|^3} < \infty$ and $\esm{X}=0$\,, and let $\sigma^2 = \esm{X^2}$.
If $|t| \le \left(\frac{3}{5}\right)^{1/3} \sigma^2/\fm_3$, then
\be
\label{eq:bound.l,.ffi.2}
\left|  \ln \ffi(t)|^{-1} + \frac{\sigma^2 t^2}{2} \right|
\le \frac{5}{12} \fm_{3} |t|^3 \,.
\ee
\ele

\vskip1mm

This general result will be used in the situation where $X(\om)$ $= a_{x} \om_x$
with $\pr{|\om_x|\le 1} = 1$ for all $x$ and $a_{x} = \fu(|x|)$, so
$\fm_3 = \esm{ a_{x}^3 |\om_x|^3} \le a_{x}^3 \, \fmbar_3$, where
\be
\label{def:mu.bar}
\fmbar_3 := \esm{|\om_\bullet|^3}\in(0,1] \,.
\ee
Notice that that the key ratio used in Lemma \ref{lem:bound.ln.ffi} reads as
$$
\frac{\fm_{2}}{\fm_3} = \frac{ \esm{ a_{x}^2 |\om_x|^2} }{ \esm{ a_{x}^3 |\om_x|^3} }
= \frac{ \overline{\fm}_2 }{ a_{x} \, \fmbar_3 }\, , \qquad
\overline{\fm}_2 := \esm{|\om_\bullet|^2}\in(0,1] \,,
$$
and $\overline{\fm}_2/\overline{\fm}_3\in(0,+\infty)$
is a fixed parameter characterizing the common probability distribution of the
IID random amplitudes $\om_x$. For example, $\overline{\fm}_2 = \overline{\fm}_3 = 1$ for the Bernoulli
distribution with atoms $\{-1, +1\}$.

\subsection{Smoothness of disorder for polynomial potentials.
Proof of Theorem \ref{thm:thermal.bath.F.V.polynom}}

\label{ssec:thermal.bath.ch.f}

The claim follows directly from Lemma \ref{lem:Main};
we only need to identify its main parameters and check the validity of its
assumptions. Fix $t>0$, which we can assume
sufficiently large, let $\DN\ni n\ge 1$ and denote
\be
\label{eq:def.cX.r.n.r}
\bal
\mcX_n &:= \big\{ \, x\in\DZ^d:\, |x|\in[n, n+1) \, \big\} \,,
\\
\fa_{n,k} &:=  n^{-A} \,,
\;\;
1 \le k \le K_n  := \big| \mcX_n \big| \,.
\eal
\ee
The random variables $X_{n,k}$ figuring in Lemma \ref{lem:Main} are now $\om_x$ with $x\in \mcX_n$, numbered
in an arbitrary order by $k\in[1, K_n]$.

Note that Lemma \ref{lem:Main} applies here, since $\om_x$ are a.s. bounded, thus have
finite  moments of all orders.
Let
$N_t = \left\lfloor C |t|^{1/A} \right\rfloor$,
where $C$ is chosen so that for $|t|$ large and any $n \ge N_t$ one has
$$
\fa_{n,k} |t| \le \fa_{N_t} |t| \sim N_t^{-A} \, |t| \sim C^{-A} |t|^{-1} \, |t| = C^{-A} \le \frac{3}{5} \fm_2^{1/2} \,,
$$
hence by Lemma  \ref{lem:bound.ln.ffi}
$$
 \ln \big| \ffi_\mu\big( \fa_{n,k} t \big) \big|^{-1} \ge  \frac{\overline{\fm}_2 \,}{4} r^{-2A} t^2\,.
$$
Now the claim follows from Lemma \ref{lem:Main}:
$\big| \ffi_S(t) \big| \le \Const\, \eu^{ - c|t|^{-d/A} }$, so $\ffi_S$
is the [inverse] Fourier transform of a probability measure on $\DR$ with density $\rho_V\in\mcC^\infty(\DR)$
relative to the Lebesgue measure.

Finally, let $F_V$ be the probability distribution function
of the measure with density $\rho_V$. Due to the a.s. boundedness of the potential $V$, we have
$\nu_* := \inf \supp \rho_V > -\infty$, and since $\rho_V|_{(-\infty, \nu_*]} \equiv 0$,
it follows from the finite Taylor expansions of all orders of  $F_V\in\mcC^\infty(\DR)$
at $\nu_*$ that $F_V(\nu_* + \lam) = \ord{|\lam|^\infty}$.
\qedhere

\section{Smoothness of disorder under exponential screening}
\label{sec:char.f.exp}

\subsection{An adaptation of Lemma \ref{lem:Main}}

In the condition on the combinatorial parameter $K_n$ in the formulation of the next general result,
$d>1$ is an arbitrary number, and in application to the smoothness of cumulative potential
in $\DZ^d$, the optimal choice for $d$ is exactly the dimension of the lattice. Technically, however,
a more pertinent parameter is $d-1$ which we need to be strictly positive. Geometrically, it is the
exponent of power-law growth of the cardinality of a "sphere" $\mcA_{n} = \{x\in\DZ^d: \, |x|\in[n,n+1)\}$
as $n\to\infty$.
In specific applications
one may want to, or have to,
restrict $\mcA_{n}$ to a subset thereof, $\mcX_n$, of smaller cardinality $n^{\rro-1}$, $0 < \rro \,< d-1$.

\ble
\label{lem:Main.exp}
Let be given a family of IID r.v.
$$
X_{n,k}(\om), \; n\in \DN, \;\; 1 \le k \le K_n \,, \;\; K_n \asymp n^{d-1}, \;\; d>1,
$$
and assume that their common characteristic function $\ffi_X(t) = \esm{ \eu^{\ii t X}}$ fulfills
\be
\label{eq:Main.Lemma.ffi.t0.exp}
\ln \, \big| \ffi_X(t) \big|^{-1} \ge C_X t^2, \;\; |t|\le s_0.
\ee
Let
$$
\bal
S(\om) &= \sum_{n\ge 1} \sum_{k=1}^{K_n} \fa_{n,k} X_{n,k}(\om), \;\; \fa_{n,k} \asymp \eu^{- a n^{\delta} } \,,
\;\; a,\delta>0,
\\
S_{M,N}(\om) &= \sum_{n=M}^N \sum_{k=1}^{K_n} \fa_{n,k} X_{n,k}(\om), \;\;  N\in[M, +\infty] \,.
\eal
$$
Denote $\gamma := \frac{d-1}{\delta}$. Then the following holds true.
\begin{enumerate}[\rm(A)]
  \item\label{item:Main.Lemma.A.exp}
There exists $C, c\in(0,+\infty)$ such that
for any $N\in[M, +\infty]$
\be
\label{eq:Main.Lemma.exp.B}
\all |t| \le \eu^{ c N^\delta} \quad \ln \big|\ffi_{M, N}(t)\big|^{-1} := \ln \left| \esm{ \eu^{\ii t S_{M,N}(\om)}} \right|^{-1}
\ge \frac{C}{a \delta } \, \ln^\gamma |t|  \,.
\ee
Consequently,
\be
\label{eq:Main.Lemma.exp.A}
\all t\in\DR  \quad \big| \ffi_S(t) \big|
\lea \eu^{ -  \frac{C}{a \delta} \ln^\gamma |t| }
\lea |t|^{ - \frac{C}{a \delta} \ln^{\frac{d-1-\delta}{\delta}} (|t|) } \,,
\ee
hence for $d>1+\delta$ the random variable $S(\om)$ has density $\rho_S\in\mcC^\infty(\DR)$.

   \item
   Assume that $d=2$ and $\delta=1$. Then there exist $a_*>0$ and $C>0$ such that
  if $0 < a \le a_*$, then $S$ has a density $\rho_S\in\mcC^Q(\DR)$ with $Q \ge C a^{-1}\ge 1$.

  \item\label{item:Main.Lemma.C.exp}
  Let $I_\eps\subset \DR$ be an interval of finite length
$|I_\eps|=\eps \ge \eu^{ -a N^\delta }$. Then
\be
\label{eq:Main.Lemma.D.exp}
\pr{ S_{M,N}(\om) \in I_\eps \cond \fF_{M,N}} \lea \eu^{a M^\delta} \,\eps
\lea \eps^{1 - \kappa_{M,N}} \,,
\ee
where $\fF_{M,N}$ is generated by $\myset{X_{n,k}, \, n\not\in[M,N]}$ and
$
\kappa_{M,N} = \left(M/N\right)^\delta \tto{N/M \to +\infty} 0\,.
$


\end{enumerate}
\ele

\proof
\textbf{(A)}
As in Lemma \ref{lem:Main}, we have, with
$\fa_{r} = \fu(r)$,
$$
\bal
\ln \big| \ffi_S(t) \big| &= \sum_{n=0}^{\infty} \, \sum_{k=1}^{K_n} \ln\big| \ffi_X( \fa_{n,k} t) \big| .
\eal
$$
To assess the range of $t$ for which $|1 - \ffi_X(\fa_{n,k} t)| \asymp \fa_r^2  t^2$,
use again Lemma \ref{lem:bound.ln.ffi}: with $\sigma_{n,k}^2 = \esm{\fa_{n,k} X_{n,k}^2}$
\be
\label{eq:sub-exp.log}
\left| \ln \ffi_{\fa_{n,k} X}(t) + \shalf \sigma_{n,k}^2 t^2 \right| \le \frac{5}{12} \mu_{n,k}^3 |t|^3.
\ee
Now we have $\mu_{n,k}^3 = \esm{ \fa_{n,k}^3 |\om_{n,k}|^3} \le \fa_r^3 = \eu^{-3 a r^\delta}$, since
$\pr{|\om_{n,k}|\le 1} = 1$.
If $|t|\le \frac{3\sigma_r^2}{5\mu_{n,k}^3}$, the RHS of \eqref{eq:sub-exp.log}
is upper-bounded by $\sigma_{n,k}^2/4$.
Taking a suitable constant $C$,
we come to the following definition of the threshold $R_t$ to be used in the sequel:
\be
\label{eq:def.R.t.exp}
 C |t| = \eu^{ a R^\delta_t} \; \Longleftrightarrow \; R_t = C_a \ln^{1/\delta}|t| = \frac{C'}{a} \ln^{1/\delta}|t| \,,
\ee
so that for $|t|>0$, $r\ge R_t$, it holds that
$\ln |\ffi_V( \fa_{n,k} t )|^{-1} \ge \sigma_{n,k}^2 t^2/4$.
It follows that
\be\label{eq:log.ffi.thermal.1}
\bal
\sum_{n \ge 1} \; \sum_{k=1}^{K_n} \ln |\ffi_V( \fa_{|x|} t)|^{-1} & \ge
\sum_{n \ge R_t} \; \sum_{k=1}^{K_n} \ln |\ffi_V( \fa_{{n,k}} t)|^{-1}
  \gea t^2 \sum_{n \ge R_t} r^{d-1} \fa_{n,\bullet}^2
\\
&
\gea t^2  \,  \sum_{r \ge R_t} r^{d-1} \eu^{ -2r^\delta}
 \gea t^2 \int_{R_t}^{+\infty} r^{d-1} \eu^{-2r^\delta} \, dr
\\
&
 = t^2 \delta^{-1}
    \int_{2R_t^\delta}^{+\infty}
s^{ \frac{d-1}{\delta} } \eu^{ - s} s^{\frac{1}{\delta} - 1}\, ds
\\
&
=  t^2 \delta^{-1} \Gamma\left( \frac{d}{\delta},\, 2R_t^\delta \right) ,
\eal
\ee
where $\Gamma(\cdot,\cdot)$ is the incomplete upper Gamma function
$\Gamma(\alpha,z) := \int_z^{+\infty} u^{\alpha-1} \eu^{-u} \, du$
having a well-known asymptotics (cf. \lcite{GraRyz07}{8.357--8.358})
$$
\lim_{z\to+\infty} \; \frac{\Gamma(\alpha,z)}{z^{\alpha-1} \eu^{-z}} =1 \,, \quad
s >-1, \; z\in\DR_+ \,,
$$
which shows that the contribution to the LHS of \eqref{eq:log.ffi.thermal.1} from a single term
with $r= R_t$ cannot be significantly improved by taking
the entire tail sum $\sum_{r \ge R_t}(\,\cdot \,)$.
Therefore, for any $R \in [R_t, +\infty]$, we have to settle for a bound of the form
$$
\bal
\sum_{n =R_t}^{R} \; \sum_{k=1}^{K_n} \ln |\ffi_V( \fa_{n,\bullet} t)|^{-1}
&
\gea \delta^{-1} t^2  \,  R_t^{ \delta\left(\frac{d}{\delta} - 1\right)} \eu^{- 2 R_t^{\delta} }
=
\delta^{-1} t^2  \,  R_t^{ d - \delta} \eu^{- 2 R_t^{\delta} }
\\
&
\gea a^{-1} \delta^{-1} t^2  \,  \left(\ln |t|\right)^{\frac{d -  \delta}{\delta}}
   \, \eu^{- 2\ln^{\delta \cdot \frac{1}{\delta} } |t| }
\\
&
\gea  a^{-1}\, \delta^{-1} \left(\ln |t|\right)^{\frac{d - \delta}{\delta} } \,.
\eal
$$
Solving $r \ge C_a \ln^{1/\delta}|t|$ for $t$ as a function of $r$,
we see that, once $R>0$ is fixed, a suitable  bound  on $\ffi_R(t)$ holds for
all $t\in[-T_R,T_R]$ with
$T_R := C \eu^{-R^\delta}$, for some $C\in(0,+\infty)$.

This proves assertion \eqref{item:Main.Lemma.A.exp}.

\vskip1mm
\noindent
\textbf{(B)} This assertion follows immediately from (A).

\vskip1mm
\noindent
\textbf{(C)}
We have to assess the integrals of the probability measure of $S(\om)$ on intervals $I_\eps$ of length
$\Ord{\eps}$.

Arguing as in the proof of Lemma \ref{lem:Main} and using again the smoothing trick with the density
\eqref{eq:def.smoothing.p.T}, we obtain for any interval $I_\eps = [a-\eps,\,a+\eps]$,
uniformly in $a\in\DR$,
$\mu_{S}(I_\eps \cond ;\om_{\mcX,\oball}^\perp) \le J_- + J_+$,
where
\beal
J_- &=  8\eps \int_{-T_M}^{T_M} \big| \ffi_S(t) \big| \, dt
\lea  \eps \int_{-T_M}^{T_M} \exp\left( -C t^2 \lea  \eps \int_{-T_M}^{T_M} \eu^{-C t^2 M^{d-1} \eu^{-aM^\delta} } \, dt \right) \, dt
\lea  \eu^{aM^\delta} \, \eps \,,
\\
%
J_+ &=  8\eps \int_{T_M \le |t| \le \eps^{-1}}
   \exp\left( - \frac{C}{a \delta} \ln^{\gamma} |t| \right)\frac{}{} \, dt \,,
\eeal
with $\gamma = \frac{d-\delta}{\delta}$. If either $d>1+\delta$,
or $d-1=\delta=1$ and $a>0$ is small enough,
then the characteristic function $\ffi_S$, upper-bounded by
$\eu^{-\frac{C}{a \delta} \ln^{\gamma} |t|}$ for large $|t|$,
is integrable on $\DR$, in which case $J_+ = \ord{\eps}$ and
\be
\label{eq:proof.B.exp.final.1}
\mu_{S_{M,N}}(I_\eps) \lea \eu^{ a M^\delta} \, \eps + \ord{\eps}
\lea  \eu^{ a M^\delta} \, \eps \,.
\ee
This completes the proof of assertion (C).
%
\qedhere

\section{Eigenvalue concentration estimates. Staircase potentials}
\label{sec:DoS.staircase}

\subsection{Smoothness of DoS}

Due to the translation invariance of the random field $\om_\bullet$, it suffices to consider the case
where $\ball = \ball_L(0)$.

For each $L,n\in\DN$, and $x\in\DZ^d$, introduce the lattice subsets
\beal
\mcA_n(x) &= \ball_{\fr_{n+1}}(x)\moins\ball_{\fr_{n}}(x)
\\
\mcX_n &= \myset{x\in\DZ^d:\, \ball_L(0) \subset \mcA_n(x)=\ball_{\fr_{n+1}}(x)\moins\ball_{\fr_{n}}(x) }
\eeal

Next, consider the following decomposition of the operator $H_\ball(\om)$:
\beal
\label{eq:staircase.H.tA.W}
H_\ball(\om) & = \tA_\ball(\om) + W_\ball(\om),
\eeal
where $W_\ball(\om)$ is the operator of multiplication by the random function
$$
y \mapsto W_\ball(y,\om) = \sum_{n\ge n_\circ} \sum_{x\in\mcX_n} \om_x \fukap(x-y)
$$
with a suitably chosen $n_\circ\ge 0$ (to be defined below), and
$\tA_\ball(\om) = -\Delta_\ball + \tW_\ball(y,\om)$ with
$$
\tW_\ball(y,\om) = \sum_{x\not\in\cup_{n\ge n_\circ}\mcX_n} \om_x \fukap(x-y).
$$
The contribution $\tW_\ball(\cdot,\om)$ to the potential energy on $\ball$ will be rendered below
non-random by conditioning on $\myset{\om_x\, x\not\in\cup_{n\ge n_\circ}\mcX_n}$, and our regularity analysis
will rely exclusively on the terms from $W_\ball(\cdot,\om)$.

The threshold $n_\circ$ is defined as follows. We need to cover $\ball$ by the plateaus of the potentials
$y \mapsto \om_x\fukap(x-y)$; in other words, we need that the potential $\om_x \fukap(x-\cdot)$
take a constant value on the entire cube $\ball_L(0)\subset\mcA_n(x)$.
By \eqref{eq:def.fukap}, for each fixed $x\in\mcX_{n}$ the potential $\om_x \fukap(x-\cdot)$
takes a constant value on an annulus $\mcA_n(x)=\ball_{\fr_{n+1}}(x)\moins\ball_{\fr_{n}}(x)$.
for $\mcX_n$ to be non-empty,
let alone  having a large cardinality, $n$ must be such that
\be
\label{eq:cond.kappa.rn.L.1D.1}
\fr_{n+1} - \fr_n \equiv (n+1)^{\varkappa} - n^\varkappa
\ge \diam\, \ball_L(y) \asymp  L \,.
\ee
Clearly, $(n+1)^{\varkappa} - n^\varkappa \asymp n^{\varkappa-1}$.
For large $L$, the index $n$ must also be large:
\be
\label{eq:kappa.n.L}
n \ge n_\circ = n_\circ(L,\varkappa) \asymp L^{\frac{1}{\varkappa-1}}
\;\; \Longrightarrow \;\; \fr_n \ge \fr_{n_\circ} \gea L^{\gamkap} \,,
\quad \gamkap := \frac{\varkappa}{\varkappa - 1} .
\ee
An elementary geometrical argument shows that
\beal
\min_{y\in\ball_L(0)} |y-x| & \equiv \dist\left(x, \ball_L(0) \right)
  \in \left[\,  |x| - C_1 L, \, |x| + C_1 L \, \right] ,
\\
\max_{y\in\ball_L(0)} |y-x| & \in \left[\, |x| - C_2 L, \, |x| + C_2 L \, \right] ,
\eeal
thus
$$
\all x\in \ball_{\fr_{n+1}-C'L}(0) \moins \ball_{\fr_{n}+C'' L}(0)
\qquad
\ball_L(0) \subset \ball_{\fr_{n+1}}(x) \moins \ball_{\fr_{n}}(x) .
$$
We can have therefore at our disposal the subsets
$\mcX_n = \ball_{C'L^{(1+\theta)\tau}}(0)\setminus \ball_{C''L^{\gamkap}}(0)$
with $L_\varkappa = L^{\gamkap}$, for some $C', C''>0$ and $\tau > \gamkap$.
The choice of the constants $C', C''$ need not be optimal, for it has virtually no impact on
the final estimates.

When \eqref{eq:cond.kappa.rn.L.1D.1} is satisfied, there exist some points $x\in\DZ^d$
such that an entire given ball $\ball_L(u)$ is covered by a plateau
of the potential $\fukap(x - \cdot)$ on which it takes the value $\fr_n^{-A}$.
Consequently,
\beal
W_\ball(y)
&
= \one_\ball(y) \sum \sum_{n\ge n_\circ} \sum_{x\in\mcX_n} \om_x \fukap(x-0)
\\
&
=: \one_\ball(y) \eta_\ball(\om) .
\eeal
We see that the random variable $\eta_\ball(\om)$ is merely the value of the cumulative
potential induced at the origin $x=0$ by the random potentials originating at all the lattice points
$x\in\cup_{n\ge n_\circ}\mcX_n$. One can also formulate it in a slightly different way:
\be
\eta_\ball(\om) = V(0,\om)- \tW_\ball(0,\om),
\ee
and the last term $\tW_\ball(0,\om)$ is rendered non-random by conditioning
on the \sigal $\fB_\ball^\perp$ generated by $\myset{\om_x, \, x\not\in \cup_{n\ge n_\circ}\mcX_n}$.
Therefore, we return to a more comfortable framework of classical regularity analysis of the
random field $V(\cdot,\om)$ obtained by the linear alloy transform \eqref{eq:def.BU.2}
of the random field $\om_\bullet$.

In the representation \eqref{eq:staircase.H.tA.W}, replace $W_\ball(y,\om)$ by
$\eta_\ball \one_\ball(y)$, condition on $\fB_\ball^\perp$ defined above.
The scalar  operator $\eta_\ball(\om) \one_\ball$ is stochastically independent of,
and commutes with, $\tH_\ball(\om)$, so we can drop the argument $\om$ in $\tW_\ball(y,\om)$
(becoming non-random by conditioning) and write
\beal
\label{eq:staircase.H.tA.W.eta.B}
H_\ball(\om) & = \tA_\ball + \eta_\ball \one_\ball,
\eeal
Labeling all the eigenvalues $\lam_j(\om)$ of $H_\ball(\om)$ in a measurable way, we infer from
the identity \eqref{eq:staircase.H.tA.W.eta.B} that
\be
\label{eq:lam.tlam.staircase}
\lam_j(\om) = \hlam_j(\om) + \eta_\ball(\om), \;\; j=1, \ldots, |\ball|,
\ee
where all r.v. $\hlam_j$ are independent of $\eta_\ball$. The regularity of the probability measure
$\mu_{\eta_\ball}$ of
$\eta_\ball \equiv V(0,\om)- \tW_\ball(0,\om)$ can therefore be established with the
help of Lemma \ref{lem:Main}; we only need to identify
its key ingredients:
$$
\bal
\mcX_n &: = \myset{x\in\DZ^d:\, \ball_L\subset \mcA_n(x) }\,,
\;\; K_n := \big|\mcX_n \big| \,,
\\
\myset{ \om_x, \, x\in\mcX_n } & \leftrightarrow \myset{ X_{n,k}, \, k=1, \ldots, K_n } ,
\;\; n\ge n_\circ\,,
\\
M &:= \fr_{n_\circ}, \;\; N = +\infty\,.
\eal
$$
By Lemma \ref{lem:Main},
the characteristic function $\ffi_{\eta_\ball}(t)$ of the measure
$\mu_{\eta_\ball}$ decays faster than any negative power of $|t|$, so $\mu_{\eta_\ball}$
admits a probability density $\rho_{\eta_\ball}\in\mcC^\infty(\DR)$.
By the identities \eqref{eq:lam.tlam.staircase}, this implies the infinite smoothness
of the IDS (hence, of the DoS) in $\ball$.
\qedhere

\subsection{Wegner estimate}

\begin{figure}
\begin{tabular}{c}
%
%
%
\begin{tikzpicture}
\begin{scope}[scale=0.15]
\clip (-7,-10) rectangle ++(100.0,22.0);

\draw[color=white!85!gray, line width = 34] (30, 0) circle (31);

\fill[color=white!70!gray] (-2, -2) rectangle ++(4, 4);
\draw (-2, -2) rectangle ++(4, 4);

\node[](An) at (0, -7) {$\mcA_n(x)$};



\node (ball) at (-0.5, 7.0) {$\ball_L(0)$};
\draw[->,bend left = 20] (ball.south) to (0, 2.5);

%


\begin{scope}
\draw[color=white!50!gray, line width = 30] (0, 0) circle (20);
\end{scope}

\draw[dotted, color=white!10!black, line width = 0.5] (30, 0) circle (27);
\draw[color=white!10!black, line width = 0.5] (30, 0) circle (35);


\fill (33, 17.5) circle (0.34);
\draw[->] (20, 0) -- (3.5,-1);
\draw[->] (20, 0) -- (-4, -5);

\node[rotate=5](r2) at (11, 0.5) {$\fr_{n}$};
\node[rotate=14](r1) at (9, -3.7) {$\fr_{n+1}$};

\node (cXn) at (19, 7.0) {$\mcX_n$};



\node[](x) at (21, 0) {$x$};
\fill(20,0) circle (0.3);

\begin{scope}
\clip (20,-10) rectangle ++(100.0,22.0);
\draw[color=white!50!gray, line width = 50] (0, 0) circle (35);

\node (cXnplus) at (35, 7.0) {$\mcX_{n+1}$};

\draw[color=white!50!gray, line width = 80] (0, 0) circle (56);
\node (cXnplustwo) at (55, 7.0) {$\mcX_{n+2}$};

\end{scope}

\end{scope}

\end{tikzpicture}

\end{tabular}

\caption{  \footnotesize\emph{Example for Section \ref{sec:DoS.staircase}.}
For each fixed $x\in\mcX_{n}$, $n\ge n_\circ(L,\varkappa)$, the potential $\om_x \fukap(x-\cdot)$
takes a constant value on an annulus
$\mcA_n(x)=\ball_{\fr_{n+1}}(x)\moins\ball_{\fr_{n}}(x)$ (the leftmost light-gray arc),
hence on the entire cube $\ball_L(0)\subset\mcA_n(x)$. Therefore the sum of such potentials
is a random constant on $\ball_L(0)$ with a smooth probability measure. The regularity of the latter
can be assessed essentially in the same way as for the individual values of the cumulative potential
$V(y,\om)$, $y\in\ball_L(0)$. The remaining potentials $\om_x \fukap(x-\cdot)$
(those which are non-constant on $\ball_L(0)$) can be rendered non-random
by conditioning.}
\end{figure}
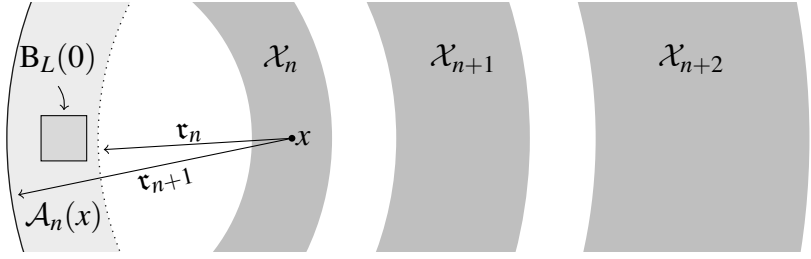%
%

The starting point for the "finite bath" eigenvalue concentration analysis is again the
representation \eqref{eq:lam.tlam.staircase} of the eigenvalues $\lam_j(\om)$
introduced in the previous subsection. The bulk of the technical work has been done in the
proof of Lemma \ref{lem:Main}, and we only have to make some adaptations due to the
specificity of the staircase potential $\fukap(r)$ and particularly the
constraints on the index $n$ in the lattice subsets $\mcX_n$ introduced in the previous subsection.
Again, we focus on the case where $\ball = \ball_L(0)$.

The union of the suitable subsets $\mcX_n$ is a large annulus
$$
\ball_{C'L^{(1+\theta)\tau}}\moins\ball_{C'' L^\gamkap}(0)
$$
from which we exclude a sequence
of thinner annuli, each of width of order $\Ord{L}$ (see the blank annuli separating
$\mcX_n$, $\mcX_{n+1}$, $\mcX_{n+2}$ on Fig~1), to avoid the potentials
$\om_x\fukap(x-\cdot)$ which are not constant on the entire cube $\ball$; let us call them unsuitable.
It is straightforward that relative density of unsuitable radii $r$ in a large interval
$\left[C'L^{(1+\theta)\tau}, \, C'' L^\gamkap \right]$ tends to $0$ as $L\to \infty$, so the exclusion
of unsuitable radii has no significant impact on the estimates given in Lemma \ref{lem:Main},
other than modification of some auxiliary constants.

By \eqref{eq:lam.tlam.staircase} each EV $\lam_j(\om)$ of the operator $H_{\ball}(\om)$ is a $j$-dependent
shift of $\eta_{\ball}(\om)$, and for the latter we can use the concentration estimates for the values
of the cumulative potential. Specifically, we can apply assertion (C) of Lemma  \ref{lem:Main} with
$M=L^\tau$ and $N=L^{\tau}$, $\tau> \gamkap$:
for any interval $I_\eps\subset\DR$ of length $|I_\eps| = \eps \ge L^{-A\tau}$,
we have
$$
\bal
\pr{ \lam_j(\om)\in I_\eps } &\lea (\fr_{n_\circ}\!)^{A - \frac{ d }{2}} \eps
\lea L^{\left( A - \frac{ d }{2} \right)\gamkap }\,  \eps \,.
\eal
$$
which proves the assertion \eqref{eq:claim.staircase.DoS.lam.j.eps.R}.
In particular, for $\eps = L^{-A\tau}$ we obtain
$$
\bal
\pr{ \lam_j(\om)\in I_{\eps} } &\lea L^{ - A\tau - \frac{ d \gamkap}{2} + A\gamkap}
&
\lea L^{ -A\tau \left(1  - \frac{\gamkap}{\tau}\left( 1 - \frac{d}{2A} \right) \right)}
\eal
$$
so the required result follows by counting the number of eigenvalues in $\ball_L(0)$:
$$
\bal
\pr{ \Sigma\big(H_\ball(\om)\big) \cap I_{\eps_R} \ne \varnothing }
\lea L^{ -A\tau \left(1  - \frac{\gamkap}{\tau}\left( 1 - \frac{d}{2A} \right) - \frac{d}{A\tau}\right)}
\eal
$$
\qedhere

\section{Eigenvalue concentration and comparison estimates. Polynomial potentials}
\label{sec:EVC.pol}

\subsection{A Bernstein-type lemma}
\label{ssec:Bernstein.lemma}

\ble
\label{lem:CLT.with.Bernstein}
Let be given random variables $X_k(\om)$, $1 \le k \le n$,  of the form
$$
X_k(\om) = \rY_k(\om) + \ry_k(\om) = a_k \om_k + \ry_k(\om), \quad \om = (\om_1, \ldots, \om_n),
$$
and assume the following.

\begin{enumerate}[\rm(1)]
  \item The family $\{\om_k, k\in\llb 1, n\rrb\}$ is independent.

  \item $|a_k| \le a \le n^{-2\beta}$ for some $\beta\in (0, 1)$, and $\| \om_k(\cdot) \|_\infty \le 1 $.

  \item $\| \ry_k(\cdot) \|_\infty \le n^{-\gamma} a$, $\gamma \ge 1$.

\end{enumerate}
Denote $\sigma_k^2 = \esm{ \rY_k^2}$ and
let $\Psi_n(t) = \esm{ \eu^{\ii t S_n(\om)}}$ where $S_n(\om) = \sum_{k=1}^n X_k(\om)$.
There
exist some $C_1, C_2\in(0,+\infty)$ such that
for all $t\in\DR$ with $|t| \le \frac{C_1}{a n^{5/12}}$ it holds that
\beal
\label{eq:lem.CLT.with.Bernstein}
|\Psi_n(t)|
&
\le C_2 \, \eu^{ - \frac{t^2}{2} \sum_{k=1}^n \sigma_k^2} .
\eeal
\ele

\proof
Gix $n\ge 1$. If $n=1$, the claim follows by a simple calculation. If $n>1$,
proceed by induction.
Suppose that for some $k\in\llb 1, n-1\rrb := [1, n-1] \cap \DZ$
$$
\bal
\Psi_n(t)
&
= \eu^{ - \frac{t^2}{2} \, \sum_{j=0}^{k-1} \sigma^2_{n-j} } \;\Psi_{n-k}(t) \,,
\\
\Psi_{n-k}(t) &= \esm{ \eu^{ \ii t \left(S_{n-k}  +  t^{-1} \delta_{n-(k-1)} \right) } } \,,
\eal
$$
where
\begin{align}
\label{eq:ind.hyp.delta}
\| \delta_{n-(k-1)} \|_\infty
&
\lea \sum_{j=0}^{k-1} \left( t \| \ry_{n-j} \|_\infty  + 2 t^2 n^{-\gamma} \sigma_{n-j}^2 + \eta_{n-j} \right),
\\
\label{eq:ind.hyp.eta}
 \| \eta_{n-j}  \|_\infty
&
\le \frac{5}{12} \cdot 2 \, \esm{|\rY_{n-j}|^3} \,  |t|^3
\lea n^{-5/4} .
\end{align}
Note that for $k=0$ there is nothing to prove, for we deal with $S_n \equiv S_n+0$.
Let
$$
\bal
\ttzeta_{n-k} &:= \ry_{n-k} + t^{-1} \delta_{n-(k-1)} \,,
\eal
$$
and denote $\esmk{j}{\cdot} = \esm{\cdot \cond \fF_j}$, where $\fF_j$ is the $\sigma$-algebra
generated by $\{X_i\,, i \le j\}$.
Then
$$
\bal
\esm{ \eu^{ \ii t \left(S_{n-k} + t^{-1} \delta_{n-(k-1)} \right)  } }
=
\esm{ \eu^{\ii t S_{n-(k+1)} }\;  \esmk{n-(k+1)}{ \eu^{\ii t\left( \rY_{n-k} + \ttzeta_{n-k} \right)} }\, } .
\eal
$$
Denote
$$
\bal
\ffi_{n-(k+1)}(t)
= \esmk{n-(k+1)}{ \eu^{\ii t (\rY_{n-k} + \ttzeta_{n-k})} } \,,
\eal
$$
then it follows from the absolutely convergent Taylor expansion of $\ln \ffi_{n-(k+1)}(t)$
(cf. \lcite{Feller66}{Section XVI.5, Eqn (5.6)}) that%
\beal
\label{eq:lemma.Bernstein.Taylor}
\eta_{n-(k+1)} & := \left| \ln \ffi(t)- \left( \ii t \esm{\ttzeta_{n-k}}
   - \frac{t^2}{2} \esm{\left(\rY_{n-k} + \ttzeta_{n-k} \right)^2 } \right) \right|
\\
&
  \le \frac{5}{12} |t|^3 \esm{|\rY_k + \ttzeta_k|^3} \,.
\eeal
Since $\| \ry_j\|_\infty = \ord{ \| \rY_j \|_\infty }$, for $n\gg 1$ we have, e.g.,
$\esm{|\rY_j + \ry_j|^3} \le 2  \esm{|\rY_j|^3}$, whence
\beal
\label{eq:ind.bound.eta.n-k}
\eta_{n-(k+1)}
&
\le \frac{5}{12} \cdot 2 \, \esm{|\rY_k|^3} \,  |t|^3
%
\le  C' |at|^3
\le  \left(\frac{C''}{n^{5/12}}\right)^3 \lea n^{-5/4}
\eeal
owing to the assumption $|t|\le C_1 a^{-1} n^{-5/12}$.
Under this assumption, we infer from \eqref{eq:lemma.Bernstein.Taylor}
$$
\bal
\ln \ffi(t)
&
= - \frac{t^2 \sigma_{n-k}^2}{2}  + \delta_{n-k}(t,\om),
\eal
$$
where $\sigma_{n-k}$ is non-random, and
$$
\delta_{n-k}(t,\om)
:= \ii t \esm{ \ry_{n-k} } + \ii \esm{ \delta_{n-(k-1)} }
- \frac{t^2}{2} \esm{ 2 \rY_{n-k} \ttzeta_{n-k} + \ttzeta_{n-k}^2} + \eta_{n-k} \,.
$$
Using the inductive hypothesis \eqref{eq:ind.hyp.delta} on $\delta_{n-(k-1)}$, we get
%
\beal
\label{eq:ind.bound.delta.n-k}
\| \delta_{n-k} \|_\infty
&
\le \sum_{j=0}^{n-k} \Big( |t| \,\| \ry_j\|_\infty + 2 t^2 n^{-\gamma} \, \sigma_j^2 + \eta_j \Big) .
\eeal
Assess the terms of the last RHS:
$$
\bal
|t| \sum_{j=0}^{n-k} \| \ry_j \|_\infty &\le n \max_j \| \ry_j \|_\infty
\le n^{-\gamma+1} \frac{C_1 a}{a n^{5/12}}
\lea n^{-\gamma + \frac{7}{12}} \,,
\\
2 t^2 n^{-\gamma} \sum_{j=0}^{n-k} \sigma_j^2
&\le 2 n^{-\gamma+1} \cdot \left(\frac{C_1a}{a n^{5/12} }\right)^2
\lea n^{-\gamma + \frac{1}{6}} \,,
\\
\sum_{j=0}^{n-k} \| \eta_j \|_\infty &\lea  \frac{n}{n^{5/4}} = n^{-1/4} \,.
\eal
$$
Recall that $\gamma>1$ by hypothesis.
Therefore, one has a uniform upper bound on $\delta_{\bullet}$:
$\max_j \|\delta_j\|_\infty \lea n^{-1/4}$.
This proves the inductive bound for $j=k$.

Further, it follows from the representation
$
\ffi_{n-(k+1)}(t)
= \eu^{ - \frac{\sigma_{n-k}^2t^2}{2} } \eu^{  \delta_{n-k}(t,\om) }
$
that
$$
\bal
\Psi_n(t)
&
= \eu^{ - \frac{t^2}{2} \, \sum_{j=0}^{k} \sigma^2_{n-j} } \;\Psi_{n-(k+1)}(t) \,,
\\
\Psi_{n-(k+1)}(t) &= \esm{ \eu^{ \ii t \left(S_{n-k}  +  t^{-1} \delta_{n-(k-1)} \right) } } .
\eal
$$
By induction in $k$ we find
$
\Psi_n(t)
= \eu^{ - \frac{t^2}{2} \, \sum_{j=1}^{n} \sigma^2_{j} } \; \esm{ \eu^{ \ii \delta_1} }
$,
$\| \delta_1 \|_\infty \le n^{-1/4}$, hence
$$
\bal
\big| \Psi_n(t) \big|
&
\le \Const \, \eu^{ - \frac{t^2}{2} \sum_{k=1}^n \sigma^2_k } \,.
\eal
$$
By hypothesis,  $a \asymp n^{-1/(2\beta)}$, so
\beal
\label{eq:def.n.cond.Bernstein.polynom}
a t \asymp n^{-5/12} \quad
\Longrightarrow  \quad
t \asymp n^{ \frac{1}{(2\beta)} - \frac{5}{12}}  \quad
\Longrightarrow  \quad
n \asymp t^{ \frac{12 \beta}{6 - 5\beta}} \,,
\eeal
thus
$$
\sum_{k=1}^n t^2 \sigma_k^2 = \sum_{k=1}^n t^2 \esm{ \rY_k^2 }
\asymp n (at)^2
\asymp n^{1 - \frac{10}{12}}
\asymp |t|^{ \frac{2\beta}{6 - 5\beta}}\,.
$$
Finally, we obtain a decay bound on the characteristic function at infinity,
crucial for the proof of infinite derivability of the probability measures of the finite-volume
Hamiltonians, given  in the next subsection,
\beal
\label{eq:bound.ffi.large.t}
\big| \Psi_n(t) \big|
&
\le \Const \, \eu^{ - c|t|^{\kappa } }  \,, \quad
\kappa =  \frac{2\beta}{6 - 5\beta}\,, \;\; c>0,
\eeal
and also a bound important in the zone of small $|t|$,
\beal
\label{eq:bound.ffi.small.t}
\sum_{k=1}^n t^2 \sigma_k^2 = \sum_{k=1}^n t^2 \esm{ \rY_k^2 }
\asymp n\, t^2 a^2 =  a^{2-2\beta} t^2 \,,
\eeal
which will be useful for the proof of the concentration bounds.

We do not choose a specific value for $\beta$: for our purposes, it suffices to know that
the main estimate holds true for some $\beta>0$. Quite probably, the technique used in the proof
can be streamlined, so it seems premature to try and optimize our bounds.
$\,$
\qedhere

\subsection{Proof of Theorem \ref{thm:Wegner.polynom.with.Bernstein}}
\label{ssec:proof.thm.Wegner.polynom.with.Bernstein}

Fix $u\in\DZ^d$, $\DN\ni L\ge 1$, $\tau>1$, $\theta>0$, $\btau>(1+\theta)\tau$,
consider the cubes $\ball_L(u)\subset \ball_{L^\btau}(u)$, and denote
$\ball = \ball_L(u)$ and $\oball = \ball_{L^\btau}(u)$.

For the sake of notational brevity,
we assume $u=0$: this can be done without loss of generality in the case of the periodic
lattice $\DZ^d$ with IID random amplitudes $\om_x$, and adaptation to a more general setting
would be quite straightforward.

In the analysis of regularity of the probability measure of the random eigenvalues of
$H_\ball(\om)$, we shall make use
of the random amplitudes $\om_x$ with $x$ selected from some subset $\mcX$ of the annulus
$\ball_{L^{(1+\theta)\tau}}(0)\moins\ball_{L^\tau-1}(0)$, so that
$r=|x|\in \big[L^\tau, \, L^{(1+\theta)\tau} \, \big]$. This annulus is obviously a union
of the form $\cup_r \myset{x:\, |x|\in [r, r+1)}$. A specific choice of the subsets
$\mcX_r := \mcX\cap \myset{x:\, |x|\in [r, r+1)}$ is irrelevant, but we shall need
some combinatorial restrictions on the cardinalities $|\mcX_r|$ in order to apply Lemma
\ref{lem:CLT.with.Bernstein}.

Introduce the following decomposition of the restriction $V|_{\ball}$:

$$
\bal
V = \tV_L + \tWin + \tWort,
\eal
$$
where
\begin{align}
\label{eq:decomp.xi}
\tV_L(y,\om) &:= \xi(\om) \one_{\ball}(y), \quad \xi(\om) := \sum_{x\in\mcX } a_x\om_x \,,
\quad a_x := \fu(x-0) \,,
\\
\label{eq:decomp.a.x}
\tWin(y,\om) &:= \sum_{x\in\mcX} c_x(y)\om_x \,,
\quad c_x(y) := \fu(x-y) - \fu(x-0) \,,
\\
\label{eq:decomp.c.x}
\tWort(y,\om) &:= \sum_{x\in\DZ^d\setminus \mcX} \fu(x-y) \om_x \,.
\end{align}
The roles of the above components of $V$ are as follows:
\begin{itemize}

  \item $\tV_L$ is obviously a random constant function on $\ball_L(u)$; the "tidal" mechanism is going
  to be the main tool in the proofs given below;

  \item $\tWin$ is the "ripple" perturbation due to approximation of the potential
  $y \mapsto \fu(x - y)\one_{\ball}(y)$ by a constant function $\fu(x-u)\one_{\ball}(y)$ on $\ball$;
  as a matter of fact, the profile of $\tWin$ is "almost flat" due to its very small amplitude;
  it can be easily obtained by Taylor expansion of the function $y \mapsto \fu(x - y)$ for $x$ located
  far from $\ball \ni y$;

  \item $\tWort$ is the cut-off error, measuring the stability of Wegner estimate eventually obtained with
  the help of $\om_x$ only in a bounded neighborhood of $\ball_L(u)$.
\end{itemize}



%

Next, consider the layers $\mcX_r = \mcX \cap \myset{x:\, |x|\in[r, r+1)}$,
and with $r$ fixed, introduce the decomposition
\beal
\label{eq:decomp.om.mcX.oball}
\om &= \om_{\mcX_r} + \om_{\mcX_r}^\perp
\\
&
= \om_{\mcX_r} + \om_{\mcX_r,\oball}^\perp + \om_{\oball}^\perp
\eeal
where we identify, as usual, a configuration $\om$ with a mapping $\om:\, \DZ^d \to [-1,1]$,
and the three above components are as follows:
\beal
\om_{\mcX_r} & = \om  \, \one_{\mcX_r} \,,
\\
\om_{\mcX_r,\oball}^\perp &= \om  \, \one_{\oball \moins \mcX_r} \,,
\\
\om_{\oball}^\perp &= \om \, \one_{\DZ^d\moins\oball} \,.
\eeal
%
Consider the random Hamiltonian
$\tH_\ball(\om) = H_\ball(\om_{\mcX_r} + \om_{\mcX_r,\oball}^\perp)$, obtained from $H_\ball(\om)$
by elimination of the remote component $\om_{\oball}^\perp$.
By construction, it is $\fF_{\oball}$-measurable. For the purposes of eigenvalue concentration analysis,
we fix $\om_{\mcX_r,\oball}^\perp$ and use conditioning on $\fF_{\oball\moins\mcX_r}$. As to the effect of
perturbation by $\BU[\om_{\oball}^\perp]$, it will be assessed at the last stage of analysis
by a simple application of the min-max principle
(cf. \eqref{eq:norm.out.oball}). Essentially, we are going to work now with
the truncated Hamiltonian $\tH_\ball(\om_\mcX)$, as the other two components of $\om$ figuring in
the decomposition \eqref{eq:decomp.om.mcX.oball} are either cut-off or fixed by conditioning.

The impacts on the eigenvalues of $\tH_\ball(\om)$ (or of $H_\ball(\om)$) of the potentials
$\om_x\fu(x-\cdot)$ cannot be defined in a unique, canonical way: if we enumerate the sites $x$ in some way
and "switch them on" one by one, the respective perturbations of a given eigenvalue do (or might)
depend on the chosen enumeration. However, one can proceed as follows:

\begin{enumerate}[$\bullet$]
  \item For each $r\in \big[L^\tau, \, L^{(1+\theta)\tau} \, \big]$, enumerate in some way
   the sites $x_{r,k}$ of $\hmcX_r$ starting with $n_r$ elements of $\mcX_r$; apart from this restriction,
   the order of $x_{r,k}$ can be arbitrary. Introduce a linear order "$\prec$" on
   $\cup_{r=L^\tau}^{L^{(1+\theta)\tau}} \,\mcX_r$ so that $x_{r,j}\prec x_{r',k}$
   iff either $r < r'$, or $r=r'$ and $j<k$.

  \item Form the Hamiltonian $H = -\Delta_\ball + W^\perp$ where $W^\perp$ is the potential induced on
  $\ball$ by all random sources $\om_x\fu(x-\cdot)$ with
  $x\not\in \cup_{r=L^\tau}^{L^{(1+\theta)\tau}} \,\mcX_r$.

  \item Fix a measurable enumeration $\lam^{\!\ball,i}$ of the eigenvalues of $H_\ball(\om)$ and pick
  some $i$; we will work with this random eigenvalue $\lam^{\!\!\ball,i}$.

  \item Denote by $\hlam^{\!\ball,i}$ the $i$-th eigenvalue of $-\Delta_\ball + W^\perp$.

  \item Introduce a sequence of potentials
\be
\label{eq:V.r.k}
V_{(r,k)}(\cdot) = \sum_{(r',j) \prec (r,k)} \om_{x_{r',j}} \fu(x_{r',j} - \cdot)
\ee
and denote by $prec(r,j)$ the $2$-index immediately preceding $(r,j)$.
(If there is no predecessor, we use the convention that a sum in \eqref{eq:V.r.k}
over an empty index set is zero.)

  \item Denote $\lam^{\!\!i}_{r,k}$ the $i$-th eigenvalue of $H + W^\perp + V_{r,k}$ and
$$
X_{r,k} = \lam^{\!\!i}_{r,k} - \lam^{\!\!i}_{prec(r,k)}.
$$
Decompose $X_{r,k} = \rY_{r,k} + \ry_{r,k}$ where $\rY_{r,k}$ is the EV perturbation induced
by the flat potential $z\mapsto \om_{x_{r,k}}\fu(x-z)\one_\ball(z)$ and
$\ry_{r,k} := X_{r,k} - \rY_{r,k} +$ is the respective correction term.
\end{enumerate}

Then we have
\beal
\label{eq:lam.X.k}
\lam^{\!\!i} = \hlam^{\!\!i} + S_L
= \hlam^i + \sum_{r=L^\tau}^{L^{(1+\theta)\tau}} X_{r,k}
= \hlam^i + \sum_{r=L^\tau}^{L^{(1+\theta)\tau}} \big( \rY_{r,k} + \ry_{r,k}  \big) .
\eeal
While the formal construction of the above decomposition of $\lam^{\!\!i}$ is rather technical,
the dominant contributions $\rY_{\!x_{r,k}}$
have a more transparent structure and can be defined in a much simpler way. It suffices to
replace each potential $\om_x\fu(x-\cdot)$ by a flat one,
$$
z \mapsto \om_x \fu(x-u)\one_{\ball}(z) \,, \;\; z\in\ball \,,
$$
then $\rY_x(\om) = \om_x \fu(x-u)$; these terms are stochastically independent, and all enumerations
of the sites $x$ are equivalent.
The above boring formalities are required only for a rigorous definition
of the perturbation terms $\ry_\bullet$ due to the terms $c_x(y)\om_x$ in \eqref{eq:decomp.c.x}.
Once they are defined, all we need for the actual calculations and estimates
is some satisfactory uniform upper bound on $|\ry_\bullet|$ (cf. \eqref{eq:proof.EVC.Bernstain.cond.gamma}).

In notations of Section \ref{ssec:Bernstein.lemma},
operating in a more abstract framework but actually tailored to
suit the needs of the present subsection, we have
by a simple calculation, for the sites $x$ satisfying
$|x|=r\in\big[L^\tau, \, L^{(1+\theta)\tau} \, \big]$,
\begin{align}
\notag
a_x &\asymp a = r^{-A} \,,
\\
\label{eq:proof.EVC.Bernstain.cond.gamma}
\max_{y\in\ball}|c_x(y)| &\asymp r^{-(A+1)}L
\le
n^{-\gamma}\,,
\\
\notag
\gamma & := \frac{1+ A^{-1}(1 - \tau^{-1})}{2\beta}>1.
%
\end{align}
The condition \eqref{eq:proof.EVC.Bernstain.cond.gamma} with $\gamma>1$ is required
for an application of Lemma \ref{lem:CLT.with.Bernstein}.

Further, let $t\in\DR$ satisfy
$$
 L^{\frac{A(6-5\beta)\, \tau}{6}} \le |t| \le L^{\frac{A(6-5\beta)\, (1+\theta) \tau}{6}}
$$
and set $R_t := |t|^{\frac{6}{A(6-5\beta)}}$. Then
$R_t \in \big[ L^{\tau}, \, L^{(1+\theta)\tau} \,\big]$,
and we have $n_{R_t} = |t|^{ \frac{12\beta}{6-5\beta}}$
as required for \eqref{eq:def.n.cond.Bernstein.polynom}.

Once a random eigenvalue $\lam = \lam^{\!\ball,i}$ is chosen, consider its characteristic function
$\ffi_{\lam}(t) = \esm{ \eu^{ \ii t S_L} }$:
$$
\bal
\big| \ffi_{\lam}(t) \big| =
\left|\esm{ \eu^{\ii t \lam} } \right| &
= \left| \esm{ \eu^{\ii t \hlam}\esm{\eu^{\ii t S_L} \cond \fF} } \right|
\le \essup \left| \esm{\eu^{\ii t S_L} \cond \fF} \right|.
\eal
$$
Here "$\essup$" refers to the random nature of the conditional expectation; in other words,
we have to upper-bound the $\rL^\infty(\Om)$-norm of that conditional expectation.
Furthermore, to assess the RHS, fix $t\ne 0$, set $R_t = |t|^{\frac{6}{A(6-5\beta)}}$,
condition on all $\om_{x_{r,k}}$ with $r\ne R_t$, and consider the conditional characteristic function
of the sum $S_t = \sum_{k=1}^{n_t} (\rY_{R_t,k} + \ry_{R_t,k})$:
$$
 \left|\esm{\eu^{\ii t S_t} \cond \fF} \right|
\le  \essup \left| \esm{\eu^{\ii t \sum_{k=1}^{n_t} (\rY_{R_t,k} + \ry_{R_t,k})} \cond \fF_t} \right| .
$$
It follows from Lemma \ref{lem:CLT.with.Bernstein}
that for some $\beta\in(0,1)$, $\ffi_{\lam}(t)$ admits the upper bounds
\beal
\big| \ffi_{\lam}(t) \big| &\le \eu^{- C a^{2 - 2\beta} t^2}
\le \eu^{- C L^{-A\tau(2 - 2\beta)} t^2} & \text{by \eqref{eq:bound.ffi.small.t}}
\\
\big| \ffi_{\lam}(t) \big| & \le \eu^{ - C |t|^{\kappa}} & \text{ by \eqref{eq:bound.ffi.large.t}}
\eeal
where the first bound is important for relatively small $|t|$ and the second one for large $|t|$.
Arguing as in Section \ref{ssec:proof.assertion.C.Berry.theorems}
and using the Parseval identity,
we infer from the analogs of \eqref{eq:J1.J1+.J1-} and \eqref{eq:proof.C.J1.J2},
with $M = L^{\tau}$ and $N = L^{(1+\theta)\tau}$,
that for any interval $I_\eps = [E-\eps, E+\eps]$ one has, uniformly in $E\in\DR$,
$\mu_{\lam}(I_\eps \cond \om_{\mcX,\oball}^\perp) \le J_- + J_+$,
where
\beal
J_- &=  8\eps \int_{-L^{\tau}}^{L^{\tau}} \big| \ffi_S(t) \big| \, dt
\lea  \eps \int_{-L^{\tau}}^{L^{\tau}} \eu^{-C t^2 L^{-A\tau(2 - 2\beta)} } \, dt
\lea L^{A\tau(1 - \beta)} \, \eps \,,
\\
J_+ &=  8\eps \int_{L^{\tau} \le |t| \le \eps^{-1}} \eu^{-c|t|^{\kappa} } \, dt = \ord{\eps} \,.
\eeal
Hence for $\eps\ge \eps_L := L^{-A(1+\theta)\tau}$ we have
\beal
\label{eq:mu.eps}
\mu_\lam(I_\eps \cond \om_{\mcX,\oball}^\perp ) \lea  L^{A(1-\beta)\tau}  \, \eps \,.
\eeal
Let $\eps = \eps_L$, $I_{2\eps_L}= |E - 2\eps_L, E+2\eps_L]$, $E\in\DR$, and introduce
an $\fF_\oball$-measurable event
\be
\label{eq:EVC.Bernstein.def.Om.mcX}
\Om_{\oball,\eps}
:= \myset{\om_\mcX:\;
\dist\left[ \Sigma\left(H_\ball\left(\om_\mcX+\om_{\mcX,\oball}^\perp\right)\right), E \right] \ge 2\eps_L} \,,
\ee
then it follows from \eqref{eq:mu.eps} that
\be
\label{eq:EVC.Bernstein}
\pr{ \Om_{\oball,\eps_L} } \lea  L^{-A(\beta + \theta) \tau }
\le  \eps_L^\fb\,,
\ee
where
\be
\label{eq:EVC.Bernstein.fb}
\fb = \frac{\beta+\theta}{1+\theta}
>0 \,.
\ee

\vskip1mm
\noindent
$\blacktriangleleft$  Observe that
the above H\"{o}lder-type estimate is valid for all sufficiently small $\eps>0$, provided
that the exponent $\tau$, which determines the size of the "finite bath" contributing to
the estimate, is large enough, and the H\"{o}lder exponent $\fb$ admits a uniform lower bound
$\fb_\theta=\frac{\theta}{1+\theta}>0$, regardless of the choice of $\beta>0$. Alternatively,
$\fb \ge \beta/2>0$ uniformly in $\theta\in[0,1]$.

It remains to assess the effect of the cut-off we made in the definition of $\tH(\om)$.
Fix any sub-configuration $\om_\oball\in \Om_{\oball,\eps}$ and consider the
Hamiltonian $H_\ball(\om_\oball + \om_\oball^\perp)$ with an arbitrary complementary sub-configuration
$\om_\oball^\perp \in [-1,1]^{\oball^\rc}$. Since $\oball \equiv\ball_{L^\btau}(u)$ and
$$
\dist\left[\ball_L(u), \, \DZ^d\moins\ball_{L^\btau}(u)  \right] \ge L^\btau - L \ge \half L^{\btau} \,,
$$
it follows by a straightforward calculation (cf. \eqref{eq:def.BU})
\be
\label{eq:norm.out.oball}
\left\| \BU\big[\om_\oball^\perp \big] \, \one_{\ball_L(u)}\right\|_\infty \le C L^{-(A-d)\btau} .
\ee
Since $\BU[\om_\oball] = \BU[\om] - \BU[\om_\oball^\perp]$,
the inclusion
$\lam\big(\om\big) \equiv \lam\big(\om_\oball+\om_\oball^\perp\big)\in I_{\eps}$
implies by the min-max principle that
$\lam\big(\om_\oball\big)\in I_{2\eps}$,
provided $\btau > \frac{A + \beta }{A-d} \tau$,
so that
$$
CL^{-(A-d)\btau} \le \eps_L \le  L^{-\left(A\theta + \beta \right)\tau} \,.
$$
Therefore, the assertion \eqref{eq:thm.Wegner.polynom.cond.E.2} of Theorem
\ref{thm:Wegner.polynom.with.Bernstein} follows from \eqref{eq:EVC.Bernstein}.
\qedhere

\subsection{Smoothness of DoS. Proof of Theorem \ref{thm:Wegner.polynom.with.Bernstein}}
\label{ssec:smooth.DoS.with.Bernstein.polynom}

In the previous subsection we established decay bounds on the characteristic functions
of all eigenvalues $\lam = \lam^{\!\!\ball,i}$ of $H_\ball(\om)$ subject to an annular "thermal bath"
of finite size; the impact of the complementary infinite area was unused
in the finite-volume regularity analysis (only upper-bounded).
Technically, $|\ffi_\lam(t)|$ could be assessed for $|t|\le T(L^\tau)$ when the external potential
used in the estimates was restricted to the annulus
$\ball_{L^{(1+\theta)\tau}}(u)\moins \ball_{L^{\tau}}(u)$, $\tau>1$.
However, it is readily seen that
for the same Hamiltonian subject to the potential induced by its entire environment the decay
bounds on $|\ffi_\lam(t)|$ can be extended to arbitrarily large $|t|$, which proves the
existence and infinite derivability of the probability measure of each random eigenvalue
$\lam^{\!\!\ball,i}(\om)$, as asserted in Theorem \ref{thm:Wegner.polynom.with.Bernstein}.
\qedhere

\subsection{Eigenvalue comparison. Proof of Theorem  \ref{thm:EVComp.with.Bernstein.polynom}}
\label{ssec:EV.compare.polynom}

In the spectral theory of random Hamiltonians with potentials featuring independence or
independence at distance, lower bounds on inter-spectral spacings (differences between
eigenvalues of Hamiltonians relative to two distant cubes) follow easily from eigenvalue
concentration estimates for a single cube: it suffices to condition on the sample
of the random potential in the other cube. A similar strategy can be applied to
long-range Hamiltonians in the case where the decay of the source potential at infinity is
sufficiently fast: faster than any power-law.
We have to face simultaneously two problems:
a polynomial decay of the source potential and strong singularity of the marginal disorder.
A preliminary analysis which we do not present here suggests that making use only of
the stochastic decoupling results in eigenvalue concentration estimates insufficient for the
purposes of the MSA. A similar problem has been encountered in the multi-particle
scaling analysis of interacting disordered quantum systems where it was realized that
apart from the conventional Wegner-type estimates for single cubes
one needs a different kind of concentration inequalities: eigenvalue \emph{comparison} estimates
based not on stochastic decoupling (which would be insufficient) but on an accurate
comparison of sensitivity of two spectra to a common family of random fluctuations.
Below we follow this general approach, but its technical implementation is different.



\vskip2mm

Fix a pair of cubes, $\ball'=\ball_L(u')$ and $\ball''=\ball_L(u'')$, with $|u'-u''| = L^\sigma$,
$\sigma>1$, embedded into a larger ambient cube $\oball = \ball_{L^\bsigma}(u')$, $\bsigma>\sigma$.
Since $L^\bsigma \gg L^\sigma$, we have (cf. Fig.~2)
$$
\dist\left[\ball'', \DZ^d\moins \oball \right] \asymp
\dist\left[\ball', \DZ^d\moins \oball \right] \asymp L^\tau \,.
$$
Now we can repeat the construction used in the previous subsection and obtain a representation
for the eigenvalues $\lam'_i = \lam^{\!\!\ball'\!,i}$ of $H_{\ball'}$ of the form \eqref{eq:lam.X.k};
idem for the eigenvalues $\lam''_j = \lam^{\!\!\ball''\!,j}$ of $H_{\ball''}$.

In this way, we obtain, as in the previous subsection,
$X'_k = Y'_k + y'_k$ for a fixed eigenvalue $\lam'_i$ of $H_{\ball'}$ and, respectively,
$X''_k = Y''_k + y''_k$ for an eigenvalue $\lam''_j$ of $H_{\ball''}$.

\begin{figure}
\begin{tabular}{c}
%
%
%
\begin{tikzpicture}
\begin{scope}[scale=0.059]
\clip (-70,-30) rectangle ++(140.0, 60.0);

\draw (-2, -2) rectangle ++(4, 4);
\fill (0,0) circle (0.2);

\draw (-8, -7) rectangle ++(4, 4);
\fill (-6,-5) circle (0.2);

\draw[line width = 0.8] (-25, -25) rectangle ++(50, 50);
\fill (0,0) circle (0.2);

\node (ballx) at (-15, 7.0) {$\ball_L(u')$};
\draw[->,bend left = 40] (ballx.east) to (-0.0, 2.4);

\node (bally) at (7, -15.0) {$\ball_L(u'')$};
\draw[->,bend left = 30] (bally.west) to (-6, -7.3);

\node (ball) at (45, -7.0) {$\ball_{L^\bsigma}(u')$};
\draw[->,bend right = 30] (ball.west) to (25.5, -0.3);



\end{scope}
\end{tikzpicture}

\end{tabular}

\caption{  \footnotesize\emph{
Example for the proof of Theorem
\ref{thm:EVComp.with.Bernstein.polynom}}.
%
}
\end{figure}
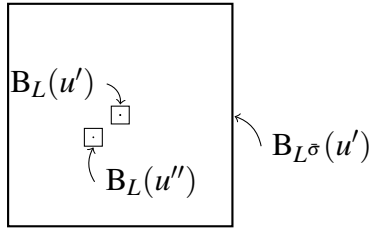%
%


Consider an inter-spectral spacing
\beal
\Lam_{i,j}  := \lam'_i - \lam''_j
&
= \hLam_{i,j}(\om_\mcX^\perp) + \sum_{k=1}^n \left(X'_k - X''_k \right)
\\
&
= \hLam_{i,j}(\om_\mcX^\perp) + \sum_{k=1}^n \left(\rY_k + \ry_k \right)
\eeal
where $\rY_k = Y'_k - Y''_k$, $\ry_k = \ry'_k - \ry''_k$. As before, $\{\rY_k, 1\le k \le n\}$ is an
independent family of random variables; explicitly,
$$
\rY_k(\om) = \rY_k(\om_{x_k}) = \fu(|x-u'|) - \fu(|x-u''|),
$$
while $\{ \ry_k \}$ are correlated, due to the nonlinear dependence of the eigenvalues
upon the  values of the cumulative potential on $\ball'$ and on $\ball''$.

Now assess the amplitudes of $Y'_k$, $Y''_k$ which are the dominant contributions of the cumulative
potential to the eigenvalues in $\ball'$ and $\ball''$.

We choose the sites $x_k$ (cf. Fig.~3) supporting the relevant source
potentials $\om_{x_k} \fu(|x_k - \cdot|)$
as follows\footnote{Some technicalities are required here: the essential components of our calculations
rely on a simple, one-dimensional geometry of the line passing through $u'$ and $u''$,
but not all the sites $x$ figuring in $\om_x\fu(x-\cdot)$ belong to this line; some of them are
at distance $\Ord{1}$ from it.}.
Once the centers $u', u''$ are fixed, take the line in $\DR^d$ passing through
$u'$ and $u''$, and consider the interval $\mcJ \subset\DR^d$ of length $L^{2\beta \tau}$ of this line.
Consider the successive unit cubes with centers in $\DZ^d$ that $\mcJ$ intersects,
and take as $x_k$ the centers of these cubes, $\ball_{1/2}(x_k)$, $k=1, 2, \ldots$.
We shall also need representatives $\ttx_k\in\ball_{1/2}(x_k)\cap\mcJ$
of the latter balls, given by the orthogonal projection of $x_k$ onto $\mcJ$.
With $R_k := |\ttx_k - u'|$ and $|u'-u''| = L^\sigma$ we obtain
$$
\bal
|x_k - u'| &= \sqrt{ |\ttx_k - u'|^2 + \Ord{1}} = R_k \left( 1 + \Ord{L^{-2\tau}} \right) \,,
\\
|x_k - u''| &= \sqrt{ |\ttx_k - u''|^2 + \Ord{1}} =
\left(R_k + L^\sigma \right) \left( 1 + \Ord{L^{-2\tau}} \right) \,.
\eal
$$
It follows that
$$
\bal
R_k^{A} |x_k-u''|^{-A} &= R_k^{A} \left( R_k + L^\sigma \right)^{-A} \left( 1 + \Ord{L^{-2\tau}} \right)^{-A}
\\
&
= \left( 1 + R_k^{-\tau} L^{\sigma} \right)^{-A}\left( 1 + \Ord{L^{-2\tau}} \right)^{-A}
\\
&
= 1 -A R^{-\tau} L^{\sigma}\left(1 + \ord{1} \right)
\\
&
= 1 -A L^{-\tau} L^{\sigma} \left(1 + \ord{1} \right) \,,
\eal
$$
whence
$$
\bal
|x-u'|^{-A} - |x-u''|^{-A}  & =
L^{-A\tau}  - \left( L^\tau + L^\sigma \right)^{-A}
\\
&
= L^{-(A+1)\tau + \sigma}\left( A + \ord{1} \right).
\eal
$$

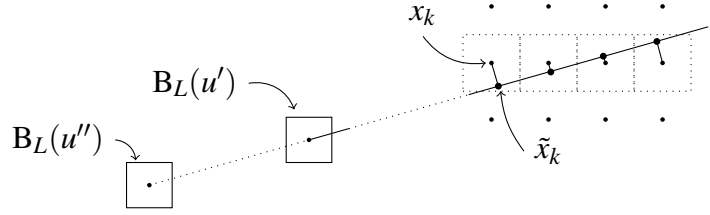
\begin{figure}
\begin{tabular}{c}
%
%
%
\begin{tikzpicture}
\begin{scope}[scale=0.15]
\clip (-20,-5) rectangle ++(70.0, 30.0);

\node (ball) at (-8, 4.0) {$\ball_L(u'')$};
\draw[->,bend left = 40] (ball.east) to (-1.2, 2.2);

\node (ball) at (4, 9.0) {$\ball_L(u')$};
\draw[->,bend left = 40] (ball.east) to (12.7, 6.7);

\node (xk) at (24, 15) {$x_k$};
\draw[->,bend right = 30] (xk.south) to (29.5, 10.8);

\node (txk) at (35, 3) {$\ttx_k$};
\draw[->,bend left = 15] (txk.west) to (30.8, 8.3);

\foreach \x in {30, 35, 40, 45}
{
  \foreach \y in {5, 10, 15}
    \fill (\x,\y+0.8) circle (0.2);
}

\draw (30, 10.8) -- (30.6,8.75);
\draw (35, 10.8) -- (35.2, 10.0);
\draw (40, 10.8) -- (39.8, 11.4);
\draw (45, 10.8) -- (44.5, 12.70);

\fill[color=black] (30.6, 8.75) circle (0.3);
\fill[color=black] (35.2, 10.0) circle (0.3);
\fill[color=black] (39.8, 11.4) circle (0.3);
\fill[color=black] (44.5, 12.70) circle (0.3);

\draw[dotted] (30-2.5, 10.8-2.5) rectangle ++(5, 5);
\draw[dotted] (35-2.5, 10.8-2.5) rectangle ++(5, 5);
\draw[dotted] (40-2.5, 10.8-2.5) rectangle ++(5, 5);
\draw[dotted] (45-2.5, 10.8-2.5) rectangle ++(5, 5);

\draw[dotted] (0,0) -- (14,4);


\draw[color=black] (14,4) -- (17.5,5);
\draw[color=black,dotted] (17.5,5) -- (28, 8);
\draw[color=black] (28, 8) -- (49, 14);

\draw (-2, -2) rectangle ++(4, 4);
\fill (0,0) circle (0.2);

\draw (12, 2) rectangle ++(4, 4);
\fill (14,4) circle (0.2);

\end{scope}
\end{tikzpicture}

\end{tabular}

\caption{  \footnotesize\emph{
Example for the proof of Theorem
\ref{thm:EVComp.with.Bernstein.polynom}. Choice of the points $x_k$ and $\ttx_k$}.
%
}
\end{figure}%
%
%
\noindent
Therefore,
\begin{align}
\rY_k \asymp L^{-(A+1)\tau + \sigma}\left( A + \ord{ 1 } \right)
&= L^{-\tA \tau} \left( A + \ord{ 1 } \right) ,
\\
\label{eq:def.tA}
\tA &:= A+1 - \tau^{-1}\sigma > A \,.
\end{align}
In \eqref{eq:def.tA} we used the condition $\tau>\sigma$.
As to the flat approximation error, we have
$$
\ry_k \asymp L^{-(A+1)\tau + 1} = \left(L^{-(A+1)\tau + \sigma} \right)^\gamma\,,
$$
where
$$
\gamma = \frac{ (A+1)\tau - 1 }{ (A+1)\tau - \sigma }
= 1 + \frac{ \sigma - 1 }{ (A+1)\tau - \sigma } >1.
$$
Thus we recover the general framework of the multiplicative variant of the Bernstein-type regularity
analysis carried in Section \ref{ssec:proof.thm.Wegner.polynom.with.Bernstein},
albeit with a modified exponent of the power law.

Recall that we work with the Hamiltonians in the cubes
$$
\ball'=\ball_L(x), \;\; \ball''=\ball_L(y)\subset \oball = \ball_{L^\btau}(u)
\,,
\quad |x-y| \asymp L^{\sigma}, \;\;  \btau>\tau>\sigma>1,
$$
subject to the potential   generated by an infinite "thermal bath"
$\DZ^d \moins \oball$,
and assess inter-spectral spacings $\Lam_{i,j}(\cdot) = \lam'_{i} - \lam''_{j}$.
Repeating the analysis from Section \ref{ssec:proof.thm.Wegner.polynom.with.Bernstein}
with $A$ replaced by $\tA = A+1 - \tau^{-1}\sigma>A$, we come to the
following conclusions.

\begin{enumerate}
  \item Each random variable $\Lam_{i,j}(\cdot) = \lam'_{i}(\cdot) - \lam''_{j}(\cdot)$ admits a density
  $\rho_{i,j} \in \mcC^\infty(\DR)$. Its Fourier transform $ \ffi_{i,j}$ satisfies
$\big| \ffi_{i,j}(t) \big| \lea \eu^{- C |t|^\kappa }$, $\kappa>0$.

  \item
Using the decomposition $\om = (\om_\oball, \om_\oball^\perp)$, one has for the probability
of the event
$$
\Om_{\ball', \ball'',\eps_L} := \myset{ \om_\oball: \;  \inf_{\om_\oball^\perp} \;
\dist\left[ \Sigma\left(H_{\ball'} \right), \, \Sigma\left(H_{\ball'} \right) \right] \le 2\eps_L } \,,
$$
with $\eps_L = L^{ -\tA(1+\theta)\tau}$, an upper bound
\beal
\label{eq:EVComp.Bernstein}
\pr{ \Om_{\ball', \ball'',\eps_L} } \lea L^{ - \left( \tA \theta + \beta \right)\tau + 2d }
&
< L^{ - \left( A \theta + \beta \right)\tau + 2d }
\\
&
\le L^{2d} \eps_L^\fb
\eeal
\end{enumerate}
with $\tA = A+1 - \tau^{-1}\sigma>A$ and $\fb > \frac{\theta}{1+\theta}$ regardless of the choice
of the parameter $\beta>0$. (Recall: one also has $\fb \ge \beta/2$ for $\theta\in[0,1]$.)

It is to be stressed again that the obtained regularity estimate relies not on a stochastic decoupling
of the eigenvalues $\lam'_{i}$ and $\lam''_{j}$ relative to two distant cubes but on
an accurate estimate of the difference of their sensitivities to the common remote random potentials.

\section{Smoothness of the finite-volume DoS. Exponential potentials}
\label{sssec:proof.smooth.exp.max-norm}

An important particularity of exponentially decaying potentials $\fu(r)$,
as compared to those decaying at a polynomial
or even fractional-exponential rate, is that $\fu(r)$ is no longer a "slowly decaying" function:
$|\fu'(r)| \asymp |\fu(r)|$. As a result, one cannot make use of the flat approximation of $\fu(r)$
at large distances. For this reason, one needs an alternative mechanism providing eigenvalue concentration
estimates and regularity of the finite-volume DoS. In the present paper, we consider two models where
the distance is integer-valued; this makes the analytic study simpler. Surprisingly, it is possible
to turn the relation $|\fu'(r)| \asymp |\fu(r)|$ to our advantage, once it is replaced by
an exact algebraic identity $\fu(r+s)| = \fu(r) \fu(s)$.

\subsection{Proof of Theorem \ref{thm:smooth.exp.sum-norm.and.max-norm} for the sum-norm model}
$\,$

Consider the interaction $(x,y)\mapsto  \eu^{-a|x-y|_1}$.
Fix a cube $\ball = \ball_L(u)$ with $u=(L, \ldots, L)$, so that $\ball\subset \DN^d \hookrightarrow \DZ^d$,
and consider the negative orthant $\cX:=(-\DN)^d$. Then for any $x\in\cX$ and any $y\in\ball$,
hence with the coordinates $y_{(j)}\ge 0$, $1\le j \le d$,
\beal
\label{eq:sum-norm.x.y}
|y - x|_1 &= \sum_{j=1}^d |y_{(j)} - x_{(j)}| = \sum_{j=1}^d \big( y_{(j)} +|x_{(j)}| \big)
= |y|_1 + |x|_1,
\eeal
so we have a factorized potential $\fu(x-y) = \eu^{-a|y|_1} \, \eu^{-a|x|_1}$, as
long as $y\in\ball$ and $x\in\cX$.

Fix $L\ge 0$. Due to the translation invariance of the random
field $\{\om_x, \, x\in\DZ^d\}$, it suffices to consider the case where $\ball = \ball_L(u_L)$,
$u_L = (L, L, \ldots, L)$, so that $\ball_L(u_L) \subset \DN^d$.
Denote $\cX = (-\DN)^d$, then we can write the random potential on $\ball$ as follows:

\beal
V(y,\om)
&= \sum_{x\in \cX} \fu(y-x) \, \om_x + \sum_{x\in \DZ^d\moins\cX} \fu(y-x) \, \om_x
\\
& =: W_\ball(y,\om) + \tW_\ball(y,\om)\,,
\eeal
where the random potential $\tW_\ball(\om)$ is measurable with respect to
the \sigal $\fF^\perp_{\cX}$ generated by $\myset{\om_x,\, x\in\DZ^d\setminus\cX}$.
Due to the identity \eqref{eq:sum-norm.x.y} and the resulting factorization, we have
\beal
W_\ball(y,\om) = \sum_{x\in \cX} \fu(x-y) \om_x &= \eu^{-a|y|_1} \, \sum_{x\in \cX} \eu^{-a|x|_1} \om_x
\\
&
= U_{\ball}(y) \, \eta_\ball(\om) \,,
\eeal
with $\eta_\ball(\om) := \sum_{x\in \cX} \eu^{-a|x|_1} \om_x$ and $U_\ball(y) = \eu^{-a|y|_1}$.
Thus we can write
\beal
\label{eq:def.H.tA.U2L}
H_\ball(\om) &= \tA_\ball(\om) + \eta_\ball(\om) \, U_{\ball}\,,
\eeal
with $\tA_\ball(\om) = -\Delta_\ball + \tW_\ball(\cdot,\om)$,
while $\eta_\ball(\om)$
is measurable with respect to the \sigal $\fF_{\!\!\cX}$ generated by
$\myset{\om_x,\, x\in \cX}$, and the multiplication operator $U_\ball$ is non-random.
For our purposes, it would suffice to assess the regularity of the conditional distributions given
$\fF^\perp_{\cX}$, so we fix a sample $\{\om_x, \,x\in\DZ^d\setminus\cX\}$. Then $\tA_\ball(\om)$ becomes
nonrandom, so we  drop its argument $\om$:
$H_\ball(\om) = \tA_\ball + \eta_\ball(\om) U_{\ball}$
with an $\fF_{\!\!\cX}$-measurable $\eta_\ball(\cdot)$.

The regularity properties of the probability measure of the random variable $\eta_\ball$ can be
established with the help of Lemma \ref{lem:Main.exp} (where $\delta$ is to be set to $1$).
Specifically,  by assertion (A), if $d>2$, then for any $a>0$ the probability measure of the
random variable $\eta_\ball$ admits a compactly
supported\footnote{Recall that the interaction potential $\fu$ is absolutely summable
on the lattice, hence $\eta_\ball\in\rL^\infty(\Om)$.}
density $\rho_\ball\in\mcC^\infty(\DR)$. Being smooth and compactly supported, $\rho_\ball$ is bounded.
For
$d=2$ and some sufficiently small $a_*>0$ and some $C\in(0,+\infty)$ the density $\rho_\ball$
exists and belongs to $\mcC^Q(\DR)$ with $Q\ge C a \ge 1$, provided $a\in(0,a_*)$.

For the rest of this Section, we always assume that one of the two above mentioned conditions holds true,
so that $\eta_\ball$ has a compactly supported bounded density
$\rho_\ball\in\mcC^1(\DR)$ (eventually, $\rho_\ball$ is more regular).

Next, introduce an analytic operator family
$$
A_\ball(\lam) = \tA_\ball + \zeta U_{\ball}, \;\; \zeta\in\DC.
$$
$\tA_\ball$ and $U_\ball$ are self-adjoint operators in a finite-dimensional Hilbert space,
hence by the Kato--Rellich theorem, all the eigenvalues $\lam_j(\zeta)$ of $A_\ball(\lam)$ are analytic.
Since $b\one \le U_\ball \le \one$ with $b = \eu^{-2aL}$, one has
\be
\label{eq:deriv.EV.zeta.sum-norm}
\forall\, \zeta\in\DR \qquad 0 < b \le \frac{\rd}{\rd \zeta} E_j(\zeta) \le 1.
\ee
Identifying $\zeta\in\DR$ with the random variable $\eta_\ball$, we see that,
conditional on $\fF_{\!\!\cX}$, the probability measure $\mu_j$ of each eigenvalue $E_j(\om)$
is the image of the measure $\mu_\ball$ of $\eta_\ball$ by the real analytic
(hence infinitely differentiable) mapping
$\zeta \mapsto \lam_j(\zeta)$, where $\{\lam_j(\zeta)\}$ are appropriately numbered eigenvalues of
$A_\ball(\zeta)$.

By \eqref{eq:deriv.EV.zeta.sum-norm},
the Radon-Nikodym derivative
of the measure $\mu_j$ of $E_j(\om)$ with respect to the Lebesgue measure is bounded:
\beal
 \frac{\rd \mu_j(s)}{\rd s} \le b^{-1} \| \rho_\ball\|_\infty
=  \eu^{2aL} \| \rho_\ball\|_\infty < \infty.
\eeal

%
$\,$
\qedhere

\subsection{Proof of Theorem \ref{thm:smooth.exp.sum-norm.and.max-norm} for the max-norm model}
$\,$

The case of the potential $\fu_\infty(r)$ can be treated in a similar way.
See the comments in Section \ref{ssec:smooth.max.norm}.

\section{Eigenvalue concentration and comparison estimates. Exponential potentials.}
\label{sec:exp.potentials.sum-norm}


\subsection{Wegner bound. Proof of Theorem \ref{thm:Wegner.exp.max-norm}
for the sum-norm model}

$\,$

Wegner estimates are often obtained by spectral averaging; cf. the original work by Simon and Wolff \cite{SW86},
examples of applications in \lcite{ComHisKlo07}{Theorem 1.1},
\lcite{St10}{Theorem 3.2}, and some abstract
functional-analytic presentation along with an extensive historical review in \cite{Sab14b}.

We focus on lattice models. The main tool for proving EVC bounds in continuum media, spectral averaging,
is well-known to work for various types of Hamiltonians. In fact, the Simon--Wolff technique
has been applied by Kotani and Simon to the continuum Schr\"{o}dinger operators \cite{KotSim87} shortly after
the publication of the paper  \cite{SW86}. An essential hypothesis was complete covering of the configuration
space by the supports of the non-negative scatterer potentials; in the long-range models, the covering
is not only complete but has infinite multiplicity. Considerable efforts were required to cope with the lack of
complete covering in later works; cf., e.g., \cite{ComHisKlo07,St10,Kle13}, a review and some abstract variants
of Wegner's estimate in \cite{Sab14b}. A detailed presentation of analytical aspects, inevitable in the
case of unbounded self-adjoint operators, would make the present work, already too long, even longer.



\bpr[Spectral averaging estimate]
\label{prop:spec.average}
Let be given two bounded self-adjoint operators $\rA, \rQ$ in a Hilbert space $\mcH$.
Suppose that
\be
\label{eq:spec.averaging.q}
\rP_I(\rA) \, \rQ \, \rP_I(\rA) \ge q\, \rP_I(\rA), \;\; q\in(0,+\infty) \,.
\ee
Further, let $\eta$ be a real-valued random variable with probability measure $\mu$
having continuity modulus $\fs_\mu$.
Then
\be
\label{eq:spec.averaging.bound}
\esm{  \trpar{ \rP_I\big(\rA + \eta(\om) \rQ\big) } }
\le C q^{-2} \, |\ball_L(u)| \; \fs_\mu(|I|)\,.
\ee
\epr

\noindent
\emph{Proof of Theorem \ref{thm:Wegner.exp.max-norm} (sum-norm model). }
%

Fix $L\in\DN$. It suffices to consider the ball $\ball_L(u_L)$, $u_L = (L, L, \ldots, L)$.
Let $\cX = (-\DN)^d$, $\mcX_r = \cX \cap \myset{x:\, |x|_1\in[L,KL]}$, $K\ge 1$.
We use again a decomposition similar to \eqref{eq:def.H.tA.U2L},
$$
\bal
H_\ball(\om) &= \tA(\om) + W_\ball(y,\om) \,,
\\
\tA(\om) &= -\Delta_\ball + \tW_\ball(\cdot,\om),
\eal
$$
but now $W_\ball$ is given by a finite sum,
\beal
W_\ball(y,\om) = \sum_{x\in\cX_L} \fu(y-x)\om_x
= U_{\ball}(y) \sum_{r=L}^{ML} \; \sum_{x\in\mcX_r} \om_x =: U_{\ball}(y) \eta_\ball(\om) \,.
\eeal
(we used again the identity \eqref{eq:sum-norm.x.y}),
and we have $U_{\ball}(y)\one_{\ball_L(u)}(y) \ge \eu^{-2aL} \one_{\ball_L(u)}(y)$. Therefore, we can can set
in \eqref{eq:spec.averaging.q} $q=\eu^{-2aL}$.

The regularity of the probability measure of $\eta_\ball(\om)$ can be treated with the help of assertion (C)
of Lemma \ref{lem:Main.exp},  which takes the following form with $\delta=1$,
$M = L$ and $N = KL$:
\be
\label{eq:proof.Wegner.exp.prob.eta.I}
\pr{ \eta_\ball(\om) \in I \cond \fF_{L}} \lea \eu^{a L}  |I| \,.
\ee
It suffices to bound $\Bigesm{ \trpar{ \rP_{I}\big( H_\ball(\om) \big)\cond \fF_L} }$.
By \eqref{eq:spec.averaging.bound} and \eqref{eq:proof.Wegner.exp.prob.eta.I},
\be
\label{eq:proof.Wegner.exp.max-norm.final}
\bal
\Bigesm{ \trpar{ \rP_{I}\big( H_\ball(\om) \big)} \cond \fF_L}
&
\lea q^{-2} |\ball| \, \pr{ \eta_\ball(\om) \in I \cond \fF_{L}}
\\
&
\lea \eu^{5aL} |\ball| \, \eu^{aL}  \,  |I| \,.
\eal
\ee
Taking expectation of the LHS of \eqref{eq:proof.Wegner.exp.max-norm.final} completes the proof:
$$
\Bigesm{ \trpar{ \rP_{I}\big( H_\ball(\om) \big)} } \lea |\ball| \,  \eu^{6aL} |I| .
$$
\qed

\subsection{Eigenvalue comparison. Proof of Theorem \ref{thm:EVC.two.balls.exp}
for the sum-norm model}
\label{ssec:proof.thm:EVC.two.balls.exp.sum-norm}


The fast decay of the potential $\fu(r) = \eu^{-ar}$ makes unnecessary an accurate comparison
of sensitivities of the eigenvalues of random Hamiltonians in two distant balls $\ball', \ball''$
of size $L$. We shall see that with $\dist\big(\ball', \ball'') \ge CL$ and appropriately chosen
constant $C>0$ the required estimate essentially follows, as in the case of independent
(or independent at distance) values of the potential $V(x,\om)$, from the regularity of the probability measure
of random eigenvalues in an isolated ball ($\ball'$ or $\ball'$). In fact, a similar observation can be made
for a large class of potentials $\fu(r)$ decaying at infinity faster than polynomially. This one of the aspects
where the power-law potentials present a particular technical challenge.

\vskip1mm

Consider two balls $\ball'=\ball_L(u')$, $\ball''=\ball_L(u'')$.
The random potential field $\{\om_x, x\in\DZ^d\}$
is invariant under the isometries of the lattice $\DZ^d$, so by an appropriate permutation of the coordinates
$(x_{(1)}, \ldots, x_{(d)})$ and a translation, we can assume without loss of generality that
$u'=(L, L, \ldots, L)$ and $u''-u'$ is in the positive orthant: $u''_{(1)}\ge L, \ldots, u''_{(d)}\ge L$
(cf. Fig~5). As in the previous subsection, we shall make use of the random amplitudes
$\om_x$ with supports $x$ located in the negative orthant $\cX=(-\DN)^d$.
This set has a unique common point $0\in\DZ^d$ with $\ball'$.
Let $y''_\circ$ be the vertex of $\ball''$ closest to the origin $0$, then it is readily seen
that for any lattice point $x\in\cX$
\beal
\forall \, y\in \ball' \quad  |y - x|_1 &= |y - 0|_1 + |0 - x|_1 \,,
\\
\forall \, y\in \ball'' \quad  |y - x|_1 &= |y - y''_\circ|_1 + |y''_\circ - 0|_1 + |0 - x|_1 \,.
\eeal
Notice that by construction, $|y''_\circ|_1 = |x'' - x'|_1$. Consequently,
\begin{align}
\label{eq:U.B1}
\forall \, y\in \ball' \quad  \fu(x - y) &= \eu^{-a|0 - x|_1} \, \eu^{ -a|y - 0|}
= \eu^{-a|x|_1} U'(y) \,,
\\
\label{eq:U.B2}
\forall \, y\in \ball'' \quad  \fu(x - y)
&= \eu^{-a|0 - x|_1} \, \eu^{-a|y''_\circ - 0|_1} \eu^{-a|y - y''_\circ|_1}
=  \eu^{ - a|x|_1 -a |x'' - x'|_1 } U''(y).
\end{align}

Next, for the purposes of the "finite bath" regularity analysis, we further restrict the choice of $x\in\cX$
to the subset $\cX_L = \cX \cap \myset{x:\, |x|_1 \in[L, KL]}$.
Decompose $\om = (\om_{\cX_L}, \om_{\cX^\perp_L})$
and fix $\om_{\cX_L}$; for the rest of the argument, we will work
with the conditional probabilities $\prcXL{\cdot} := \pr{\cdot \cond \fF_{\!\!\cX_L}^\perp}$.

Owing to the additional factor $\eu^{- a |y''_\circ|_1}$ in \eqref{eq:U.B2} as compared to \eqref{eq:U.B1},
the amplitude of the random potential remaining non-constant after conditioning on $\fF_\cX^\perp$
is uniformly much smaller on $\ball''$ than on $\ball'$:
$$
\forall\, x\in\cX \qquad
\frac{ \min_{y\in\ball'} \fu(x-y) }{ \max_{y\in\ball''} \fu(x-y) }
\le \eu^{- a\left( |y''_\circ|_1 -2dL\right) } = \eu^{-C a L} \,,
$$
so $C$ can be made arbitrarily large by taking $|x'-x''|_1/L$ large.

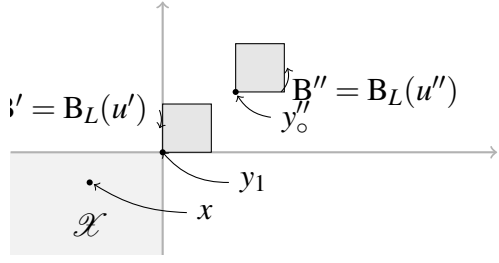
\begin{figure}
\begin{tabular}{c}
%
%
%
\begin{tikzpicture}
\begin{scope}[scale=0.08]
\clip (-25,-17) rectangle ++(140.0, 42.0);

\fill[color=white!90!gray] (-30, -30) rectangle ++(30, 30);

\draw[line width = 0.8,->,color=white!70!black] (-25, 0) -- ++(80, 0);

\draw[line width = 0.8,->,color=white!70!black] (0, -25) -- ++(0, 50);

\fill[color=white!80!gray] (0, 0) rectangle ++(8, 8);
\draw (0, 0) rectangle ++(8, 8);

\fill[color=white!80!gray] (12, 10) rectangle ++(8, 8);
\draw (12, 10) rectangle ++(8, 8);
\fill (12,10) circle (0.5);


\node (ballx) at (-15, 7.0) {$\ball'=\ball_L(u')$};
\draw[->,bend left = 20] (ballx.east) to (-0.50, 4.0);

\node (bally) at (35, 10.0) {$\ball''=\ball_L(u'')$};
\draw[->,bend right = 30] (bally.west) to (20.5, 14);

\node (ypp) at (22, 5.8) {$y''_\circ$};
\draw[->,bend left = 25] (ypp.west) to (12.3, 9.4);

\fill (0, 0) circle (0.5);
\node (yone) at (15, -5) {$y_1$};
\draw[->,bend left = 25] (yone.west) to (0, 0);

\fill (-12, -5) circle (0.5);
\node (x) at (7, -10) {$x$};
\draw[->,bend left = 15] (x.west) to (-11.3, -5.3);

\node (cX) at (-12, -12) {$\cX$};



\end{scope}
\end{tikzpicture}

\end{tabular}

\caption{  \footnotesize\emph{
Example for the proof of Theorem
\ref{thm:EVC.two.balls.exp}}.
%
}
\end{figure}%
%


Pick any pair of eigenvalues $\lam'_\ra$, $\lam''_\rb$ of the respective operators
$H_{\ball'}(\om)$ and $H_{\ball''}(\om)$. With $\om_\mcX^\perp$ fixed, we have
\beal
\lam'_\ra(\om) &= c'_\ra(\om_\mcX^\perp) + \eta'_\ra(\om_\mcX) \,,
\\
\lam''_\rb(\om) &= c''_\rb(\om_\mcX^\perp) + \eta''_\rb(\om_\mcX) \,,
\eeal
whence, with $\om_\cX^\perp$ fixed and omitted from notation,
\be
\xi_{\ra,\rb}(\om) := \lam'_\ra(\om) - \lam''_\rb(\om)
= \lam'_\ra(\om)  - c''_\rb - \eta''_\rb(\om_\mcX).
\ee
With $\eps \ge \eu^{-CaL}$ (this lower bound is required in the last line of
the following equation),
\beal
\label{eq:sum-norm.perturb.EV.prob}
\prcXL{ |\xi_{\ra,\rb}| \le \eps}
&
= \prcXL{ |\lam'_\ra(\om)  - c''_\rb - \eta''_\rb(\om_\mcX)| \le \eps }
\\
&
\le \prcXL{ |\lam'_\ra(\om)  - c''_\rb | \le \eps + \|\eta''_\rb\|_\infty}
\\
&
\le \prcXL{ |\lam'_\ra(\om)  - c''_\rb | \le \eps + \eu^{-C a L} }
\\
&
\le \prcXL{\lam'_\ra(\om)\in I_{2\eps}  } \,,
\eeal
where $I_{2\eps} = [-c_{\ra,\rb} -\eps, c_{\ra,\rb} + \eps]$; the center of this interval is rendered
non-random by fixing $\om_\cX^\perp$.
By Wegner estimate from Theorem \ref{thm:Wegner.exp.max-norm}
for the single cube $\ball_L(0)$ surrounded by the annulus $\ball_{10L}(u)\!\setminus\!\ball_{L}(0)$,
\beal
\prcXL{ \lam'_\ra(\om) \in I_{2\eps} \le 2\eps   } & \lea \eu^{8aL} |I_{2\eps}| \lea \eu^{8aL} \, \eps \,,
\eeal
so the claim follows by counting the number of pairs of eigenvalues $(\lam'_\ra(\om), \lam''_\rb(\om))$.
$\,$
\qedhere



\subsection{Proof of Theorem \ref{thm:smooth.exp.sum-norm.and.max-norm} for the max-norm model}
\label{ssec:smooth.max.norm}

The eigenvalue concentration and comparison analysis for the potential
$\fu_\infty(x) = \exp\left( - a |x|_\infty \right)$ (cf. \eqref{eq:def.fu.max-norm})
is quite similar to that for the sum-norm model; only some geometrical adaptations
are required here, so we focus on the latter without repeating all stages of the technical analysis
carried out in the previous Section.

By translation invariance of the random field $\om_\bullet$, it suffices to consider the case of
a ball $\ball=\ball_L(0)$.
We will use for the coordinates of the points $x\in\DZ^d$ the notation $(x_{(1)}, \ldots x_{(d)})$.
The potential induced on $\ball_L(u)$ by $y \mapsto \fu_\infty(x-y)$
is proportional to a fixed function $y\mapsto U_{\ball}(y) = \eu^{-a|y_{(1)} - L|}$:
$$
\fu_\infty(x-y) = \fa_r U_{\ball}(y),  \;\; \fa_r := \eu^{ - a(r-2L)} \,.
$$
This potential,  considered as a function of $x$ (the support of the
"scatterer" $\om_x\fu(x-\cdot)$),
takes constant values on the $(d-1)$-dimensional sections
of the cube $\ball_L(0)$ orthogonal to the basis
vector $(1, 0, \ldots, 0)$ (cf. Fig.~5).
To see this, write $x = (x_{(1)}, x_\perp)$, $y = (y_{(1)}, y_\perp)$, where
$x_\perp, y_\perp\in\DR^{d-1}$, then we have a general identity
\beal
\label{eq:why.max-norm.x-y.1}
|x-y|_\infty =
\max \left[ \big|x_{(1)} - y_{(1)} \big|, \,  \big|x_\perp - y_\perp \big|_\infty \right]
\eeal
yielding in the particular case where $\big|x_\perp - y_\perp \big|_\infty \le \big|x_{(1)} - y_{(1)} \big|$
\beal
\label{eq:why.max-norm.x-y}
|x-y|_\infty =\big|x_{(1)} - y_{(1)} \big| \,  .
\eeal
Obviously, the RHS of \eqref{eq:why.max-norm.x-y} as a function of $x$
takes constant values on the hyperplanes
orthogonal to the basis vector $(1, 0, \ldots, 0)$. Respectively,
the LHS of \eqref{eq:why.max-norm.x-y.1}--\eqref{eq:why.max-norm.x-y}
is constant in $x$ on sections delimited by the angular sectors
$|x_\perp - y_\perp | \le |x_{(1)} - y_{(1)} |$
(cf. the vertical gray stripes on Fig.~5, where $d=2$ and
$|x_\perp - y_\perp |_\infty \equiv |x_{(2)} - y_{(2)} |$).

Further, let (cf. Fig.~5)
\be
\label{eq:exp.max-norm.cXr}
\mcX_r = \myset{x\in\DZ^d:\, x_{(1)}=r, \,
|x_\perp - y_\perp | \le |x_{(1)} - y_{(1)} -L |  }, \; r\ge 2L \,.
\ee
Then $|\mcX_r| \ge (r-L)^{d-1} \ge r-L \ge L$, so we have indeed a framework for application of
limit theorems, as in the analysis of slower decaying potentials in previous sections. Notice also
that for $y_{(1)} \le L < x_{(1)}$ one has
$|x_{(1)} - y_{(1)}| = \big(x_{(1)} - L\big) + \big(L - y_{(1)}\big)$, hence
$$
\fu_\infty(x-y) = \eu^{-a(x_{(1)} - L)} \, \eu^{-a(L-y_{(1)})} .
$$
Therefore, the random potential induced on $\ball_L(0)$
by the sources at
$x  \in \cX := \cup_{r\ge 2L} \mcX_r$
has a factorized form:
\be
\label{eq:max.norm.factorized}
W_{\ball}(y,\om) = \sum_{r\ge 2L} \sum_{x\in\mcX_r} \fu_\infty(x-y) \om_x
= U_{\ball}(y) \sum_{r\ge 2L} \big|\mcX_r\big| \eu^{-a(r-L)} \om_x
:= U_{\ball}(y) \eta_\ball(\om),
\ee
where the first factor is nonrandom, while the probability distribution of the second one
is the convolution of the independent random amplitudes
$\myset{ \fa_r\om_x, \, x\in\mcX_r}$.

Now we can follow the general path of the proof of Theorem \ref{thm:smooth.exp.sum-norm.and.max-norm}
for the sum-norm model (cf. Section \ref{ssec:proof.thm:EVC.two.balls.exp.sum-norm} )
with the help of the representation of the (cumulative) random potential
\beal
V(y,\om)
&= \sum_{x\in \cX} \fu_\infty(x-y) \, \om_x + \sum_{x\in \DZ^d\moins\cX} \fu_\infty(x-y) \, \om_x
\\
& =: W_\ball(y,\om) + \tW_\ball(y,\om)\,,
\eeal
with $\fF^\perp_{\cX}$-measurable random potential $\tW_\ball(\om)$ and $W_\ball(y,\om)$
defined by \eqref{eq:max.norm.factorized}.
Thus we can write
\beal
\label{eq:def.H.tA.U2L.max-norm}
H_\ball(\om) &= \tA_\ball(\om) + \eta_\ball(\om) \, U_{\ball}\,,
\eeal
with $\tA_\ball(\om) = -\Delta_\ball + \tW_\ball(\cdot,\om)$,
while $\eta_\ball(\om)$
is measurable with respect to the \sigal $\fF_{\!\!\cX}$ generated by
$\myset{\om_x,\, x\in \cX}$, and the multiplication operator $U_\ball$ is non-random.

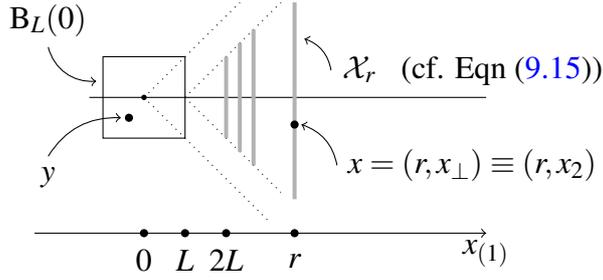
\begin{figure}
\begin{tabular}{c}
%
%
%
\begin{tikzpicture}
\begin{scope}[scale=0.18]
\clip (-17,-14) rectangle ++(70.0, 21.0);
\draw (-3, -3) rectangle ++(6, 6);

\draw (-4, 0) -- (25, 0);
\fill (0,0) circle (0.2);

\draw[->] (-8, -10) -- (25, -10);

\node (xone) at (25, -11.5) {$x_{(1)}$};

\fill (0,-10) circle (0.3);
\fill (3,-10) circle (0.3);
\fill (6,-10) circle (0.3);
\fill (11,-10) circle (0.3);

\draw[dotted] (0, 0) -- (25, 25);
\draw[dotted] (0, 0) -- (9, -9);

\draw[dotted] (3, 0) -- (22, 19);
\draw[dotted] (3, 0) -- (10, -7);

\draw[color=white!70!black,line width = 1.5] (6, -3) -- (6, 3);
\draw[color=white!70!black,line width = 1.5] (7, -4) -- (7, 4);
\draw[color=white!70!black,line width = 1.5] (8, -5) -- (8, 5);

\draw[color=white!70!black,line width = 1.5] (11, -7.5) -- (11, 8);

\node (cXr) at (24, 2) {$\mcX_r$ \; (cf. Eqn \eqref{eq:exp.max-norm.cXr})};
\node (cXrp) at (10.9, 4.5) {};
\draw[->,bend right = 30] (cXr.west) to (cXrp);

\node (r) at (0, -12) {$0$};
\node (r) at (3, -12) {$L$};
\node (r) at (6, -12) {$2L$};
\node (r) at (11, -12) {$r$};

\node (ball) at (-7, 5.7) {$\ball_L(0)$};
\draw[->,bend right = 40] (ball.south) to (-3.5, 1.2);


\fill (-1.1,-1.5) circle (0.3);
\node (y) at (-7, -5.9) {$y$};
\draw[->,bend left = 30] (y.north) to (-1.7, -1.5);

\fill (11, -2) circle (0.3);
\node (x) at (24, -5) {$x=(r,x_\perp)\equiv(r, x_2)$};
\draw[->,bend right = 30] (x.west) to (11.5, -2);

\end{scope}
\end{tikzpicture}
%

\end{tabular}

\caption{  \footnotesize\emph{Example for Section \ref{ssec:smooth.max.norm}}. Here $d=2$
and $x_\perp \equiv x_{(2)}$.
%
}
\end{figure}%
%


Now we can follow the general path of the proof of Theorem
\ref{thm:smooth.exp.sum-norm.and.max-norm} for the sum-norm model.




\subsection{Eigenvalue comparison. Proof of Theorem \ref{thm:EVC.two.balls.exp}
for the max-norm model}
\label{ssec:max.norm.EVComp}

Using permutations of the coordinates and reflection symmetries $x_{(i)}\mapsto -x_{(i)}$,
we can reduce the analysis
to the case where
$$
|u' - u''|_\infty = |u'_{(1)} - u''_{(1)} | = u'_{(1)} - u''_{(1)} >0,
$$
and by translation invariance of the random potential, we can assume without loss of generality that
$u'=0$. Then for any $y'\in\ball_L(0)$, $y''\in\ball_L(u'')$ and $x\in\mcX_r$ with $r\ge 2L$
\beal
| x-y' |_\infty &= x_{(1)} - y'_{(1)} = \big(x_{(1)} - L\big) + \big(L-y'_{(1)}\big)
\\
| x-y'' |_\infty &= x_{(1)} - y''_{(1)} = \big(x_{(1)} - L)
   + \big(u'_{(1)} - u''_{(1)}) \big) +   \big(u''_{(1)} +L - y'_{(1)}\big)
\eeal
Denote $R:=u'_{(1)} - u''_{(1)} \equiv |u' - u''|$, then we have the following
analogs of \eqref{eq:U.B1}--\eqref{eq:U.B2}:
\begin{align}
\label{eq:max-norm.U.B1}
\forall \, y\in \ball' \quad  \fu_\infty(x - y) &
%
= \eu^{-a(x_{(1)}-L)} \, U_{\ball'}(y) \,,
\\
\label{eq:max-norm.U.B2}
\forall \, y\in \ball'' \quad  \fu_\infty(x - y)
&
%
=  \eu^{-aR} \eu^{-a(x_{(1)}-L)} \, U_{\ball''}(y).
\end{align}
With $R\ge CL$ we obtain
\beal
\max_{y\in\ball''} \;\fu_\infty(x-y)
\le \eu^{-a(C-4)L} \, \min_{y\in\ball'} \;\fu_\infty(x-y)
\eeal

Now it is clear that one can obtain the required eigenvalue comparison bound arguing
as in the proof of Theorem \ref{thm:EVC.two.balls.exp} in Section
\ref{ssec:proof.thm:EVC.two.balls.exp.sum-norm}. Again, this comparison bound follows
easily from a single-volume eigenvalue concentration bound, since the potentials
$\om_x\fu(x-\cdot)$ are much weaker on $\ball''$ than on $\ball'$, when $x\in\mcX_r\subset\cX$; cf. \eqref{eq:sum-norm.perturb.EV.prob}.

%

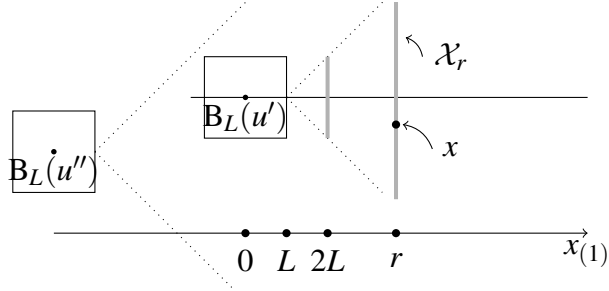
\begin{figure}
\begin{tabular}{c}
%
%
%
\begin{tikzpicture}
\begin{scope}[scale=0.18]
\clip (-25,-14) rectangle ++(70.0, 21.0);
\draw (-17, -7) rectangle ++(6, 6);
\fill (-14,-4) circle (0.2);

\draw[dotted] (-11, -4) -- (0, 7);
\draw[dotted] (-11, -4) -- (0, -15);

\draw (-3, -3) rectangle ++(6, 6);

\draw (-4, 0) -- (25, 0);
\fill (0,0) circle (0.2);

\draw[->] (-14, -10) -- (25, -10);

\node (xone) at (25, -11.5) {$x_{(1)}$};

\fill (0,-10) circle (0.3);
\fill (3,-10) circle (0.3);
\fill (6,-10) circle (0.3);
\fill (11,-10) circle (0.3);


\draw[dotted] (3, 0) -- (22, 19);
\draw[dotted] (3, 0) -- (10, -7);

\draw[color=white!70!black,line width = 1.5] (6, -3) -- (6, 3);

\draw[color=white!70!black,line width = 1.5] (11, -7.5) -- (11, 8);

\node (cXr) at (15, 3) {$\mcX_r$ };
\node (cXrp) at (10.9, 4.5) {};
\draw[->,bend right = 30] (cXr.west) to (cXrp);

\node (r) at (0, -12) {$0$};
\node (r) at (3, -12) {$L$};
\node (r) at (6, -12) {$2L$};
\node (r) at (11, -12) {$r$};

\node (ballone) at (-0, -1.5) {$\ball_L(u')$};

\node (balltwo) at (-14, -5.5) {$\ball_L(u'')$};

\fill (11, -2) circle (0.3);
\node (x) at (15, -3.8) {$x$};
\draw[->,bend right = 30] (x.west) to (11.5, -2);

\end{scope}
\end{tikzpicture}
%

\end{tabular}

\caption{  \footnotesize\emph{Example for Section \ref{ssec:max.norm.EVComp}}.
%
}
\end{figure}%
%

\section{Initial Length Scale (ILS) estimates}
\label{sec:ILS}

\subsection{ILS localization at low energies}

The first scenario leading to the onset of Anderson localization is more universal and robust than the
one considered in the next subsection; here we do not make any assumption on the magnitude of the potential
and do not attempt to achieve a global bound on the entire spectrum (which is usually possible
in discrete systems or in one dimension). This will result in ILS estimates easily
adapted to the continuous alloy models in $\DR^d$, $d\ge 1$, as well as to a large class of quantum graphs,
with tempered underlying combinatorial graphs of coupling vertices. Note that this is the scenario explored
by Bourgain and Kenig \cite{BK05} in the case of the Bernoulli potential, and later extended by Germinet and
Klein \cite{GK13}  to general alloy potentials with arbitrary marginal distributions not concentrated on a single
point\footnote{Recall that the crucial eigenvalue concentration bound used in \cite{GK13} was proved
in \cite{AGKW09}, with the help of a result by Kolmogorov \cite{Kolm58}.}.

To obtain a simple upper bound on the Green functions at the energies sufficiently close
to the lower edge of the spectrum, we assume that the interaction potential is non-negative, and that
the support of the probability measure for $\om_\bullet$ is a subset of $\DR_+$ containing $0$.
The latter can always be achieved by a shift $\om_x \rightsquigarrow \om_x + \const$. (Although such a shift
breaks the convenient condition $\esm{\om_x} \equiv 0$, the concentration inequalities remain unchanged.)
Therefore, by a standard Weyl-type argument, $\inf\, \Sigma(H(\om)) = 0$ with probability $1$.
For brevity, below we often use notations
\be
\ball = \ball_L(0), \;\; \oball = \ball_{L^\btau}(0),
\ee
where the values of $u\in\DZ^d$ and $L\in\DN$ are usually clear from the context.

Further, given a lattice subset $\Lam$,
we often make use, as before, of the decomposition $\om = (\om_\Lam, \om_\Lam^\perp)$
where $\om_\oball$ and $\om_\Lam^\perp$ are the restrictions of the configuration $\om$
on $\Lam$ and on its complement $\DZ^d\moins\Lam$, respectively.
Identifying these restrictions with their zero-extensions to $\DZ^d$, we can also write
$(\om_\Lam, \om_\Lam^\perp)= \om_\Lam+\om_\Lam^\perp$. As before,
the cumulative random potential is given by
$V(x,\om) = \BU[\om](x) = \BU[\om_\Lam](x)+ \BU[\om_\Lam^\perp](x)$.

\ble[Stable ILS estimate]
\label{lem:ILS.edges}
Let $\fu$ be one of the polynomially decaying potentials figuring in
Theorems \ref{thm:thermal.bath.F.V.polynom} and \ref{thm:smooth.DoS.pol.staircase}.
Let be given any $g>0$ and $q\in(0,1)$.
There exist $L_*\in\DN$ and $h, m, c>0$ such that for any $L_0\ge L_*$,
\be
\label{eq:lem.ILS.edges.thin.tails.V.1}
\pr{ \om_\ball: \, \inf_{\om_\ball^\perp} \;
\min_{x\in\ball}\;  g\BU\left[\om_\ball + \om_\ball^\perp\right](x) \le h L_0^{-q} }
\le \eu^{ - m L_0^{c}}.
\ee
Therefore, by positivity of $H_0$, the operator $H_0 + gV$ in $\ball_{L_0}(0)$ satisfies
\be
\label{eq:lem.ILS.edges.thin.tails.V.2}
\pr{  \om_\ball: \, \inf_{\om_\ball^\perp} \;
\dist\left[ \Sigma\left( H(\om_\ball + \om_\ball^\perp)\right), 0\right] \le h L_0^{-q} }
\le \eu^{ - m L_0^{c}}.
\ee

\ele

\proof
First of all, note that by non-negativity of the interaction potential,
$$
\all x\in\ball \;\quad
 \inf_{\om_\ball^\perp} \; \BU\left[\om_\ball + \om_\ball^\perp\right](x) =
 \BU[\om_\ball + 0](x)
$$
so the claim would follow from the estimate where $\om_\ball^\perp$ in the LHS of
\eqref{eq:lem.ILS.edges.thin.tails.V.1}--\eqref{eq:lem.ILS.edges.thin.tails.V.2}
is replaced by the zero complementary configuration $\om_\ball^\perp$; we focus on this
particular case.

Fix any $g>0$ and $q\in(0,1)$.
Let be given an integer $L_0$, and pick a smaller integer $l_0 = L_0^{q'}$, $0<q'<1$; a suitable value
$q'$ will be fixed a bit later.
There exists $h>0$ and $m'>0$ such that
$$
\pr{\max\left[ \om_y, \, {y\in\ball_{l_0}(0)} \right] < h } \le \big( \eu^{-m'} \big)^{l_0^d}
= \eu^{- m' L_0^{q'd} } \,.
$$
If $\max\left[ \om_y, \, {y\in\ball_{l_0}(x)} \right] \ge h$, then $g\BU[\om_\ball](x) \ge gh l_0^{-A}
= gh L_0^{-q'A}$, so
$$
\pr{\om_\ball:\,  \min_{x\in\ball_{L_0}(0)} g\BU[\om_\ball](x) < gh L_0^{-q'A} }
\le C L_0^d \eu^{- m' L_0^{q'd} } \,.
$$
For $L_0$ large enough and with $m = m'/2$, $q' = 2qd/A$, $c = qd/A$,
\be
\pr{\om_\ball:\,  \min_{x\in\ball_{L_0}(0)} \; \BU[\om_\ball](x)  < h L_0^{-q} }
\le \eu^{- m L_0^{c} } \,.
\ee
Now the claim follows from the non-negativity of $H_0$.
$\,$
\qedhere
\vskip2mm

Naturally, the usual Lifshitz tails type estimate also applies, since
$\sum\limits_{x\in\ball} V(x) \ge \sum\limits_{x\in\ball} \om_x$, where $\{\om_x\}$ are IID.
This covers the case of the potentials $\fu(x) = \eu^{-m |x|_p}$, $p\in [1,+\infty]$ where
the elementary approach of Lemma \ref{lem:ILS.edges} would be inefficient.

The next result fully exploits the "smooth tails" phenomenon inside a finite ball $\ball_L(0)$
without using a large annular area around it.

\ble
\label{lem:ILS.edges.2}
Let the potential $\fu(x)$,  $x\in\DZ^d$ with $d\ge 2$,
be of one of the following classes:
either $\fu(x)=\eu^{-a\ln^2 |x|}$ or
$\fu(x) = \eu^{-a|x|^\delta_p}$ with $a>0$, $p\in[1,+\infty]$, and $\delta\in(0,1)$.
Next, let be given any $g>0$ and $q>0$. For any $b>0$ there exist
$L_*\in\DN$ and $C>0$ such that for any $L\ge L_*$ and $\ball = \ball_{L_0}(0)$
\be
\label{eq:lem.ILS.edges.thin.tails.V.3}
\pr{ \om_\ball: \, \inf_{\om_\ball^\perp} \;
\min_{x\in\ball}\;  g\BU[\om_\ball + \om_\ball^\perp](x) \le L_0^{-q} }
\lea L^{-b}.
\ee
Therefore, by positivity of $H_0$, the operator $H_0 + gV$ in $\ball_{L}(0)$ satisfies
\be
\label{eq:lem.ILS.edges.thin.tails.V.4}
\pr{  \om_\ball: \, \inf_{\om_\ball^\perp} \;
\dist\left[ \Sigma\left( H\big(\om_\ball + \om_\ball^\perp\big)\right), \,  0\right] \le L_0^{-q} }
\lea L^{-b}.
\ee
\ele

\proof
Fix $b>0$.
As in the proof of the previous lemma, we start with concentration bounds for the individual
values $V(x,\om)$. In the LHS of \eqref{eq:lem.ILS.edges.thin.tails.V.3} we want to rely only
on the random amplitudes $\om_y$ with $y\in\ball$, so for fixed $x\in\ball$,
we may not have at our disposal all potentials $\om_y\fu(y-x)$ with $y$ running through entire ball
of size $\Ord{L}$ centered at $x$. For example, $x$ may be at or near the boundary of $\ball$.
However, one can always find an octant $\cO_x$ with origin $x$ delimited by hyperplanes parallel
to the coordinate hyperplanes and such that
$$
\cO_{x,L} := \cO_x \cap \ball_{L/2}(x) \subset \ball .
$$
As in the previous proof, owing to the
positivity of $\fu$, the assertion \eqref{eq:lem.ILS.edges.thin.tails.V.3}
would follow from its counterpart with $\om_\ball$ replaced with $\om_{\cO_x}$
and $\om_{\ball}^\perp$ replaced with $0$. Clearly, for any $u\in\DZ^d$
\be
\label{eq:lem.ILS.edges.thin.tails.V.4.proof}
\pr{ \om_{\cO_x}: \,  \;
\min_{x\in\ball}\;  \BU[\om_{\cO_x}](x) \le L_0^{-q} }
\le
\big| \ball \big| \cdot \max_{x\in\ball}\; \pr{ \om_{\cO_x}: \; g\BU[\om_{\cO_x}](x) \le L_0^{-q} }
\ee
so we focus on $\pr{ \om_{\cO_x}: \; \BU[\om_{\cO_x}](x) \le L_0^{-q} }$.

Denote
$\mcX_r := \myset{y\in \cO_x:\, |y-x|_p\in[r,r+1)}$, then one has $|\mcX_r| \asymp r^{d-1}$,
and therefore we can apply Lemma \ref{lem:Main} and conclude that the random variable
$\BU[\om_{\cO_x}](x)$ has probability density $\rho_x\in\mcC^\infty(\DR)$.

Decompose
$\om_{\cO_x} = \om_{\cO_{x,L}} +  \om_{ \cO_{\!\!x,L} }^\perp
$.
For the random function
$\BU[\om_{ \cO_{\!x,L} }^\perp]$ we have a deterministic sup-norm bound
$$
\big\| \BU[\om_{ \cO_{\!x,L} }^\perp] \big\|_\infty
\le  \sum_{y\in\cO_x\setminus\ball_{L/2}(x)} \fu(y-x) \, \om_y
\lea
\left\{
  \begin{array}{ll}
   \eu^{-C a \ln^2 L} , & \hbox{if $\fu(x) = \eu^{-a\ln^2 |x|}$,} \\
   \eu^{-CaL^\delta} , & \hbox{if $\fu(x) = \eu^{-a|x|^\delta_p}$.}
  \end{array}
\right.
$$
In both cases, we have an upper bound by $L^{-b(L)}$ with $b(L)\uparrow +\infty$ as $L\to\infty$.
By linearity of the alloy transform,
$\BU[\om_{\cO_{x,L}}] = \BU[\om_{\cO_x}] - \BU[\om_{\cO_{x,L}}^\perp]$, so
for all $\eps>0$
\beal
\label{eq:prob.eps.L}
\pr{ \BU[\om_{\cO_{x,L}} ] \le \eps  }
&
\le \pr{ \BU[\om_{\cO_{x}}] \le \eps + \big\| \BU[\om_{\cO_{x,L}}^\perp] \big\|_\infty }.
\eeal
With $\eps = L^{-q}$ and $L$ large enough, $ \eps + \| \BU[\om_{\cXbar_{x,L}}] \|_\infty\le 2\eps$.
It follows from \eqref{eq:prob.eps.L} and the infinite derivability of the density $\rho_x$
at $0 \le \inf\, \supp \rho_x$
that $\rho_x(E) = \ord{|E-0|^\infty}$.
Thus for $L$ large enough and \emph{any} $b'>0$
$$
\pr{ \BU[\om_{\cO_{x}}] \le 2\eps} \lea \eps^{b'} .
$$
Finally,
$$
\pr{ \min_{x\in\ball} \BU[\om_{\cO_{x}}] \le L^{-q} }  \le
\big| \ball \big| \cdot \max_{x\in\ball} \; \pr{ \BU[\om_{\cO_{x}}] \le L^{-q}} \lea L^d \, L^{-b'q} \,,
$$
and so the claim follows by setting $b' = (d+b)/q$.
$\,$
\qedhere

\bco
In the general setting of Lemma \ref{lem:ILS.edges.2}, let $d=2$ and $\delta=1$.
Then by virtue of assertion (C) of Theorem \ref{thm:thermal.bath.F.V.exp}, the results
\eqref{eq:lem.ILS.edges.thin.tails.V.3}--\eqref{eq:lem.ILS.edges.thin.tails.V.4} still hold
for $a>0$ small enough, but with $b=b(a)\uparrow+\infty$ as $a\downarrow 0$.
\eco

\subsection{ILS localization under strong disorder}

Increasing the size of the potential is well-known to increase the size of the energy zone
where Lifshitz tails asymptotics or a similar behaviour of the disordered system at hand
results in the onset of localization,
even in a continuous configuration space; see, e.g., the thorough analysis by Germinet and Klein \cite{GK03}.
We focus however on the discrete case, where the complete localization can occur, to see if
it actually does occur in infinite-range alloys with a structural, combinatorial disorder.

The ILS estimates for the strongly disordered discrete systems usually (viz., in the models with a short-range interaction
potential) do not require the initial
scale length $L_0$ to be large (contrary to the extreme energy case where the large deviations estimates
become efficient only for $L_0 \gg 1$). In fact, the smaller $L_0$, the better is usually for
the strong-disorder variant of the ILS. In a model with purely discrete structural disorder,
without continuous degrees of freedom at each source site $x\in\DZ^d$,
the uniform upper bound on the probabilities
of the events
$$
\myset{\om:\, \ball_{L_0}(0) \text{ is not $(E, \eu^{-mL_0})$-NS} }
$$
requires at least a surrogate Wegner-type estimate, hence a certain degree of continuity
of the IDS in the ball $\ball_{L_0}(0)$. Were we interested only in the eigenvalue concentration estimate for
an individual ball, it would suffice to set $L_0=0$, i.e., restrict the analysis to
single-point "balls", apply the previously established universal continuity of the
single-point effective potential (be it a.c. or s.c.), and take the coupling constant large enough.
Then the usual, very robust argument going back to \cite{DK89} would prove that
$$
\lim_{g\to \infty} \, \pr{ \ball_{L_0}(0) \text{ is not $(E,\eu^{-mL_0})$-NS, for $H_g = H_0 + gV$} } = 0.
$$
However, we need more than that: an eigenvalue concentration bound for the Hamiltonian in $\ball_{L_0}(0)$,
in order to fit into the subsequent scale induction, must
\begin{itemize}
  \item rely only on the disorder in a relatively small neighborhood of $\ball_{L_0}(0)$, and

  \item be stable under the fluctuations outside the above mentioned neighborhood of $\ball_{L_0}(0)$
        carried inside $\ball_{L_0}(0)$  by the non-local scatterer potential.
\end{itemize}
Thus some additional technical analysis is in order. Clearly, all one needs is a probabilistic bound on the event
"\emph{the minimal difference $|V(x;\om) - E|$ for $x\in\ball_{L_0}(0)$ is strictly positive}",
no matter how small that quantity actually is,
for once it is nonzero, multiplying it by $g\gg 1$ one can make it arbitrarily large.
The following result is relatively weak, but applies to a large class of potentials $\fu$.
We list only those potentials which have been examined in the previous sections, although this list can obviously
be enlarged. The bound given in Lemma \ref{lem:ILS.strong.disorder} suits
a bootstrap-type MSA scheme (cf. \cite{GK01}) based on scale-free initial probabilistic bounds.
Since we use a different approach in this paper, the proof will be omitted. Lemma
\ref{lem:ILS.strong.disorder.thin.tails.exp} is more specific, but provides stronger (scale-dependent) bounds.

\ble
\label{lem:ILS.strong.disorder}
Let $x\mapsto \fu(x)$ be one of the potentials $|x|_p^{-A}$, $\fukap(|x|_p)$, $\eu^{-a|x|_p^\delta}$ with
$p\in[1,+\infty]$.
For any $m>0$ and $\fp_0>0$ there exist $\btau \ge \tau \ge 1$, $L=L(m,\fp_0)\in\DN$ and $g_0(m,\fp_0)>0$
such that for $|g|\ge g_0(m,\fp_0)$
\be
\label{eq:lem.ILS.strong.disorder.E.V.1}
\sup_{E\in\DR} \;
\BIGpr{ \min_{x\in\ball_{L}(0)}\big| gV(x;\om) - E \big| \le \eu^{mL} \; \cond \fF_{L}(0) } \le \fp_0 \,,
\ee
where the $\sigma$-algebra $\fF_{L}(0)$ is generated by all
$x \not\in \ball_{C'L^{\btau}}(0)\setminus\ball_{C''L^{\tau}}(0)$ with some $C', C''$.
Consequently, by virtue of the min-max principle,
\be
\label{eq:lem.ILS.strong.disorder}
\sup_{E\in\DR} \;
\BIGpr{ \dist\big[ \Sigma(H_0 + gV), E \big] \le \eu^{mL} - \|H_0\| \; \cond \fF_{L}(0) } \le \fp_0 \,.
\ee

\ele

The analysis in Section \ref{sec:two-point}
allows one to establish an even stronger eigenvalue concentration estimate, based on the "regularity"
of the two-point correlation measure for the pairs $\big(V(x;\om), V(y;\om)\big)$, $x\ne y$, and
establish an analog of the result easy to prove for the IID potentials with arbitrary continuous
marginal distribution, no matter how singular:
\be
\label{lem:ILS.two.point}
\lim_{\eps\to 0} \,
\BIGpr{ \min_{\substack{x,y\in\ball_{L_0}(0)\\x\ne y}} \big| V(x,\om) - V(y,\om) \big| \le \eps } = 0.
\ee
Taking $\eps >\|H_0\|$, one can infer from \eqref{lem:ILS.two.point} a lower bound on the
spectral spacings in the ball $\ball_{L_0}(0)$, making use of the random fluctuations coming
from a finite annulus around it.

\ble[ILS estimate for exponential potentials]
\label{lem:ILS.strong.disorder.thin.tails.exp}
For any $m>0$ there exist $L_*, M,N\in\DN$ such that
for $L\ge L_*$ and some $g_0(m,L)>0$, for any $|g|\ge g_0(m,p_0)$ and some $C>0$,
denoting $\ball = \ball_{L}(0)$ and $\oball = \ball_{ML}(0)$, one has
\be
\label{eq:lem.ILS.strong.disorder.E.V.2}
\sup_{E\in\DR} \;
\pr{\om_\oball:\;  \inf_{\om_\oball^\perp} \;
\min_{x\in\ball}\big| g\BU[\om_\oball + \om_\oball^\perp](x) - E \big| \le \eu^{mL} }
    \le \eu^{- C L} \,,
\ee
where the $\sigma$-algebra $\fF_{ML,NL}(0)$ is generated by all
$x \not\in \mcA_{L, ML} := \ball_{ML}(0)\setminus\ball_{L}(0)$.
Consequently, by virtue of the min-max principle, we have for $H_\ball = -\Delta_\ball + gV_\ball$
and $L$ large enough
\be
\label{eq:lem.ILS.strong.disorder.2}
\sup_{E\in\DR} \;
\pr{\om_\oball: \; \inf_{\om_\oball^\perp} \;
\dist\big[ \Sigma\big(H_\ball\big(\om_\oball + \om_\oball^\perp\big)\big), E \big] \le \eu^{mL} }
\le \eu^{- C L} \,.
\ee

\ele

\proof
Fix $E\in\DR$ and $x\in\ball$, and note that $\ball_{(M-1)L}(x)\subset \ball_{ML}(0)\equiv \oball$.
As in the proof of Lemma \ref{lem:ILS.edges.2}, define an octant $\cO_x$ and let
$\cO_{x,L} = \cO_x\cap \oball \supset \cO_x\cap \ball_{(M-1)L}(x)$. Since in Lemma \ref{lem:ILS.edges.2}
we did not assume anything about the amplitude $g$ of the potential, except that $g\ne 0$, we still have
existence and derivability of the compactly supported probability density $\rho_x$ of
the random potential $\BU[\om_\oball](x)$, hence $\|\rho_x\|_\infty \le C' < +\infty$.
Further, with $x$ fixed (and omitted in some formulae) we have
\be
gV(x,\om) = g\BU[\om_\oball](x) + g\BU[\om_\oball^\perp](x)
=: W(\om) + \zeta(\om) \,,
\ee
where $W(\cdot)$ is $\fF_{\oball}$-measurable, and the  $\fF_{\oball}$-independent
random variable $\zeta_L$ obeys
\be
\| \zeta \|_\infty \equiv \| \zeta(\om) \|_{\rL^\infty(\Om)}
\lea \eu^{-aML}.
\ee
Take $M$ so that $C' \eu^{-aML} = \eu^{-mL}$.
$$
\bal
\sup_E \; \pr{  \inf_{\om_\oball^\perp} \; |gV(x,\om) - E| \le \eu^{mL} }
&
= \sup_E \; \pr{  \inf_{\om_\oball^\perp} \; |W(\om_\oball) - (g^{-1}E - \zeta_L)| \le g^{-1} \eu^{mL} }
\\
&
\le \sup_{E'}\;  \pr{  \inf_{\om_\oball^\perp} \; \left|\big(W(\om_\oball) - E' \big) - \zeta(\om)\right| \le g^{-1}\eu^{mL} }
\\
&
\le \sup_{E} \pr{  \left| W(\om_\oball) - E \right|
  \le g^{-1}\eu^{mL} + \| \zeta \|_\infty}
\eal
$$
In the last line we drop $\inf_{\om_\oball^\perp}\left[ \, \cdot \, \right]$, since its argument
is $\fF_\oball$-measurable, hence constant in $\om_\oball^\perp$.
Let $M\ge 2ma^{-1}$, $g\ge g_0 := \eu^{2mL}$,
then
$$
 g^{-1}\eu^{mL} + \| \zeta_L \|_\infty \le \eu^{-mL} + \eu^{-aML} \le 2\eu^{-mL} ,
$$
whence
\be
\label{eq:gS.gV.exp}
\bal
\sup_E \;
\pr{\om_\ball:\,  \inf_{\om_\oball^\perp} \;
\left|gV\big(x,\om_\oball+ \om_\oball^\perp\big) - E \right| \le \eu^{mL} }
&
\le \sup_{E} \pr{  \left| W(\om) - E \right| \le 2\eu^{-mL} }
\\
&
\le C \eu^{-mL} ,
\eal
\ee
where $C=C(x)$ is uniformly bounded in $x$,
since the random variables $W(\cdot)=W(\cdot,x)$ have uniformly bounded densities $\rho_x$,
and so
\be
\label{eq:rho.x.eps}
 \sup_{E\in\DR} \; \pr{\om_\ball:\,  \inf_{\om_\oball^\perp} \;
       \min_{x\in\ball_{L}(0)} |V(x,\om) - E| \le 2\eu^{-mL} }
    \le C'' |\ball_{L}(0)| \, \eu^{-mL}.
\ee
It follows that, e.g., for $C=m/2$ and $L$ large enough
$$
\sup_E \;  \pr{\om_\ball:\,  \inf_{\om_\oball^\perp} \;
      \min_{x\in\ball_{L}(0)}  |gV(x,\om) - E| \le 2\eu^{mL} } \le \eu^{- C L} \,.
$$
This proves the assertion \eqref{eq:lem.ILS.strong.disorder.E.V.2}
which implies  \eqref{eq:lem.ILS.strong.disorder.2} by a simple application of the min-max principle,
provided $L$ is large enough so that $2\eu^{mL} - \|\Delta\| \ge \eu^{mL}$.
$\,$
\qedhere

\ble[ILS for polynomial potentials]
\label{lem:ILS.strong.disorder.thin.tails.polynom}
Consider the potential $\fu(r) = r^{-A}$.
For any $b>0$ there exist $L_*\in\DN$, $\tau>1$  such that
for $L\ge L_*$ and some $g_0(b,L)>0$, for any $|g|\ge g_0(m,b)$ and some $C>0$
\be
\label{eq:lem.ILS.strong.disorder.E.V.3}
\sup_{E\in\DR} \;
\pr{\om_\ball:\,  \inf_{\om_\oball^\perp} \;
   \min_{x\in\ball_{L}(0)}\big| gV(x;\om) - E \big| \le 2\eu^{mL}  }
    \le L^{- b} \,,
\ee
Consequently, by virtue of the min-max principle,
\be
\label{eq:lem.ILS.strong.disorder.2.a}
\sup_{E\in\DR} \;
\pr{\om_\ball:\,  \inf_{\om_\oball^\perp} \;
    \dist\big[ \Sigma\big(H_{0,\ball_{L}(0)} + gV\big), E \big] \le \eu^{mL} }
\le L^{-b} \,.
\ee

\ele

\proof
One can  adapt the proof of the previous lemma as follows, starting with Eqn \eqref{eq:gS.gV.exp}.
We have now $\|  \zeta_L \|_\infty \le CL^{-A\tau+1}$; let $g\ge g_0 = \eu^{mL}\|\zeta_L\|^{-1}_\infty$,
so for any fixed $x\in\ball_L(0)$
\be
\label{eq:gS.gV.polynom}
\bal
\pr{  |gV_L(x) - E'| \le \eu^{mL}}
&
= \pr{  |gV(x,\om) - g\zeta_L(x,\om) - E'| \le \eu^{mL}}
\\
&
\le  \pr{   |V(x,\om) - g^{-1}E'| \le g_0^{-1} \eu^{mL} + \|  \zeta_L \|_\infty }
\\
&
\le \sup_{E} \pr{   |V(x,\om) - E| \le 2\|\zeta_L\| } \le C'' L^{-A\tau+1} \,.
\eal
\ee
By assertion (A) of Theorem \ref{thm:thermal.bath.F.V.polynom}, the r.v. $V(x,\om)$ have uniformly bounded densities
$\rho_x$, so using again Eqn \eqref{eq:rho.x.eps},
and letting $\tau\ge (b+d+1)/A$, it follows that for $L$ large enough
$$
\sup_E \; \pr{ \om_\ball:\,  \inf_{\om_\oball^\perp} \;
    \min_{x\in\ball_{L}(0)}  |gV(x,\om) - E| \le \eu^{mL} \cond \fF_{L,L^\tau} } \le L^{-b} \,.
$$
\qedhere

\subsection{Dilute alloys}

Now I would like to briefly mention a third scenario: a substitution alloy of two (possibly more) kinds of atoms
of which one (say, type I) creates lowest potential values $U^{(\rm I)}$ well-separated from all the others
and has a sufficiently low concentration. Here the discrete nature of the disorder induced by
the type I atoms plays an important role in the estimates and arguments,
while in the other scenarios a discrete (e.g., Bernoulli) disorder was merely tolerated.
Variational arguments imply in such a situation the existence of a spectral band
(or better to say, zone, for its gapped or contiguous nature requires a more thorough analysis)
emerging around  $U^{(\rm I)}$ and separated by a gap from the rest of the spectrum.
The size of the gap depends both on the magnitude of the hopping terms in the kinetic energy operator and
on the rate of decay (perhaps, also on the precise profile) of the scatterer potential; the latter
can be viewed as a perturbation of the local alloy potential with non-overlapping supports
of the individual scatterers.

More generally, one could consider several types of atoms and have type I atoms, again of low concentration,
with the potential values inside the spectrum but still well-separated from the remaining
spectral bands.

\section{Two-point correlation measures of the cumulative potential}
\label{sec:two-point}

The analysis carried out in this section is not directly applied to the
MSA schemes used in previous sections, so a number of minor technical details will be omitted
and only sketches of proofs will be given. Here we shall make only a small step towards
a functional CLT.

Consider the joint probability distribution of the values of the potential $V$ at two arbitrary distinct lattice
points; by translation invariance, we can shift one of them to $0\in\DZ^d$; let the other have the form
$\rho \Bx$ with $\rho>0$ and $\| \Bx \|=1$.

Next, consider the potential values induced at $-\rho\Bx$ and at $\rho\Bx$ by two scatterers at two
opposite lattice points
$$
\pm r\Bu, \;\; r>0, \; \|\Bu\|=1, \;\; \text{ with } (\Bu, \Bx) =: \cos \tth_{\Bu,\Bx} \ne 0, \;\;
\tth_{\Bu,\Bx} \in [0,\pi]\setminus \left\{\frac{\pi}{2} \right\}, \;\; \kap := \rho/r >0 \,.
$$
The condition $\cos \tth_{\Bu,\Bx} \ne 0$ assumes asymmetry of $\pm r\Bu$ w.r.t. the sites
$\pm\rho\Bx$ in the sense that one of the scatterers at $\pm r\Bu$ is closer to $0$
while the other is closer to $\rho\Bx$. This is necessary for the non-degeneracy of the two-point distribution.

The general strategy is as follows:

\begin{itemize}
  \item The goal is to establish continuity (which will in fact be absolute continuity)
   of the probability distribution of the random vector $\big(V(x), V(y)\big) \in\DR^2$.

  \item Varying the scatterer's position $r\Bu$ in a large annulus
 $$
\{\Bz:\,  \|\Bz - 0\| \sim \|\Bz - \rho\Bx| \in [c_1 r, c_2 r], \, r \gg \rho\}
$$
  and fixing the potential induced by all remaining scatterers,
  we can obtain a large sample of values of the vector $\big(V(x), V(y)\big)$; with $r\gg 1$, they will
  concentrate around the vector produced by the fixed (i.e., conditioned) scatterers; for simplicity
  we simply ignore this "background" potential, which results in the shift of the expectation but does not
  affect adversely regularity of the joint distribution: convolution with the independent "background" potential
  can only enhance regularity.

\item Since the active, non-conditioned random scatterer potentials have comparable amplitudes (this is why the
annulus is introduced), their convolution should obey a CLT, i.e. have an asymptotically Gaussian
distribution, albeit of small variance. For the respective covariance matrix to be non-degenerate, one
needs at least three non-aligned values in $\DR^2$, and to this end we consider pairs of
asymmetric scatterers, providing in the Bernoulli case four non-aligned points in $\DR^2$.

\end{itemize}

We focus on the Bernoulli case for simplicity, but it will be clear from the calculations
that an extension to arbitrary nontrivial distributions would not require such a radical
modification of the general approach as in \cite{AGKW09} compared to the combinatorial
argument by Bourgain and Kenig \cite{BK05} based on Sperner's lemma.

Note that
$$
\cos \tth_{-\Bu,\Bx} = - \cos \tth_{\Bu,\Bx}
$$
For brevity we omit the subscript $\Bx$ and write $\tthu$.
Let
$$
\bal
\zeta_0 &= \om_{r\Bu}(0) + \om_{-r\Bu} V_{-r\Bu}(0) \,,
\\
\zeta_{\rho \Bx} &= \om_{r\Bu} V_{r\Bu}(\rho\Bx) + \om_{-r\Bu} V_{-r\Bu}(\rho \Bx) \,,
\eal
$$
Denote $a = A/2$, $c = \cos \tthu$. Then
$$
\bal
\frac{V_{r\Bu}( r\Bu )}{V_{r\Bu}( 0)} &= \frac{r^{-2a} \|\Bu - \kap \Bx\|^{-2a} }{ r^{-2a} }
= \|\Bu - \kap \Bx\|^{-a} = \big(1 - 2\kap\cos \tth_\Bu + \kap^2 \big)^{-a}
\\
& = 1 - a (-2 c\kap + \kap^2) + \frac{a(a+1)}{2}\big( -2c\kap + \kap^2\big)^2 + O(\kap^3)
\\
& = 1 + 2ac \kap + a\big( -1 + 2(a+1)c^2 \big) \kap^2 + O(\kap^3)
\\
& = 1 + \alpha\kap + \bp\kap^2 \,, \;\qquad
  \text{ with } \alpha := 2ac = Ac, \; \bp := \big(-1 + 2(a+1)c^2\big).
\eal
$$
We have
$$
\bal
(\Bu,\Bx) &\mapsto -(\Bu,\Bx)  \Rightarrow c_\Bu \mapsto c_\Bu, \; \; \alpha \mapsto \alpha \,,
\\
(\Bu,\Bx) &\mapsto \pm(\Bu,-\Bx) \Rightarrow c_\Bu \mapsto -c_\Bu, \; \; \alpha \mapsto -\alpha \,,
\eal
$$
Thus neglecting the terms $\Ord{\kap^3}$ and denoting
$$
\bal
\alpha &:= 2ac = Ac \,,
\\
\beta &:= -1 + 2(a+1)c^2 \,,
\eal
$$
and using the symmetry $V_{r\Bu}( 0) = V_{-r\Bu}( 0)$, we obtain
$$
\bal
\frac{V_{r\Bu}( \rho\Bx)}{V_{r\Bu}( 0)} &= 1 + \alpha\kappa + \beta \kap^2
\\
\frac{V_{-r\Bu}( -\rho\Bx)}{V_{r\Bu}( 0)}
\equiv \frac{V_{-r\Bu}( -\rho\Bx)}{V_{-r\Bu}( 0)} &= 1 + \alpha\kappa + \beta \kap^2
\\
\frac{V_{r\Bu}( -\rho\Bx)}{V_{r\Bu}( 0)} &= 1 - \alpha\kappa + \beta \kap^2
\\
\frac{V_{-r\Bu}( \rho\Bx)}{V_{r\Bu}( 0)}
\equiv \frac{V_{-r\Bu}( \rho\Bx)}{V_{-r\Bu}( 0)} &= 1 - \alpha\kappa + \beta \kap^2
\eal
$$
The quantity $|V_{r\Bu}( 0)| = |a_r|$ defines therefore the common scale in which fluctuations of order $\Ord{\kap}$
or $\Ord{\kap^2}$ occur.
Let
$$
\bal
\zp &= \om_{-r\Bu} V_{-r\Bu}(+\rho\Bx) + \om_{r\Bu} V_{r\Bu}(+\rho\Bx)
\\
&
= a_r \Big[   \om_{-r\Bu} (1 - \alpha\kappa + \beta \kap^2) + \om_{r\Bu}(1 + \alpha\kappa + \beta \kap^2 )  \Big] ,
\\
\zm &=  \om_{-r\Bu}V_{-r\Bu}(-\rho\Bx) + \om_{r\Bu} V_{r\Bu}(-\rho\Bx)
\\
& = a_r \big[  \om_{-r\Bu} (1 + \alpha\kappa + \beta \kap^2) + \om_{r\Bu} (1 - \alpha\kappa + \beta \kap^2)    \big]
\eal
$$
Then the $a_r$-scaled values of $\zeta_\pm$ determined by $\om_{\pm r\Bu}$ are as follows:
\vskip7mm
\begin{center}
\begin{tabular}{|c|c|c|c|}
  \hline
\rule{0pt}{2ex}
  $\om_{-r\Bu}$ & $\om_{r\Bu}$ & $\zm/ | a_r | $ & $\zp/ | a_r | $
\rule[-1ex]{0pt}{0pt}
\\
\hline
\rule{0pt}{3ex}
  $ +1$ &  $+1$ &  $2(1 + \beta \kap^2) =: \fa$ &  $ \fa $
\rule[-1ex]{0pt}{0pt}
\\
\hline
\rule{0pt}{2ex}
   $+1$ & $-1$ &    $ 2\alpha \kap  =: \fb$   &    $  -\fb  $
\rule[-1ex]{0pt}{0pt}
\\
\hline
\rule{0pt}{2ex}
  $-1$ &  $+1$ &   $ -\fb $   &      $ \fb  $
\rule[-1ex]{0pt}{0pt}
\\
\hline
\rule{0pt}{2ex}
  $-1$ & $-1$ & $ -2\fa $ & $ -2\fa $
\rule[-1ex]{0pt}{0pt}
\\
  \hline
\end{tabular}
\end{center}
Now calculate the covariance matrix:
$$
\bal
\esm{ \zm^2} &= \frac{1}{4} \big( \fa^2 + \fb^2 + \fb^2 + \fa^2  \big) = \half( \fa^2 + \fb^2 )
\\
\esm{ \zp^2} & = \half( \fa^2 + \fb^2 )
\\
\esm{ \zm \zp } &= \frac{1}{4} \big( \fa^2 - \fb^2 - \fb^2 + \fa^2  \big) = \half( \fa^2 - \fb^2 )
\eal
$$

$$
\rC_{\zeta} = \frac{1}{2}
\left(
  \begin{array}{cc}
    \fa^2 + \fb^2 & \fa^2 - \fb^2 \\
    \fa^2 - \fb^2 & \fa^2 + \fb^2 \\
  \end{array}
\right)
\sim \left(
  \begin{array}{cc}
    \fa^2  & 0 \\
    0 &  \fb^2 \\
  \end{array}
\right)
$$

$$
\bal
\det \rC_{\zeta} = \fa^2 \fb^2
& =  4 \alpha^2 \kap^2 \cdot 4 \big( 1 + (2(a+1)c^2 - 1)\kap^2 \big)^2
\\
& = 64\, a^2 c^2 \kap^2 \cdot  \big( 1 + \Ord{\kap^2} \big)^2
\\
& = 16\, A^2 \kap^2 \cos^2 \tthu \cdot  \big( 1 + \Ord{\kap^2} \big) >0
\eal
$$
for $\cos \tthu \ne 0$ and $\kap$ small enough. For example, with $|\cos \tthu|\ge 1/\sqrt{2}$
and small $\kap$,
$$
\det \rC_{\zeta} \ge 4 A^2 \kap^2 >0 \,.
$$
For the characteristic functional, under the same restriction
$|\tthu|\le \pi/4$ and with $a_r \|\Bt\| \le \half$, $\kap = \rho/r$, we get
$$
\bal
\left| \esm{ \eu^{\ii a_r (\Bt, \Bzeta)} } \right| & \approx 1 - \frac{a_r^2}{2} \rC_{\zeta} (\Bt, \Bt)
\le 1 - 2 a_r^2 A^2  \kap^2\|\Bt \|^2
=  1 - \frac{2\rho^2}{ r^{2A+2} }\|\Bt \|^2 \,.
\eal
$$
Qualitatively, we thus have  regularity properties of the joint two-point probability distribution
similar  to those for their single-point counterparts: on spatial scale relatively large compared
to $r^{-(A+2)/2} = r^{-a-1}$, it is, approximately, at least as regular as a mixture of Gaussian
measures of variance of order $\Ord{r^{-A-2}}$, while on much smaller scales its discrete, singular
nature cannot be neglected.

The above asymptotic formula for the characteristic functional implies of course regularity
bounds for any linear functional of the random vector $\big(V(\rho\Bx,\om), V(\rho\Bx,\om)\big)$,
but one can derive the one for the difference $\eta := V(\rho\Bx,\om) - V(\rho\Bx,\om)$ directly from
the previous calculations: with precision quadratic in $\kap$,
$$
\bal
\eta &\approx
\big[ \om_{r\Bu} V_{r\Bu}(\rho\Bx) + \om_{-r\Bu} V_{-r\Bu}(\rho\Bx) \big] -
\big[ \om_{r\Bu} V_{r\Bu}(-\rho\Bx) + \om_{-r\Bu} V_{-r\Bu}(-\rho\Bx) \big]
\\
&
= \om_{r\Bu} \big[ V_{r\Bu}(\rho\Bx)  - V_{r\Bu}(-\rho\Bx)  \big]
  + \om_{-r\Bu} \big[ V_{-r\Bu}(\rho\Bx)  - V_{-r\Bu}(-\rho\Bx)  \big]
\\
&
= \om_{r\Bu} \big[ (1 + \alpha \kap + \beta \kap^2)  - (1 - \alpha \kap + \beta \kap^2)  \big]
  + \om_{-r\Bu} \big[ (1 - \alpha \kap + \beta \kap^2)  - (1 + \alpha \kap + \beta \kap^2)  \big]
\\
&
= 2\alpha \kap \big( \om_{r\Bu}  - \om_{-r\Bu}  \big)
\eal
$$
whence, again with quadratic precision in $\kap$,
$$
\bal
\ffi_\eta(t) = \esm{ \eu^{ \ii a_r\eta t}} &\approx \esm{ 1 + \ii a_r\eta t - \half a_r^2 \eta^2 t^2 }
 = 1 - 2 \cdot \frac{1}{4} \cdot \half  a_r^2 \alpha^2 \kap^2 t^2
= 1 - \frac{ A^2 c^2 t^2 }{ r^{2A+2}}
\eal
$$
With $|\tthu|\le \pi/4$, hence $c^2 = \cos^2 \tthu \ge 2$, $\kap = \rho/r \le \half$,
$$
 - \ln |\ffi_\eta(t)| \ge - \ln \left| 1 - \frac{A^2 t^2}{2 r^{2A+2}} \right|
\ge \frac{A^2}{4} \left( \frac{ t}{ r^{A+1} } \right)^2
$$

This is still good enough for some satisfactory lower bounds in probability
for the spectral spacings at an initial scale $L_0$, in a ball $\ball_{L_0}(u)$, where
in the strong disorder regime the EVs are essentially given by the values of the effective random
potential $V(\cdot\,, \om)$.

\section{Concluding remarks}
\label{sec:concluding}

\subsection{Random dipoles and random displacements}

The simplest (binary) random displacement model, with two possible scatterer's positions per cell,
is close in spirit to a random dipole model: moving the source from position $a_1$ to position $a_2$
does not change the total charge but changes the orientation of a "dipole". At a remote target
point $x$, not located on the median hyperplane for $a_1$ and $a_2$, the fluctuation of the
registered potential is due to the non-flatness of the potential amplitude $y \mapsto \fu(|y-x|)$,
and essentially equivalent to the variation of the potential at either of the source points
$a_1, a_2$, so we are still in the general framework of a Bernoulli disorder.

Naturally, any finite number of admissible source locations per cell gives rise to a similar
situation, and any continuity of the probability distribution of the source points is
very welcome for the regularity of the IDS and for localization.

\subsection{Non-homogeneous models of disorder} Such models (Delone--Anderson Hamiltonians \cite{GerMulRoM15,RoMVes13},
crooked/trimmed Hamiltonians \cite{Kle13,ElKl14}) have been quite popular in the last few years. The lack of ergodicity
is often the major technical problem in such models. Whenever the particle--media interaction
potential has infinite range, particularly in the case of slowly decaying functions (in the same sense as above),
tempered non-homogeneity has but a weak effect on the statistical properties of the
mapping $\Bfq \mapsto \BU[\Bfq](\cdot)$. The Gaussian micro-scaling remains valid (the attraction
class remains the same) unless one is allowed to put arbitrarily large number of sources per unit volume
(the same remark concerns the random displacements models).

\subsection{Random magnetic fields}

There is a vast literature on local random magnetic fields; a review can be found in \cite{C16c},
but here I would like to point out a paper by Erd\"{o}s and Hasler \cite{ErHas12} where an interesting
technique was proposed to efficiently control the contribution of random fluctuations
of magnetic fields to the local energy levels. Taking a full account of magnetic fluctuations from
remote sources seems an interesting problem. Shielding from electromagnetic fields by thin
metal films, and even by a tight grid (often preferred for its weight and flexibility) is a well-known and
widely used technique (that's why people buy these days insulated wallets for contactless cards),
but, firstly, the penetration "skin" still has a finite width; secondly, it has been demonstrated
above that "exponentially small" $\ne $ "zero"; and thirdly, the surface layer is exposed to
external long-range fields. The final word here belongs to physicists.

\subsection{Correlated sources and statistical mechanics}

The random amplitudes $\om_x$ have been assumed independent to make the probabilistic analysis simpler.
More realistically, they should be considered in the framework of a DLR (Dobrushin--Lanford--Ruelle) measure,
hence correlated
in general. Recall in this connexion that von Dreifus and Klein \cite{DK91} pointed out this class of models
and explained that a strong form of decay of correlation, emerging in spin models from the
Dobrushin--Shlosman
complete analyticity conditions, leads to Anderson localization in correlated potentials. However, it appears
that in 1980s the complete analyticity could be established only in the models with a finite spin
space \cite{Shlosman},
hence with a discrete local disorder, while the MSA on a lattice requires at least log-H\"{o}lder continuity
of finite-volume EV distributions. It seems to be a natural further step to explore the regularity
of the DoS for the Gibbsian random fields $\{\om_x, \, x \in\DZ^d\}$.

\subsection{On higher smoothness in the "frozen bath" approximations}

The asymptotical smoothness of probability distributions of properly normalized sums of random variables,
$
M_n = Z_n^{-1} \sum_{i=1}^n X_i \,,
$
has been an area of active research in probability theory since
very long time. Asymptotic expansions of the PDF and, where appropriate, probability density
using the Che\-by\-shev--Hermite polynomials go back to the 19th century (cf. Chebyshev \cite{Cheb1887}).
In early 20th century this direction was further developed by Edgeworth \cite{Edgew1905}, Bruns \cite{Bruns1906},
Charlier \cite{Charlier1905}.
Edgeworth expansions
and their counterparts for the quantiles (so-called  Cornish-Fisher expansions \cite{CF38,CF60})
are frequently used in statistics and in risk management.
A good introduction can be found in the books by Cram\'{e}r \cite{Cram37}, Feller \cite{Feller66},
and Gnedenko and Kolmogorov \cite{GKolm54}.
A classical condition allowing one to achieve a higher
accuracy of approximation of a sample distribution by the Gaussian law,
one of several bearing the name of H. Cram\'{e}r
(usually called condition (C)),
is that
$$
\limsup_{ |t| \to \infty} \big| \ffi_V(t) \big| < 1.
$$
The analysis in Sections \ref{sec:char.f.exp}--\ref{sec:char.f.polynom}
evidences that any uniform decay at infinity of the Fourier transform
$\ffi_V$ of the marginal measure $\mu_V$ would be even more welcome. However, it is known that such conditions
require a fair amount of continuity on the part of $\mu_V$. A milder condition is
that $\mu_V$ is \emph{not} supported by any affine sub-lattice $\mcA = T \DZ + a$,
$T>0$; the respective measures are called "non-lattice" distributions, and many asymptotic results
for the large sums of IID r.v.
are proved either for the lattice or for non-lattice distributions. Observe that
the non-lattice condition is already fulfilled for a measure supported by three rationally incommensurate
points $\lam_1, \lam_2, \lam_3\in\DR$.

In any event, the wealth of knowledge
accumulated in this area of probability and statistics may shed some light on accurate, asymptotically
sharp results on regularity of the DoS in finite volumes subject to the non-local fluctuations
occurring in the ambient thermal bath, which transgress the usual limits of the boundary
conditions for local finite-difference or differential operators, or even on regularity of the DoS
in a finite volume with the thermal bath temporarily "frozen" for some technical reasons (broken heater).
Although we focused here on the results one can establish even for the most singular form of
marginal disorder (Bernoulli), in a unified way and without reduction to the Bernoulli case,
it is certainly worth investigating quantitatively, to what extent a discrete, e.g. substitution
disorder is enhanced in realistic systems
by additional, even relatively weak in amplitude, sources of very "shy",
just barely continuous disorder.

\subsection{Thin tails from  neighbors }

In an old, well-known story a boy asks his father:

\vskip1mm

- \emph{Dad, a cubic meter of wood, is it much?}

- \emph{To chop, yes, quite a lot. To heat the house ... not really.}

\par\vskip2mm\noindent
Likewise, a question "\emph{Is a two-channel wire ``quasi''-one-dimensional or pretty much two-dimensional?}"
is not philosophical but quantitative, to be asked for each particular problem and application.
In this paper, I focused on the qualitative regularity properties of the cumulative potential induced
by the most singular site-wise disorder, through multiple linear convolutions, as well as on
its impact on the DoS measure. It has been demonstrated that there is a fairly universal and strong tendency
to an extreme form of regularization -- infinite derivability, and the main mechanism may only break down
in \emph{bona fide} 1D systems subject to the strongest -- exponential -- form of screening.
From \emph{this particular point of view}, even a nanotube, composed of several chains of atoms, may or
may not qualify as "sufficiently one-dimensional". Add to this the requirement of exponential screening,
and we probably end up with a purely mathematical curiosity; being not a physicist myself, I cannot
be sure such systems exist at all. If they do, it would be very interesting indeed; if not,
this would give to the regularization mechanism studied here an even greater universality. The Wegner
estimate \cite{W81},
put in simple terms, says the regularity of a short-range IID disorder is preserved in the IDS
measure. In the particular case of a bounded probability density $\rho$ of the disorder,
the DoS itself is bounded exactly by $\|\rho\|_\infty$. Mathematical works extended this
to arbitrary continuous measures $\mu$ of the IID disorder: the IDS  has (up to some constants)
the same continuity modulus $\fs_\mu$ as $\mu$.
We have seen that non-local interactions significantly improve the regularity of the underlying disorder,
even when the latter is extremely singular.
In turn, this gives rise to the "thin" tails at band edges,  complementing
the usual Lifshitz tails phenomenon.

It is well-known that for a very efficient and elegant approach to Anderson localization
developed by Aizenman and Molchanov \cite{AM93},
local regularity \emph{after} conditioning on the complement of a finite ball is simply vital. In this regard,
the present paper, unfortunately, does \emph{not} shed any light on the main hypothesis of the FMM
approach (not yet, anyway).

\vskip1mm

However, dismissing regular underlying disorder is apparently unwarranted, since some additional regularity may come from
the Gibbs distribution on the configurations of "scatterers". Note in this connection that
in the case where the marginal distribution of the scatterer amplitudes is itself Lip\-schitz continuous,
for whatever reason,
and compactly supported, the convolution effect on the edge decay of probability density for the cumulative
potential $V$ is much more immediate. For example, suppose $\om_y \sim \Unif([0,1])$, and $\fu$ has range
$\sqrt{d}$, so that only the nearest cubic "sphere" surrounding a site $x$ affects $V(x;\om)$.
Apparently, one has to have a very fertile imagination even to look for a strong screening mechanism
between nearest neighbors.
We still have
$8$ neighbors in $\DZ^2$,  so the cumulative potential has $\mcC^6$-density and edge decay $\Ord{|E - E_*|^6}$.
Respectively, in $\DZ^3$ each site has $26$ neighbors  and $\mcC^{24}$-density. Also, in a double-layer,
quasi-2D sample each site has $17$ neighbors, which shows that the dimensionality parameter
figuring in various estimates is to be determined carefully. In practice, genuinely
$2$-dimensional, mono-layer samples are rare, as well as truly single-channel linear chains. Needless to say,
the screening effects do not necessarily "switch on" sharply beyond one or two atomic distances, so the role
and universality of "thin" tails cannot be just discarded in many applications, even if for some
reasons one decided to neglect strongly (say, exponentially) screened potentials at large distances.

It seems interesting to explore possible local effects of weak screening at moderate distances
undergoing a cross-over to a much stronger one beyond some typical radius. The paradigm of an infinite
media brought to life a number of deep mathematical results and techniques,
but the recent wake of interest in physics and technology
to microscopical systems suggests one should not neglect such models either; chances are this preprint
is visualized by the reader on a quantum dot based screen.
\par\vskip7mm
\noindent
\textbf{Acknowledgements}
\par
\vskip3mm
\noindent
It is a pleasure to thank
Misha Goldshtein,
G\"{u}nter Stolz and Ivan Veseli\'c for stimulating discussions of localization mechanisms in correlated potentials,
and David Khmelnitskii for fruitful discussions of physical mechanisms of screening phenomena.

\vskip1mm

Special thanks to the organizers of the semi-annual programs \emph{Mathematics and Physics of Anderson
Localization: 50 Years After} (2008) and \emph{Periodic and Ergodic Spectral Problems} (PEP, 2015) held
at the Isaac Newton Institute, Cambridge, UK, and to the INI team for the warm hospitality and opportunity
to work in the unique, stimulating atmosphere of the Institute. The project
outlined here has been initiated during the PEP program.

Resources of the \emph{Betty and Gordon Moore Library}, of the Cambridge University, have been
invaluable for the present project as well as for the one on multi-particle Anderson localization,
initiated by Yuri Suhov and myself in 2003, also at the INI.
I am greatly indebted to the founders and generous sponsors of the Library.








\end{document}